\documentclass[11pt]{article}

\usepackage{xcolor}
\usepackage{euscript}
\usepackage{amsfonts}
\usepackage{amsbsy}
\usepackage{epsfig}
\usepackage{amsthm}
\usepackage{amscd}
\usepackage{amstext}
\usepackage{verbatim}
\usepackage{cancel}
\usepackage{capt-of}
\usepackage{subcaption}
\usepackage{empheq}
\usepackage{multirow}

\usepackage[nosort]{cite}
\usepackage{bm}
\usepackage{authblk}
\usepackage{esint}
\usepackage[T1]{fontenc}
\usepackage{mathdots}
\usepackage[centertableaux]{ytableau}

\usepackage[utf8]{inputenc}
\usepackage{graphicx}
\usepackage{physics}
\usepackage{mathtools}
\usepackage{bbold}

\usepackage{setspace}
\usepackage{colortbl}

\usepackage{amsmath}
\makeatletter
\renewcommand*\env@matrix[1][\arraystretch]{%
  \edef\arraystretch{#1}%
  \hskip -\arraycolsep
  \let\@ifnextchar\new@ifnextchar
  \array{*\c@MaxMatrixCols c}}
\makeatother
\usepackage{amssymb}
\usepackage{mathrsfs}
\usepackage{mathtools}
\usepackage{physics}
\usepackage{booktabs}
\usepackage{bbm}
\usepackage{float}
\usepackage{placeins}

\usepackage{pgfplots}
\pgfplotsset{compat=1.15}
\usepackage{tikz}
\usetikzlibrary{external}
\usetikzlibrary{arrows}

\usepackage{subcaption}
\usepackage{graphicx}
\usepackage{mathtools}
\usepackage[many]{tcolorbox}
\tcbset{shield externalize}
\usepackage{xcolor, colortbl}
\usepackage[a4paper]{geometry}
\usepackage[hidelinks,linktocpage]{hyperref}
\hypersetup{
    colorlinks=true,
    linkcolor=blue,
    citecolor=blue,
    }

\usepackage{mathrsfs}

\def\ben{\begin{equation}}
\def\een{\end{equation}}

\let\l=\lambda

\def\be{\begin{equation}}
\def\ee{\end{equation}}
\def\beq{\begin{equation}}
\def\eeq{\end{equation}}
\def\ba{\begin{array}}
\def\ea{\end{array}}

\def\dalemb#1#2{{\vbox{\hrule height .#2pt
       \hbox{\vrule width.#2pt height#1pt \kern#1pt
               \vrule width.#2pt}
       \hrule height.#2pt}}}

\newcommand{\bea}{\begin{eqnarray}}
\newcommand{\eea}{\end{eqnarray}}

\makeatletter
\newcommand*\bigcdot{\mathpalette\bigcdot@{.5}}
\newcommand*\bigcdot@[2]{\mathbin{\vcenter{\hbox{\scalebox{#2}{$\m@th#1\bullet$}}}}}
\makeatother

\renewcommand{\Bar}[1]{\overline{#1}}

\usepackage{tocloft}

\title{Rotational instability and quadrupolar order\\ in BFSS matrix theory}
\author{Jun~Liu and Jorge~E.~Santos}
\affil{\it Department of Applied Mathematics and Theoretical Physics, \\
\it University of Cambridge, Wilberforce Road, Cambridge CB3 0WA, UK
}
\date{}

\begin{document}

\maketitle

\begin{abstract}
We study the effect of angular momentum on thermal states of the BFSS matrix theory in its gravitational and high-temperature classical regimes. We focus on ensembles with only one non-zero angular momentum. On the gravity side, the rotating ensemble is described by the singly-spinning D$0$ black hole. Beyond a critical rotation, we find that this solution becomes unstable to a non-axisymmetric quadrupolar perturbation. Following this mode non-linearly, we find a new branch of non-axisymmetric rotating black holes that is thermodynamically favoured over the axisymmetric branch in both the microcanonical and canonical ensembles. On the other hand, in the classical matrix limit, we find no evidence for such collective quadrupolar order over the matrix sizes and angular momenta studied. Instead, rotation accelerates the separation of a single D$0$-brane from the clump of branes initially prepared. The separated brane carries a large fraction of the angular momentum by orbiting the rest of the clump at a large in-plane radius. Both the angular momentum share and the orbital radius relative to the clump size increase with the matrix size. These results distinguish a collective quadrupolar deformation in the gravity limit from angular-momentum transfer to a single brane in the classical matrix limit.
\end{abstract}

\newpage

\tableofcontents

\section{Introduction}\label{sec:intro}\label{sec:review}
The BFSS matrix quantum mechanics was proposed as a non-perturbative description of M-theory~\cite{Banks:1996vh}. It also describes the low-energy dynamics of coincident D$0$-branes, and its thermal states have a gravitational description in terms of black D$0$-branes at large $N$ and strong coupling~\cite{Itzhaki1998}.\footnote{For the original AdS/CFT correspondence and the prescription for boundary correlation functions, see~\cite{Maldacena:1997re,Gubser:1998bc,Witten:1998qj}.} This provides a concrete setting in which to study black-hole physics directly in a quantum theory. Here we study the phase structure of thermal states with non-zero angular momentum. In particular, we distinguish a collective deformation of a rotating bound state from the separation of a single brane carrying part of its angular momentum. We study both effects using gravity at strong coupling and classical matrix dynamics at high temperature, with the quadrupole as a common observable.

Recent work has made the relation between BFSS phases and higher-dimensional black objects more systematic~\cite{DiasLowEnergy2025,Dias:2025jxx}. A Carrollian transformation relates eleven-dimensional vacuum black objects with Kaluza--Klein asymptotics to geometries with the D$0$-brane asymptotics appropriate to BFSS, with the compact identification and normalisations chosen to fix the matrix-theory parameters. Applied to static seeds, this construction yields non-uniform and localized BFSS phases. The uniform BFSS phase also develops a Gregory--Laflamme instability~\cite{Gregory:1993vy}, established directly in the D$0$-brane geometry~\cite{Dias:2025jxx}. This suggests using the known phase structure of Kaluza--Klein gravity to guide the search for further BFSS states, including those carrying angular momentum.

A natural rotating seed is a ten-dimensional singly spinning Myers--Perry black hole~\cite{MyersPerry1986} times a circle. In six or more spacetime dimensions, singly spinning Myers--Perry black holes can carry arbitrarily large angular momentum at fixed mass. Their horizons spread out in the rotation plane at large spin, suggesting a relation to the instability of black membranes~\cite{EmparanMyers2003}. Stationary axisymmetric perturbations were found at the ultraspinning onset, indicating bifurcations to new families with deformed horizons~\cite{Dias2009,Dias2010Ultraspinning}. Non-axisymmetric bar-mode instabilities of singly spinning black holes were found in numerical evolutions~\cite{Shibata:2010wz}; a linear calculation subsequently determined their quasinormal spectrum and the timescales of the bar-mode and ultraspinning instabilities~\cite{Dias:2014eua}. Equal-spin Myers--Perry black holes also admit growing gravitational modes~\cite{Dias2010Instability}, with non-axisymmetric bar modes becoming unstable before the axisymmetric modes~\cite{Hartnett2013}. Nonlinear evolutions of singly spinning black holes exhibit substantial deformation and gravitational-wave emission~\cite{Bantilan2019}. These results motivate extending the BFSS construction to rotating backgrounds and asking whether analogous bifurcations occur.

The map between stationary backgrounds does not imply an equivalence of their dynamical stability problems. It mixes time and the longitudinal coordinate, so an unstable mode of the Kaluza--Klein seed need not respect the periodicity of the BFSS circle~\cite{DiasLowEnergy2025}. The Myers--Perry instabilities therefore motivate a search, but their thresholds and growth rates cannot simply be carried over. We must determine which perturbations satisfy the BFSS boundary conditions and whether they admit a nonlinear continuation with the same asymptotic sources.

We work with the interacting $SU(N)$ sector of BFSS, with the free centre-of-mass multiplet removed. Its action is \cite{Banks:1996vh}
\be\label{eq:BFSS}
S = \frac{N}{\lambda} \int  {\rm d}t \, \text{Tr} \left(\frac{1}{2} D_t X^i D_t X_i + \frac{1}{4} [X^i, X^j][X_i, X_j] + \frac{1}{2} \Psi^\alpha D_t \Psi_\alpha + \frac{i}{2} \Psi^\alpha \gamma^{i}_{\alpha \beta} [X_i, \Psi^\beta]\right),
\ee
where $X_i$ are nine $N$-by-$N$ Hermitian traceless matrices and $\Psi$ sixteen matrices of the same type but with Grassmann-odd entries. The action has a global $SO(9)$ symmetry under which $X^i$ and $\Psi^\alpha$ transform as a vector and a Majorana spinor, respectively. Both $i,j = 1,\ldots,9$ and $\alpha, \beta = 1,...,16$ are raised and lowered by the Kronecker delta. $D_t \equiv \partial_t + i [A, \cdot]$, with $A$ being the gauge potential associated with the $SU(N)$ gauge symmetry. The 't Hooft coupling $\lambda$ has dimension $[\text{energy}]^3$. We denote the thermal energy by $E$, with the supersymmetric ground-state energy set to zero. At zero angular momentum and finite temperature $T$, the only continuous parameter is a dimensionless temperature (or equivalently an inverse coupling) $\tau = T/\lambda^{1/3}$. The effective type IIA supergravity description emerges in the strongly-coupled 't Hooft limit: taking $N \to \infty$ keeping $\tau$ fixed, and then $\tau \ll 1$. On the opposite end, $N \to \infty$ with $\tau$ fixed and $\tau \gg 1$ is the weakly coupled regime. The matrix dynamics is, to leading order, well-described by the classical dynamics of the bosonic matrices. In the microcanonical ensemble, a similar limit should be taken with the temperature replaced by the energy scale.

We restrict to angular momentum in the $(12)$-plane. It is the conserved charge $J=J^{12}$ of the corresponding $SO(2)$ subgroup of $SO(9)$; its bosonic contribution is
\begin{equation}
 J_{\rm bos}=\operatorname{Tr}(X^1P^2-X^2P^1),
 \qquad P_i=\frac{N}{\lambda}D_tX_i.
 \label{eq:introangularmomentum}
\end{equation}
The full quantum charge also contains the fermionic spin contribution. On the gravity side, $J$ corresponds to rotation in the transverse space. To describe a collective deformation, we use the gauge-invariant symmetric traceless tensor
\begin{equation}
 \mathcal O^{ij}=\operatorname{Tr}\!\left(X^iX^j-\frac{\delta^{ij}}9X^kX_k\right),
 \label{eq:quadrupoletensor}
\end{equation}
which transforms in the $\boldsymbol{44}$ of $SO(9)$~\cite{Bovon2025}. Its charge-two component in the rotation plane is the complex quadrupole
\begin{equation}
 \mathcal O_2=\mathcal O^{11}-\mathcal O^{22}+2{\rm i}\mathcal O^{12}
 =\operatorname{Tr}\!\left[(X^1+{\rm i}X^2)^2\right].
 \label{eq:quadrupole}
\end{equation}
Here the rotation is the spatial $SO(2)$ transformation in the $(12)$-plane, generated by $J^{12}$. It acts on the matrix-valued coordinates as
\[
\begin{aligned}
 X^1&\mapsto \cos\vartheta\,X^1-\sin\vartheta\,X^2,\\
 X^2&\mapsto \sin\vartheta\,X^1+\cos\vartheta\,X^2,
\end{aligned}
\]
with $X^3,\ldots,X^9$ unchanged. This mixes the spatial labels $1,2$ and leaves the $N\times N$ matrix indices unaffected. Consequently, $X^1+{\rm i}X^2\mapsto e^{{\rm i}\vartheta}(X^1+{\rm i}X^2)$ and $\mathcal O_2\mapsto e^{2{\rm i}\vartheta}\mathcal O_2$. Physically, this rotates the brane configuration within the plane carrying the angular momentum. A non-zero expectation value in a co-rotating frame selects an orientation and leaves rotations by $\pi$ unbroken. The other scalar retained below is neutral under this rotation. Up to an overall normalisation, its dual operator is
\begin{equation}
 \mathcal O_0=\operatorname{Tr}\!\left[(X^1)^2+(X^2)^2
 -\frac27\sum_{i=3}^9(X^i)^2\right].
 \label{eq:neutralquadrupole}
\end{equation}
It measures the relative extent of the rotation plane and the seven transverse directions, and can be non-zero even in an axisymmetric state. In contrast, $\mathcal O_2$ detects the loss of rotational symmetry within the plane. These charged expectation values refer to a large-$N$ saddle with a chosen orientation; averaging over orientations restores the symmetry. On the matrix side we therefore measure a rotationally invariant quadrupole magnitude. A separated brane can also make this observable large, even if the remaining clump is nearly axisymmetric. To distinguish this from a collective deformation, we analyse single-clump and emitted configurations separately and measure the angular momentum carried by the outlier.

The gravity dual is given by type IIA string theory in the near-horizon limit of a near-extremal black hole carrying $N$ units of D$0$ charge. With zero angular momenta, working with the uplift of the geometry to eleven-dimensional supergravity, the metric is \cite{Costa:2014wya}
\be
\dd s^2_{\rm 11D} = - \frac{r^7}{\mathcal{R}^7}f \dd t^2 + \frac{\dd r^2}{f} + \frac{\mathcal{R}^7}{r^7} \left(\dd z + \frac{r^7}{\mathcal{R}^7} f \dd t \right)^2 + r^2 \dd \Omega_8^2,
\ee
with 
\be
f = 1 - \frac{2 M_0}{r^7}, \quad z \sim z + 2 \pi L_{11}.
\ee
We use $L_{11}$ for the radius of the M-theory circle, whose circumference is $2\pi L_{11}$. In this convention, the matching between matrix and gravity parameters is
\begin{align}
 \lambda&=\frac{(2\pi)^{5/3}}{240}\frac{\mathcal R^7(2\pi L_{11})^5}{G_{11}^{5/3}},
 \qquad N=\frac{\pi^2}{15}\frac{\mathcal R^7(2\pi L_{11})^2}{G_{11}},\nonumber\\
 \tau^{9/5}&=\frac{7^{9/5}30^{3/5}}{(2\pi)^{14/5}}
 \frac{(2M_0)^{9/14}G_{11}}{\mathcal R^{21/2}(2\pi L_{11})^3}.
 \label{eq:BFSSdictionary}
\end{align}
where $G_{11}$ is the eleven-dimensional Newton's constant. The non-rotating geometry is locally equivalent to a black string with a ten-dimensional Schwarzschild factor~\cite{DiasLowEnergy2025,Dias:2025jxx}. With one non-zero angular momentum, the axisymmetric uplift is locally a ten-dimensional singly spinning Myers--Perry black hole times a line, with the compact identification transformed as well. It is just the 11D uplift of the 10D spinning D$0$-brane solution in the decoupling limit~\cite{Harmark:1999xt,Anabalon2013} and its metric is given explicitly in Appendix~\ref{app:11d}. The dual of $\mathcal{O}_2$ is then an eleven-dimensional metric mode with charge two under the $U(1)$ rotation generated by the one non-zero angular momentum.

To follow this deformation nonlinearly, we use a two-dimensional sector of $SO(9)$ gauged supergravity~\cite{Ortiz2013,Anabalon2013} containing the metric, dilaton, a neutral scalar, a gauge field, and a charged scalar. The gauge field is dual to $J$, while the neutral and charged scalars are dual to $\mathcal O_0$ and $\mathcal O_2$, respectively. We label each angular harmonic by an integer $m_\varphi$, defined by the phase $e^{{\rm i}m_\varphi\vartheta}$ it acquires under a rotation by $\vartheta$. The charged scalar uplifts to the quadrupolar harmonic, $m_\varphi=2$. The two-dimensional description retains its nonlinear interactions while reducing the construction to a radial boundary-value problem. Although the condensate is stationary in the co-rotating two-dimensional frame, its eleven-dimensional uplift depends on the local Myers--Perry time and longitudinal coordinate through a helical phase. The nonlinear branch must therefore be constructed with BFSS boundary conditions; it is not obtained by simply mapping an ordinary stationary Kaluza--Klein black hole. We emphasize that the nonlinear solutions still preserve translation symmetry along the M-theory circle in the BFSS frame, so the solution can be reduced to a solution within type IIA supergravity \footnote{Below a temperature scale $\tau \sim N^{-10/21}$ at fixed dimensionless spin and deformation, the resulting solution is not validly described by classical type IIA supergravity due to large dilaton values at the horizon. We choose to always work in eleven-dimensions, so that our results also hold in the regime where the 11D uplift is valid but not type IIA.}.

To describe spontaneous quadrupolar order in the undeformed BFSS model, we must allow a non-zero expectation value of $\mathcal O_2$ without an external source for this operator. We therefore determine which coefficients in the large-radius scalar expansions represent sources and which determine expectation values. For both scalars, the source is the coefficient of the faster-decaying term, while the slower-decaying term determines the expectation value. This choice is known as alternative quantisation. We give the two-dimensional equations and their eleven-dimensional uplift allowing time dependence in all fields. The boundary terms used to define finite energy and angular momentum also allow time-dependent boundary data. For the stationary black holes, these charges satisfy the first law.

\begin{samepage}
At fixed energy, we find a quadrupolar zero mode at a critical value of the dimensionless angular momentum
\begin{equation}
 \mathfrak j\equiv\frac{|J|/N^2}{\left[E/(N^2\lambda^{1/3})\right]^{9/14}},
 \qquad \mathfrak j_{\rm c}\simeq0.455059.
 \label{eq:introcriticalspin}
\end{equation}
\end{samepage}
A time-dependent calculation with ingoing behaviour at the horizon and no external quadrupolar source shows that this mode grows exponentially in time for $\mathfrak j>\mathfrak j_{\rm c}$. We also perform a gravitational perturbation calculation directly in eleven dimensions. Among the sectors $m_\varphi=2,\ldots,7$ that we have resolved, the quadrupole has the lowest onset spin. We continue it into the nonlinear regime and find that the deformed solutions have higher entropy than the axisymmetric solution at the same energy and angular momentum.\footnote{A similar bifurcation from rotating plasma balls to non-axisymmetric lobed configurations was found in~\cite{Cardoso:2009bv,Cardoso:2009nz}. This is an analogy between different systems: those works use the hydrodynamic description of Scherk--Schwarz holographic plasmas, whereas here the order parameter measures a quadrupole in the transverse matrix directions.} The entropy excess reaches $6.43\%$ over the computed branch, with no observed turnover. The same branch has lower free energy than the matched axisymmetric solution at fixed temperature and angular momentum, but does not dominate over the lower-spin axisymmetric reference at fixed temperature and angular potential, i.e. the grand canonical ensemble. These comparisons concern the computed families; determining the dynamical stability and global dominance of the new branch remains a separate problem.

In the high-temperature classical regime, we find a different way to redistribute angular momentum. The normalised quadrupole magnitude of the single-clump samples decreases with $N$, with no evidence for a collective quadrupolar condensate over the matrix sizes and spins studied. Once a brane separates, the quadrupole becomes large and is dominated by the outlier. The separated brane carries an increasing fraction of the total angular momentum as $N$ increases, reaching approximately $90\%$ in the samples with $N=50$. Its radial separation and angular momentum are consistent with an orbiting brane rather than a collective deformation of the clump. These measurements allow a comparison of the quadrupolar response in the two regimes, while keeping the contribution of a separated brane distinct from that of the bound state.

To measure the time scale for separation, we evolve thermal single-clump initial conditions in real time. Each trajectory conserves its energy and angular momentum, and we record its first sustained exit from the single-clump region. The survival probabilities are consistent with an exponential law over the observed interval, and the lifetime decreases rapidly with angular momentum. This first-emission time is a different observable from the growth rate of the collective gravitational mode. The results do not establish a transition between the two mechanisms as the coupling is varied: the classical calculation uses finite $N$, and the physical parameters and order of limits must be matched before making such a comparison. For completeness, we also use a scalar wave carrying momentum along the M-theory circle to probe D0-brane emission from the gravitational background. Some frequencies allow the wave to extract energy from the rotating horizon and propagate to infinity, a process known as superradiance. Even in this range, however, the effective radial potential can strongly suppress transmission at large $N$.

The remainder of the paper is organised as follows. In Section~\ref{sec:gravity} we present the two-dimensional theory, its uplift, the quadrupolar zero mode, and the holographic renormalisation. Section~\ref{sec:gravityresults} presents the growing quadrupolar mode, the nonlinear gravitational branch, its thermodynamic comparisons, horizon geometry and numerical convergence. Section~4 describes the classical matrix ensemble and the measurement of lifetime with results shown in Section~\ref{sec:matrixresults}. The eleven-dimensional perturbation calculation, the classical matrix equation of state, metadata and bias tests of the matrix results, and the gravitational transmission problem are included in the Appendix.

\section{The gravity side}\label{sec:gravity}
\subsection{The two-dimensional theory and choice of frame}
We use mostly plus signature and denote the gauge coupling by $g=\mathcal{R}^{-1}$. We find it convenient to distinguish two conformal frames. Let $g^{\rm C}_{ab}$ be the metric in which the dilaton has no explicit kinetic term, and let
\begin{equation}
 g^{\rm D}_{ab}=\rho^{-4/9}g^{\rm C}_{ab}
 \label{eq:frames}
\end{equation}
be the dual-frame metric, which approaches AdS$_2$. Unless indicated otherwise, contractions in this section use $g^{\rm C}_{ab}$. The action is
\begin{multline}
 S_{2\mathrm D}=\frac{1}{2\kappa_{2\mathrm D}^2}
 \int_{\mathcal M}{\rm d}^2x\sqrt{-g^{\rm C}}\,\rho\left[
 \frac14 R^{\rm C}-\frac9{14}(\nabla v)^2
 -\frac{1+\lambda_\Phi}{2}\mathcal D_a\Phi(\mathcal D^a\Phi)^\dagger\right.
 \\
 \left.+\frac18\frac{2+\lambda_\Phi}{1+\lambda_\Phi}(\nabla\lambda_\Phi)^2
 -\frac1{16}\rho^{4/9}e^{4v}F_{ab}F^{ab}
 -\rho^{-4/9}V(v,\lambda_\Phi)\right],
 \label{eq:action2D}
\end{multline}
where $F={\rm d}A$, $\lambda_\Phi=|\Phi|^2$, $\mathcal D_a\Phi=(\nabla_a+2{\rm i}gA_a)\Phi$, and
\begin{equation}
 V=-\frac{g^2}{8}e^{-4v}
 \left[35e^{36v/7}-16\lambda_\Phi(1+\lambda_\Phi)
 +28e^{18v/7}(1+2\lambda_\Phi)\right].
 \label{eq:potential}
\end{equation}
The notation $\lambda_\Phi$ distinguishes the scalar norm from the matrix-model 't Hooft coupling $\lambda$. The dynamical fields are $g^{\rm C}_{ab}$, $\rho$, $v$, $A_a$, and $\Phi$. Although it appears explicitly in the action, $\lambda_\Phi$ is not an additional field.

To give the full field equations without fixing the charged-scalar phase, write $\Phi=\psi_1+{\rm i}\psi_2$, with $\lambda_\Phi=\psi_i\psi_i$. The real doublet has covariant derivative
\begin{equation}
 \mathcal D_a\begin{pmatrix}\psi_1\\\psi_2\end{pmatrix}
 =\nabla_a\begin{pmatrix}\psi_1\\\psi_2\end{pmatrix}
 +2gA_a\begin{pmatrix}-\psi_2\\\psi_1\end{pmatrix}.
 \label{eq:realscalarderivative}
\end{equation}
Its kinetic term is $-\rho\mathsf M_{ij}\mathcal D_a\psi_i\mathcal D^a\psi_j/2$, where
\begin{equation}
 \mathsf M_{ij}=(1+\lambda_\Phi)\delta_{ij}
 -\frac{2+\lambda_\Phi}{1+\lambda_\Phi}\psi_i\psi_j.
 \label{eq:scalartargetmetric}
\end{equation}
It is convenient to define
\begin{equation*}
 \mathcal K_{ab}=\frac97\nabla_av\nabla_bv
 +\mathsf M_{ij}\mathcal D_a\psi_i\mathcal D_b\psi_j,
 \qquad \mathcal K=g_{\rm C}^{ab}\mathcal K_{ab}.
\end{equation*}
Varying Eq.~\eqref{eq:action2D} with respect to the metric, dilaton, neutral scalar, charged scalar and gauge field gives, respectively,
\begin{subequations}\label{eq:field2D}
\begin{equation}
 0=g^{\rm C}_{ab}\Box\rho-\nabla_a\nabla_b\rho
 -2\rho\left(\mathcal K_{ab}-\frac12g^{\rm C}_{ab}\mathcal K\right)
 -\frac12\rho^{13/9}e^{4v}
 \left(F_{ac}F_b{}^c-\frac14g^{\rm C}_{ab}F^2\right)
 +2\rho^{5/9}g^{\rm C}_{ab}V,
 \label{eq:metric2D}
\end{equation}
\begin{equation}
 0=R^{\rm C}-2\mathcal K
 -\frac{13}{36}\rho^{4/9}e^{4v}F^2
 -\frac{20}{9}\rho^{-4/9}V,
 \label{eq:dilaton2D}
\end{equation}
\begin{equation}
 0=\frac97\nabla_a(\rho\nabla^av)
 -\frac14\rho^{13/9}e^{4v}F^2
 -\rho^{5/9}\partial_vV,
 \label{eq:neutral2D}
\end{equation}
\begin{equation}
 0=\mathcal D_a\!\left(\rho\mathsf M_{ij}\mathcal D^a\psi_j\right)
 -\frac\rho2\partial_{\psi_i}\mathsf M_{jk}
       \mathcal D_a\psi_j\mathcal D^a\psi_k
 -\rho^{5/9}\partial_{\psi_i}V,
 \label{eq:charged2D}
\end{equation}
\begin{equation}
 \nabla_a\!\left(\rho^{13/9}e^{4v}F^{ab}\right)
 =8g\rho(1+\lambda_\Phi)
 \operatorname{Im}\!\left(\Phi^*\mathcal D^b\Phi\right).
 \label{eq:Maxwell2D}
\end{equation}
\end{subequations}
Here $F^2=F_{ab}F^{ab}$, $\Box=\nabla_a\nabla^a$, and all contractions and covariant derivatives use the canonical-frame metric. The derivative in Eq.~\eqref{eq:charged2D} acts on the real doublet as in Eq.~\eqref{eq:realscalarderivative}, and $\partial_{\psi_i}V=2\psi_i\partial_{\lambda_\Phi}V$. Although the Einstein tensor vanishes identically in two dimensions, Eq.~\eqref{eq:metric2D} is non-trivial because of the dilaton-curvature coupling. These equations allow arbitrary dependence on both two-dimensional coordinates. Linearising Eq.~\eqref{eq:charged2D} about $\Phi=0$ gives the charged-scalar equation used below.

Under Eq.~\eqref{eq:frames}, the bulk action becomes, up to a boundary term,
\begin{multline}
 S_{2\mathrm D}=\frac{1}{2\kappa_{2\mathrm D}^2}
 \int_{\mathcal M}{\rm d}^2x\sqrt{-g^{\rm D}}\,\rho
 \left[\frac14R^{\rm D}+\frac{(\nabla\rho)^2}{9\rho^2}
 -\frac9{14}(\nabla v)^2
 -\frac{1+\lambda_\Phi}{2}|\mathcal D\Phi|^2\right.
 \\
 \left.+\frac18\frac{2+\lambda_\Phi}{1+\lambda_\Phi}(\nabla\lambda_\Phi)^2
 -\frac{e^{4v}}{16}F^2-V\right].
 \label{eq:dualaction}
\end{multline}
The different powers of $\rho$ in the Maxwell and potential terms are consequences of the frame transformation. In the following we use the canonical frame for the radial equations, boundary terms and uplift.

For the stationary sector we take
\begin{equation}
 {\rm d}s_{\rm C}^2=-f(r){\rm d}t^2+\frac{{\rm d}r^2}{p(r)},
 \qquad A=A_t(r){\rm d}t,\qquad \Phi=\Phi^\dagger=\Phi(r).
 \label{eq:stationary}
\end{equation}
The phase of the scalar has been fixed by a gauge transformation. It is convenient to use
\begin{equation}
 H=e^{-18v/7},\qquad
 f=\frac{g^7r^7G}{H^{7/9}},\qquad
 p=\frac{G}{H^{2/9}},\qquad
 \rho=\frac{g^{9/2}r^{9/2}\mathcal Z}{\sqrt H}.
 \label{eq:radialgauge}
\end{equation}
Here $\mathcal Z(r)$ is a radial dilaton function, distinct from the angular function appearing below. This parametrisation fixes the radial coordinate while allowing the dilaton to respond to the charged scalar.

\subsection{The eleven-dimensional uplift}
We now give the uplift for a general two-dimensional solution of Eq.~\eqref{eq:action2D}. We impose neither stationarity nor a choice of coordinates on the two-dimensional metric. The uplift is purely gravitational, with $F_4=0$, and preserves an $SO(7)$ action on a unit $S^6$. It is the corresponding sector of the general $SO(9)$ uplift of~\cite{Bossard:2023jid}.

All Hodge duals below use the canonical-frame metric $g^{\rm C}_{ab}$. We choose $\mathrm{vol}_2$ so that $\star_2\mathrm{vol}_2=-1$. Define
\begin{equation}
 H=e^{-18v/7},\qquad
 s_\Phi=\sqrt{1+\lambda_\Phi},\qquad
 c_\Phi=1+2\lambda_\Phi,\qquad
 \vartheta={\rm d}\phi+gA,
 \label{eq:covariantupliftfields}
\end{equation}
where $0\leq x\leq1$ and $\phi\sim\phi+2\pi$ are internal coordinates. The combinations
\begin{equation}
 U_\phi=\operatorname{Re}\!\left(e^{-2{\rm i}\phi}\Phi\right),
 \qquad
 V_\phi=-\operatorname{Im}\!\left(e^{-2{\rm i}\phi}\Phi\right)
 \label{eq:covariantangularscalar}
\end{equation}
are invariant under $A\mapsto A+{\rm d}\Lambda$, $\Phi\mapsto e^{-2{\rm i}g\Lambda}\Phi$ and $\phi\mapsto\phi-g\Lambda$. The angular functions are
\begin{align}
 \Delta&=x^2+(1-x^2)H\left(c_\Phi-2s_\Phi U_\phi\right),\nonumber\\
 Z&=(1-x^2)H+x^2\left(c_\Phi+2s_\Phi U_\phi\right).
 \label{eq:angularwarp}
\end{align}
The metric takes the form
\begin{equation}
 {\rm d}s_{11\mathrm D}^2=
 \frac{\Delta}{H^{2/9}}{\rm d}s_{\rm C}^2
 +\frac{(1-x^2)\rho^{4/9}\Delta}{g^2H^{7/9}Z}\vartheta^2
 +\frac{\rho^{4/9}Z}{g^2(1-x^2)H^{7/9}}E_x^2
 +\frac{H^{2/9}}{\Delta\rho^{14/9}}E_z^2
 +\frac{x^2H^{2/9}\rho^{4/9}}{g^2}{\rm d}\Omega_6^2,
 \label{eq:uplift}
\end{equation}
with
\begin{equation}
 E_x={\rm d}x+
 \frac{2x(1-x^2)s_\Phi V_\phi}{Z}\vartheta.
 \label{eq:Ex}
\end{equation}
To specify the circle fibre, introduce the electric scalar and charged one-form
\begin{equation}
 \mathcal Q=\frac{\rho^{13/9}}{2g^2H^{14/9}}\star_2F,
 \qquad
 \mathcal J_\Phi=c_\Phi\mathcal D(s_\Phi\Phi)
              -s_\Phi\Phi\,{\rm d}c_\Phi,
 \label{eq:upliftcurrent}
\end{equation}
where $\mathcal D(s_\Phi\Phi)=({\rm d}+2{\rm i}gA)(s_\Phi\Phi)$. Then
\begin{equation}
 E_z={\rm d}z+\mathcal B+(1-x^2)\mathcal Q\,\vartheta
 -\frac{x^2\rho}{2g}\star_2{\rm d}\log H
 -\frac{(1-x^2)\rho}{g}
 \operatorname{Re}\!\left(e^{-2{\rm i}\phi}\star_2\mathcal J_\Phi\right),
 \label{eq:Ez}
\end{equation}
where the two-dimensional one-form $\mathcal B$ obeys
\begin{equation}
 {\rm d}\mathcal B=
 -g\rho^{5/9}H^{5/9}
 \left[7c_\Phi-8H\lambda_\Phi(1+\lambda_\Phi)\right]\mathrm{vol}_2
 -g\mathcal Q F.
 \label{eq:circlecurvature}
\end{equation}
This first-order equation introduces no additional propagating field. Locally, $\mathcal B\mapsto\mathcal B+{\rm d}\eta$ is accompanied by $z\mapsto z-\eta$. The metric is invariant under both this circle-coordinate transformation and the two-dimensional $U(1)$ gauge transformation. The fields $g^{\rm C}_{ab}$, $\rho$, $v$, $A$ and $\Phi$ can all depend on both two-dimensional coordinates.

For completeness, the scalar dictionary with the $SL(9)/SO(9)$ description is
\begin{equation}
 \begin{gathered}
 \mathsf m=\operatorname{diag}\!\left(e^{2v}\mathsf P^{-1},
                            e^{-4v/7}\mathbbm 1_7\right),\\
 \mathsf P=\begin{pmatrix}
 c_\Phi-2s_\Phi\operatorname{Re}\Phi & -2s_\Phi\operatorname{Im}\Phi\\
 -2s_\Phi\operatorname{Im}\Phi & c_\Phi+2s_\Phi\operatorname{Re}\Phi
 \end{pmatrix}.
 \end{gathered}
 \label{eq:upliftdictionary}
\end{equation}
We have $\det\mathsf P=\det\mathsf m=1$, and all three-form axions vanish. Substitution into the scalar kinetic term and potential of the parent theory reproduces Eq.~\eqref{eq:action2D}, after eliminating the auxiliary electric field and making the frame transformation in Eq.~\eqref{eq:frames}.

For the stationary ansatz~\eqref{eq:stationary}, $U_\phi=\Phi\cos2\phi$, $V_\phi=\Phi\sin2\phi$ and
\begin{equation}
 \mathcal J_\Phi=
 \frac{\Phi'}{\sqrt{1+\Phi^2}}\,{\rm d}r
 +2{\rm i}gA_t\Phi\sqrt{1+\Phi^2}(1+2\Phi^2)\,{\rm d}t.
 \label{eq:stationaryupliftcurrent}
\end{equation}
The stationary expressions used in our numerical construction follow by setting
\begin{equation}
 \mathcal B=(B_1-gA_t\mathcal Q)\,{\rm d}t,\qquad
 z=z_{\rm old}-\frac12(1-x^2)B_2\sin2\phi,
 \label{eq:stationarycirclegauge}
\end{equation}
with
\begin{align}
 B_1'&=\frac{8gA_t^2\rho\Phi^2(1+\Phi^2)}{\sqrt{fp}}
 +g\sqrt{\frac fp}\,\rho^{5/9}H^{5/9}
 \left[7+14\Phi^2-8H\Phi^2(1+\Phi^2)\right],\nonumber\\
 B_2'&=\frac{4A_t\rho\Phi\sqrt{1+\Phi^2}(1+2\Phi^2)}{\sqrt{fp}}.
 \label{eq:Bfunctions}
\end{align}
Equation~\eqref{eq:circlecurvature} reduces to the equation for $B_1$ upon using the Maxwell equation. The function $B_2$ is absorbed into the circle coordinate in the covariant gauge.

The time-dependent quadrupole also follows directly from these formulae. At first order about $\Phi=0$, the other two-dimensional fields are unperturbed and $\mathcal J_\Phi=\mathcal D\Phi$. For $\Phi=\epsilon e^{-{\rm i}\omega t}\tilde\varphi(r)$, the circle perturbation is
\begin{equation}
 \delta E_z=\epsilon\frac{(1-x^2)\rho}{g}
 \operatorname{Re}\!\bigg\{e^{-{\rm i}\omega t-2{\rm i}\phi}
 \left[\sqrt{fp}\,\tilde\varphi'\,{\rm d}t
 -\frac{{\rm i}(\omega-2gA_t)}{\sqrt{fp}}\tilde\varphi\,{\rm d}r\right]\bigg\}.
 \label{eq:timefibre}
\end{equation}
The remaining metric components follow by expanding Eqs.~\eqref{eq:angularwarp}--\eqref{eq:Ex}. The radial term in Eq.~\eqref{eq:timefibre} is required by the eleven-dimensional equations. Thus the time-dependent instability can be uplifted explicitly, including its complex frequency and scalar phase. The nonlinear metric retains the discrete transformation $\phi\mapsto\phi+\pi$.

The same construction can be extended to four independent rotation planes. The resulting sector has four gauge fields, four neutral scalars, and four charged scalars, with
\begin{equation}
 \mathcal D\Phi^{(i)}=\nabla\Phi^{(i)}+2{\rm i}gA^{(i)}\Phi^{(i)},
 \qquad i=1,\ldots,4,
\end{equation}
without summation over $i$. We restrict to a single non-zero rotation in the following.

\subsection{The axisymmetric rotating solution}\label{sec:symmetric}
The solution with no quadrupole condensate is
\begin{align}
 H&=\left(1+\frac{a^2}{r^2}\right)^{-1},&
 G&=1-\frac{2M_0H}{r^7},& \mathcal Z&=1,\nonumber\\
 A_t&=-\frac{H(r_+)-H(r)}{a}\sqrt{2M_0g^5},&
 B_1&=g^7(r^7-r_+^7),& \Phi&=B_2=0.
 \label{eq:background}
\end{align}
We denote the mass parameter by $M_0$ to distinguish it from an azimuthal integer. The largest root of $G$ is the horizon, and
\begin{equation}
 2M_0=r_+^5(r_+^2+a^2).
 \label{eq:massparameter}
\end{equation}
The gauge $A_t(r_+)=0$ makes the horizon regularity conditions particularly simple. In the dual frame the corresponding metric functions are
\begin{equation}
 f_{\rm D}=g^5r^5G H^{-5/9},\qquad
 p_{\rm D}=g^2r^2G H^{-4/9}.
 \label{eq:dualbackground}
\end{equation}
The $a\to0$ limit recovers the non-rotating black D$0$-brane.

The uplift is locally related to a ten-dimensional singly spinning Myers--Perry black hole times a line. To expose this geometry, set
\begin{equation}
 \mathcal A=\frac{\mathcal R^{7/2}}{\sqrt{2M_0}},\qquad
 \Omega_H=\frac{a}{r_+^2+a^2},
 \qquad t=\mathcal A(T+Z),\quad z=\frac Z{\mathcal A},
 \quad\phi=\psi-\Omega_H(T+Z).
 \label{eq:Carroll}
\end{equation}
The transformation is local: any compact identification must also be transformed when interpreting the Kaluza--Klein circle. In particular, a non-zero $\Phi$ depends on $\psi-\Omega_H(T+Z)$ and preserves the two helical Killing fields
\begin{equation}
 K_T=\partial_T+\Omega_H\partial_\psi,\qquad
 K_Z=\partial_Z+\Omega_H\partial_\psi.
 \label{eq:helical}
\end{equation}
Thus a stationary two-dimensional condensate need not uplift to a geometry invariant under $\partial_T$ and $\partial_Z$ separately. We emphasize that solutions in this work uplifted to eleven dimensions in the BFSS frame with coordinates $t, \phi, z$ are still homogeneous along the M-theory circle parametrised by $z$, since $\partial_z=\mathcal A(\partial_Z-\partial_T)=\mathcal A(K_Z-K_T)$.

We define $G_2$ by $\kappa_{2\mathrm D}^2=2\pi G_2$, so that the curvature term is $\rho R/(16\pi G_2)$. In the time coordinate $t$, the temperature and Wald entropy~\cite{Iyer:1994ys} are
\begin{equation}
 T=\frac{\sqrt{f'(r_+)p'(r_+)}}{4\pi},\qquad
 S=\frac{\rho(r_+)}{4G_2}.
 \label{eq:TS}
\end{equation}
Matching this entropy to the eleven-dimensional horizon area gives
\begin{equation}
 G_2=\frac{G_{11}g^8}{2\pi L_{11}\operatorname{Vol}(S^8)},
 \qquad \operatorname{Vol}(S^8)=\frac{32\pi^4}{105}.
 \label{eq:Newtondictionary}
\end{equation}
For Eq.~\eqref{eq:background}, writing $\alpha=a/r_+$, these become
\begin{equation}
 T=\frac{g^{7/2}r_+^{5/2}(7+5\alpha^2)}{4\pi\sqrt{1+\alpha^2}},\qquad
 S=\frac{g^{9/2}r_+^{9/2}\sqrt{1+\alpha^2}}{4G_2}.
 \label{eq:backgroundTS}
\end{equation}
The thermodynamic quantities in this section use the lower-dimensional time and gauge-field normalisation; the horizon angular velocity $\Omega_H$ in Eq.~\eqref{eq:Carroll} uses the Myers--Perry time $T$. For the axisymmetric solution, these thermodynamic quantities are the same as those obtained from the spinning D$0$ black hole in type IIA supergravity in the decoupling limit \cite{Harmark:1999xt}. 

\subsection{The quadrupolar instability}
Since the action is even in $\Phi$, a perturbation $\Phi=\epsilon e^{-{\rm i}\omega t}\tilde\varphi(r)$ about the axisymmetric background decouples from the other fields at first order. Its radial equation is
\begin{equation}
 \frac1{\rho\sqrt{f/p}}\frac{{\rm d}}{{\rm d}r}
 \left(\rho\sqrt{fp}\,\tilde\varphi'\right)
 +\left[\frac{(\omega-2gA_t)^2}{f}
 -2\rho^{-4/9}V_{\lambda_\Phi}(v,0)\right]\tilde\varphi=0.
 \label{eq:linearphi}
\end{equation}
The sign of the frequency shift follows from $\mathcal D_t=\partial_t+2{\rm i}gA_t$. On substituting Eq.~\eqref{eq:background}, this becomes
\begin{equation}
 \frac{{\rm d}}{{\rm d}r}\left(\frac{r^8G}{H}\tilde\varphi'\right)
 +\left[2r^6(7-2H)
 +\frac{r}{g^7G}(\omega-2gA_t)^2\right]\tilde\varphi=0.
 \label{eq:zeroODE}
\end{equation}
At $\omega=0$, the last term reduces to $8M_0r[H(r_+)-H(r)]^2/(a^2G)$, recovering the stationary equation. The expression has a smooth non-rotating limit.

We impose an ingoing condition at the future horizon. In the gauge $A_t(r_+)=0$, the horizon exponent is $-{\rm i}\omega/(4\pi T)$. It is convenient to factor it out and use
\begin{equation}
 \begin{aligned}
 y&=1-\frac{r_+}{r},\qquad
 \tilde\varphi=\frac{r_+^2}{r^2}
 \left[1-\left(\frac{r_+}{r}\right)^4\right]^s q(y),\\
 s&=-\frac{{\rm i}\sqrt{1+\alpha^2}\,\hat\omega}{7+5\alpha^2},
 \qquad\hat\omega=\frac{\omega}{g^{7/2}r_+^{5/2}}.
 \end{aligned}
 \label{eq:ingoingquadrupole}
\end{equation}
The remaining function $q$ is regular at $y=0$. Near infinity,
\begin{equation}
 \tilde\varphi=\frac{\phi_{(2)}}{r^2}
 -\frac{a^2\phi_{(2)}}{r^4}+\frac{\phi_{(5)}}{r^5}+\cdots.
 \label{eq:timefalloffs}
\end{equation}
The frequency term in Eq.~\eqref{eq:zeroODE} is suppressed by five powers of $r$ relative to the indicial operator. The first frequency-dependent term in the scalar expansion is $-(\omega-2gA_t(\infty))^2\phi_{(2)}/(10g^7r^7)$, so neither the indicial roots nor the independent $r^{-5}$ coefficient changes. The source-free condition remains $\phi_{(5)}=0$, with the quantisation justified below. The horizon factor in Eq.~\eqref{eq:ingoingquadrupole} differs from unity only at order $r^{-4}$, so it does not change the $r^{-5}$ coefficient. Writing $q=q_0-\alpha^2q_0(1-y)^2+q_3(1-y)^3+\cdots$, we have $\phi_{(5)}=r_+^5q_3$. Thus the boundary condition on the regular function is simply
\begin{equation}
 q^{(3)}(1)=0,
 \label{eq:timeboundarycondition}
\end{equation}
where derivatives of $q$ are with respect to $y$.

At $\omega=0$, writing $\varphi=\tilde\varphi|_{\omega=0}$, horizon regularity requires
\begin{equation}
 \varphi'(r_+)=-\frac{2\left[7-2/(1+\alpha^2)\right]}
 {r_+(7+5\alpha^2)}\varphi(r_+).
 \label{eq:linearhorizon}
\end{equation}
One overall amplitude is arbitrary. The horizon and source-free conditions make $\alpha$ an eigenvalue for a stationary mode, and the first solution occurs at
\begin{equation}
 \alpha_{\rm c}\simeq0.991384\,,
 \label{eq:onset}
\end{equation}
where we recall that $\alpha=a/r_+$. Solving Eq.~\eqref{eq:zeroODE} at $\omega=0$ independently reproduces the threshold extracted from the eleven-dimensional calculation. Allowing a complex frequency determines which side of this bifurcation is unstable: with our convention $e^{-{\rm i}\omega t}$, a mode grows when $\operatorname{Im}\omega>0$. We present this calculation in Section~\ref{sec:gravityresults}.

\subsection{Boundary conditions and holographic renormalisation}
We formulate the asymptotic variational problem without imposing stationarity. This determines whether the quadrupole is a response of the undeformed matrix model or a response to an imposed deformation. At a timelike radial cutoff, let $h_{tt}$ be the induced metric, $D_t$ its covariant derivative and $n^a$ the outward spacelike unit normal. We write the radial ADM metric as
\begin{equation*}
 {\rm d}s_{\rm C}^2=\tilde N^2{\rm d}r^2
 +h_{tt}({\rm d}t+\tilde N^t{\rm d}r)^2,
\end{equation*}
where $\tilde N$ is the radial lapse and $\tilde N^t$ is the radial shift. In the zero-shift radial gauge, with the D$0$-brane falloffs, we allow
\begin{equation}
 \sqrt{-h}=(gr)^{7/2}n_{(0)}(t)[1+o(1)],\qquad
 \rho=(gr)^{9/2}\rho_{(0)}(t)[1+o(1)],
 \label{eq:timedependentsources}
\end{equation}
where $n_{(0)}$ and $\rho_{(0)}$ are smooth positive functions. We also allow a general finite boundary value of $A_t$ and time-dependent scalar coefficients. The stationary normalisation used elsewhere in this paper has $n_{(0)}=\rho_{(0)}=1$.

Both $v$ and $\Phi$ satisfy, to leading order,
\begin{equation}
 \frac{\partial}{\partial r}(r^8\partial_r\chi)+10r^6\chi=0,
 \qquad \chi\in\{v,\Phi\}.
 \label{eq:indicial}
\end{equation}
The indicial roots are $2$ and $5$. Boundary derivatives are suppressed by five powers of $r$ relative to this operator. Nonlinear terms generate intermediate powers fixed by the slower coefficients, so it is useful to write
\begin{align}
 v&=\frac{v_{(2)}(t)}{r^2}
 +\text{locally determined terms}+\frac{v_{(5)}(t)}{r^5}+\cdots,\nonumber\\
 \Phi&=\frac{\phi_{(2)}(t)}{r^2}
 +\text{locally determined terms}+\frac{\phi_{(5)}(t)}{r^5}+\cdots.
 \label{eq:falloffs}
\end{align}
For example, $\Phi$ has an $r^{-4}$ term when the neutral scalar response is non-zero. Such a term is not an independent boundary datum. The AdS$_2$ appearance of the dual frame does not by itself give the source assignment of a scalar with a constant kinetic coefficient: the running dilaton also enters the radial canonical momenta. We follow the non-conformal holographic prescription of~\cite{Kanitscheider2008} and use alternative quantisation for both scalars, as required by the BFSS operator dictionary~\cite{Ortiz:2014aja,Bovon2025}. The relation between the radial falloffs and the scaling of matrix operators is discussed in~\cite{Biggs:2023sqw}. The undeformed theory therefore requires
\begin{equation}
 v_{(5)}(t)=\phi_{(5)}(t)=0,
 \label{eq:sourcefree}
\end{equation}
while $v_{(2)}$ and $\phi_{(2)}$ encode responses. The charged response is proportional, with the operator normalisation fixed by the reduction, to $\langle\mathcal O_2\rangle$.

In the canonical frame the Gibbons--Hawking term is
\begin{equation}
 S_{\rm GH}=\frac1{4\kappa_{2\mathrm D}^2}
 \int_{\partial\mathcal M}{\rm d}t\sqrt{-h}\,\rho K.
 \label{eq:GH}
\end{equation}
The zero-derivative counterterm is
\begin{equation}
 S_{\rm ct}^{(0)}=-\frac1{2\kappa_{2\mathrm D}^2}
 \int_{\partial\mathcal M}{\rm d}t\sqrt{-h}\,U,
 \qquad U=g\rho^{7/9}w,
 \label{eq:counterterm}
\end{equation}
with
\begin{equation}
 w=\frac74e^{4v/7}+\frac12e^{-2v}(1+2|\Phi|^2).
 \label{eq:superpotential}
\end{equation}
This choice subtracts the vacuum energy of the D$0$-brane domain wall. Using the real doublet and scalar metric in Eqs.~\eqref{eq:realscalarderivative} and~\eqref{eq:scalartargetmetric}, the constraint at zero boundary-derivative order is
\begin{equation}
 2U\partial_\rho U-\frac7{18\rho}(\partial_vU)^2
 -\frac1{2\rho}(\mathsf M^{-1})_{ij}
       \partial_{\psi_i}U\partial_{\psi_j}U
 +\rho^{5/9}V=0.
 \label{eq:covariantHJ}
\end{equation}
The function in Eq.~\eqref{eq:superpotential} satisfies this identity exactly, including the scalar self-interactions. Since $U$ depends on the charged scalar only through its norm, no choice of scalar phase is required. At a point where $\Phi$ is real, the same identity becomes
\begin{equation}
 \frac{14}{9}w^2-\frac7{18}(\partial_vw)^2
 -\frac{1+\Phi^2}{2}(\partial_\Phi w)^2
 +\frac{V}{g^2}=0.
 \label{eq:HJ}
\end{equation}
The electric-flux contribution to the radial constraint is subleading for a finite boundary gauge potential and finite charge, and does not require a divergent counterterm.

\paragraph{Boundary derivative terms.}
Equation~\eqref{eq:counterterm} is insufficient when the leading dilaton source depends on time. The dilaton-curvature coupling contributes $\Box_h\rho/2$ to the radial Hamilton--Jacobi constraint. Setting the scalars to zero at leading order, define $U_0=9g\rho^{7/9}/4$ and consider the two-derivative functional
\begin{equation}
 S_{\rm HJ}^{(2)}=\frac1{2\kappa_{2\mathrm D}^2}
 \int{\rm d}t\sqrt{-h}\,f_2(\rho)(D\rho)^2,
 \qquad (D\rho)^2=h^{tt}(\partial_t\rho)^2.
 \label{eq:derivativeHJ}
\end{equation}
The constraint at this order reduces to
\begin{equation}
 \left(\frac12-4U_0f_2\right)\Box_h\rho
 -2\partial_\rho(U_0f_2)(D\rho)^2=0.
 \label{eq:derivativeHJconstraint}
\end{equation}
Both coefficients vanish for $f_2=1/(8U_0)$. Thus we must add
\begin{equation}
 S_{\rm ct}^{(2)}=-\frac1{2\kappa_{2\mathrm D}^2}
 \int_{\partial\mathcal M}{\rm d}t\sqrt{-h}\,
 \frac{\rho^{-7/9}}{18g}(D\rho)^2.
 \label{eq:derivativecounterterm}
\end{equation}
Under Eq.~\eqref{eq:timedependentsources}, its density is $O(r^2)$ when $\partial_t\rho_{(0)}\ne0$, compared with the $O(r^7)$ zero-derivative density. Each additional pair of boundary derivatives lowers the pure-dilaton weight by five powers of $r$, so four-derivative terms vanish as $r^{-3}$ at the cutoff. The temporal scalar kinetic densities are $O(r^{-3})$ in the bulk and are radially integrable.

A possible finite two-derivative term linear in $v$ also needs to be excluded. Up to integration by parts, the candidate terms are $v f_1(\rho)(D\rho)^2+h_0(\rho)D\rho\cdot Dv$. Since $\partial_vU|_{v=\Phi=0}=0$, their Hamilton--Jacobi recursion has coefficients $-2U_0h_0$ for $\Box_hv$ and, after setting $h_0=0$, $-4U_0f_1$ for $D\rho\cdot Dv$. Hence $h_0=f_1=0$. Gauge invariance excludes terms linear in $\Phi$, and terms quadratic in either scalar decay at the cutoff. We use the local branch analytic in the scalars and respecting the scaling covariance of the bulk action, with $S_{\rm ct}=S_{\rm ct}^{(0)}+S_{\rm ct}^{(2)}$ and the vacuum subtraction fixed above.

The derivative term changes the renormalised metric and dilaton momenta. For example, with $\mathcal A=\tfrac12\log(-h_{tt})$, its contributions, with the common factor $1/(2\kappa_{2\mathrm D}^2)$ stripped off, are
\begin{equation}
 \frac{\pi_{\mathcal A}^{\rm ct,(2)}}{\sqrt{-h}}
 =f_2(D\rho)^2,\qquad
 \frac{\pi_\rho^{\rm ct,(2)}}{\sqrt{-h}}
 =2f_2\Box_h\rho+\partial_\rho f_2(D\rho)^2.
 \label{eq:derivativecountertermmomenta}
\end{equation}
It has no explicit scalar dependence and therefore does not directly change either scalar momentum.

\paragraph{The alternative variational principle.}
The finite scalar variation of $S_{\rm D}=S_{2\mathrm D}+S_{\rm GH}+S_{\rm ct}^{(0)}+S_{\rm ct}^{(2)}$, with the metric, dilaton, and gauge source held fixed, is
\begin{equation}
 \delta S_{\rm D}\big|_{\rm scalars}
 =\frac{g^8}{2\kappa_{2\mathrm D}^2}\int{\rm d}t\,
 n_{(0)}\rho_{(0)}^{7/9}
 \left[\frac{27}{7}v_{(5)}\delta v_{(2)}
 +3\operatorname{Re}\left(\phi_{(5)}^*\delta\phi_{(2)}\right)\right].
 \label{eq:scalarvariation}
\end{equation}
The factor $n_{(0)}\rho_{(0)}^{7/9}$ follows from the leading radial lapse, $\tilde N=\rho_{(0)}^{2/9}+o(1)$, in the radial gauge of Eq.~\eqref{eq:timedependentsources}. Thus the required alternative variational principle is obtained by the finite Legendre transform
\begin{equation}
 S_{\rm alt}=S_{\rm D}
 -\frac{g^8}{2\kappa_{2\mathrm D}^2}\int{\rm d}t\,
 n_{(0)}\rho_{(0)}^{7/9}
 \left[\frac{27}{7}v_{(2)}v_{(5)}
 +3\operatorname{Re}\left(\phi_{(2)}^*\phi_{(5)}\right)\right].
 \label{eq:Legendre}
\end{equation}
Its scalar variation fixes the fast coefficients pointwise in time. In particular, setting those coefficients to zero is compatible with a non-zero time-dependent quadrupole response. Variations of the metric and dilaton sources must also include the variation of the prefactor in Eq.~\eqref{eq:Legendre}. Total time derivatives are discarded for periodic boundary data or fixed endpoint data.

For the stationary solutions, $S_{\rm ct}^{(2)}$ and its first variation vanish, and the scalar formulas reduce to those with $n_{(0)}=\rho_{(0)}=1$. The Legendre term vanishes on source-free solutions, but remains essential for their interpretation and for variations with non-zero scalar sources. The linear time-dependent charged mode likewise keeps the leading dilaton source fixed. Its temporal kinetic density falls as $r^{-3}$, and its frequency-dependent boundary variation falls as $r^{-2}$. The general prescription therefore retains $\phi_{(5)}=0$ as the source-free condition used in the frequency calculation.

\subsection{Conserved charges and nonlinear boundary data}
We now specialise to stationary solutions with the leading metric and dilaton sources fixed to $n_{(0)}=\rho_{(0)}=1$. For general time-dependent sources, the boundary responses are obtained by varying $S_{\rm alt}$, including Eq.~\eqref{eq:derivativecountertermmomenta} and the Legendre term. The radial constraints give the corresponding boundary Ward identities; time-dependent sources can exchange energy and angular momentum with the system. The thermodynamic expressions below use the stationary specialisation.

The charge conjugate to the boundary value of $A_t$ can be obtained from the radial canonical momentum. With outward normal $n^a$, define
\begin{equation}
 \mu=A_t(\infty)-A_t(r_+),\qquad
 Q=\frac1{8\kappa_{2\mathrm D}^2}
 \lim_{r\to\infty}\rho^{13/9}e^{4v}\sqrt{\frac pf}\,A_t'.
 \label{eq:charge}
\end{equation}
This is the boundary charge; when $\Phi\ne0$, scalar charge outside the horizon means that the radial electric flux need not be constant. We choose the orientation for which positive $a$ gives positive $Q$ and $\mu$. Reversing the gauge field and the angular orientation reverses both signs and leaves the first law unchanged. With a $2\pi$-periodic azimuthal angle, the angular velocity and angular momentum in the time coordinate $t$ are $\Omega=g\mu$ and $J=Q/g$, so that $\mu\,{\rm d}Q=\Omega\,{\rm d}J$.

For source-free scalars, the renormalised energy is
\begin{equation}
 E=\frac1{2\kappa_{2\mathrm D}^2}
 \lim_{r\to\infty}\sqrt f\left(U-\frac12\sqrt p\,\rho'\right).
 \label{eq:energy}
\end{equation}
This formula uses the fixed leading metric source in Eq.~\eqref{eq:radialgauge}. Together with Eqs.~\eqref{eq:TS} and \eqref{eq:charge}, it gives quantities that can be evaluated directly from radial solutions.

For clarity, the asymptotic data can also be organised in the compact coordinate $y=1-r_+/r$, with $s=1-y$. Write
\begin{equation}
 G=(1-s^7)q_1(y),\quad
 \mathcal Z=q_2(y)^{9/2},\quad
 H=(1+s^2q_3(y))^{-1},\quad
 A_t=(gr_+)^{5/2}yq_4(y),\quad
 \Phi=s^2q_5(y).
 \label{eq:compactansatz}
\end{equation}
Let the source-free expansion have
\begin{align}
 q_1&=1+\varepsilon_7s^7+\cdots,&q_2&=1+O(s^9),\nonumber\\
 q_3&=c-2b^2s^2+\cdots,&q_5&=b-cbs^2+\cdots,\nonumber\\
 q_4&=\mu_0+\mu_0s+\eta_2s^2+\eta_2s^3+\cdots.
 \label{eq:asymptoticq}
\end{align}
The scalar source conditions eliminate independent $s^3$ terms in $q_3$ and $q_5$. Their leading values $c$ and $b$ are free responses. The nonlinear $-2b^2s^2$ term in $q_3$ and $-cbs^2$ term in $q_5$ agree with the direct asymptotic expansion of the radial field equations. In this notation the charges are
\begin{equation}
 E=\frac{9g^8r_+^7}{32\pi G_2}(1-\varepsilon_7),\qquad
 Q=\frac{g^{11/2}r_+^{9/2}}{8\pi G_2}(\mu_0-\eta_2),\qquad
 \mu=(gr_+)^{5/2}\mu_0.
 \label{eq:asymptoticcharges}
\end{equation}
The corresponding horizon expressions are
\begin{equation}
 T=\frac{7g^{7/2}r_+^{5/2}}{4\pi}
 q_1(0)\sqrt{1+q_3(0)},\qquad
 S=\frac{g^{9/2}r_+^{9/2}}{4G_2}
 q_2(0)^{9/2}\sqrt{1+q_3(0)}.
 \label{eq:nonlinearTS}
\end{equation}
These boundary and horizon formulae agree with an independent calculation of the conserved charges using covariant phase space~\cite{Iyer:1994ys}.

For the axisymmetric branch, they reduce to
\begin{equation}
 E=\frac{9g^8r_+^7(1+\alpha^2)}{32\pi G_2},\qquad
 Q=\frac{g^{11/2}r_+^{9/2}\alpha\sqrt{1+\alpha^2}}{8\pi G_2},\qquad
 \mu=\frac{(gr_+)^{5/2}\alpha}{\sqrt{1+\alpha^2}}.
 \label{eq:backgroundcharges}
\end{equation}
Using Eqs.~\eqref{eq:BFSSdictionary} and~\eqref{eq:Newtondictionary}, the dimensionless angular momentum in Eq.~\eqref{eq:introcriticalspin} is
\begin{equation}
 \mathfrak j=\left(\frac{2^{21}5^2}{3^{16}7^5}\right)^{1/14}
 \frac{\alpha}{(1+\alpha^2)^{1/7}}.
 \label{eq:spinenergyconversion}
\end{equation}
In particular, $J/S=\alpha/(2\pi)$ on this branch, so the quadrupolar onset corresponds to $J/S\simeq0.15778366$. Direct differentiation verifies
\begin{equation}
 {\rm d}E=T\,{\rm d}S+\mu\,{\rm d}Q,\qquad
 14E=9(TS+\mu Q).
 \label{eq:firstlaw}
\end{equation}
The second identity follows from radial scaling and applies to source-free stationary branches in the same normalisation.

Equations~\eqref{eq:compactansatz}--\eqref{eq:nonlinearTS} specify the boundary problem for continuing the quadrupolar branch. One imposes a regular horizon, fixes the leading metric and gauge sources, and sets the two scalar sources to zero. Close to the bifurcation, the charged scalar is odd in its amplitude while the metric, neutral scalar, and gauge field first respond at quadratic order. At fixed $(T,Q)$ the appropriate comparison is the Helmholtz free energy $E-TS$; at fixed $(T,\mu)$ it is the Gibbs free energy $E-TS-\mu Q$. We carry out these thermodynamic comparisons in Section~\ref{sec:gravityresults}.

\section{Gravitational results}\label{sec:gravityresults}
\subsection{Linear quadrupolar instability}
We first consider the axisymmetric black hole. Increasing $\alpha=a/r_+$, we find a source-free quadrupolar zero mode at
\begin{equation}
 \alpha_{\rm c}=0.991384,\qquad
 \left.\frac JS\right|_{\rm c}=0.157784.
\end{equation}
The radial equation in Section~\ref{sec:gravity} and the direct eleven-dimensional gravitational calculation give the same onset. In the latter calculation, the next azimuthal sector, $m_\varphi=3$, bifurcates at $\alpha\simeq1.401369$, and the thresholds increase with $m_\varphi$ over the sectors we have studied; see Table~\ref{tab:thresholds}. Thus the quadrupole is the first non-axisymmetric mode to bifurcate in this sector. This comparison concerns the onset spins; it does not order the growth rates of different azimuthal sectors away from their thresholds.

\begin{figure}[htbp]
\centering
\includegraphics[width=\textwidth]{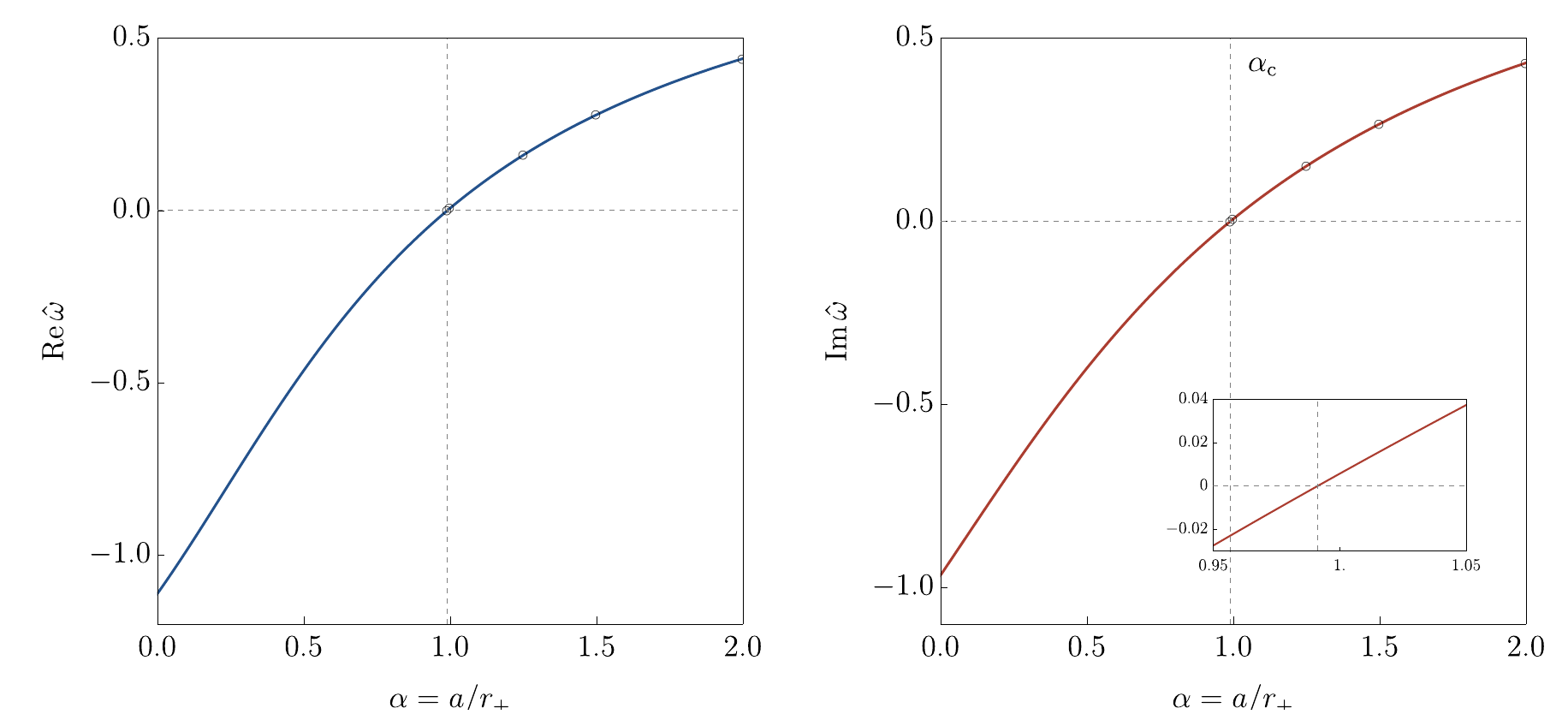}
\caption{Complex frequency of the quadrupolar mode with ingoing horizon behaviour and vanishing scalar source. We use $\hat\omega=\omega/(g^{7/2}r_+^{5/2})$. The dashed vertical line marks $\alpha_{\rm c}\simeq0.991384$, and the inset resolves the change of sign of $\operatorname{Im}\hat\omega$ near the onset. The curves use 101 collocation points; open circles show checks with 201 collocation points. Positive imaginary part corresponds to exponential growth.}
\label{fig:quadrupolarfrequency}
\end{figure}

We now follow this mode with non-zero complex frequency, imposing the ingoing condition and the physical source-free boundary condition in Eq.~\eqref{eq:timeboundarycondition}. The result is shown in Fig.~\ref{fig:quadrupolarfrequency}. Both parts of the frequency vanish at the stationary onset. Below $\alpha_{\rm c}$ the mode is damped; above it, $\operatorname{Im}\omega>0$, establishing a quadrupolar instability of the axisymmetric BFSS black hole. Away from the onset the mode has both a real and an imaginary frequency, so the perturbation oscillates as it grows in the BFSS frame used here. Along the branch shown, the mode remains unstable up to $\alpha=2$. Damping below the threshold concerns this mode; it does not establish stability in other perturbation sectors. For example,
\begin{equation}
 \hat\omega(1)\simeq0.006126+0.005603\,{\rm i},\qquad
 \hat\omega(2)\simeq0.440395+0.432175\,{\rm i}.
 \label{eq:quadrupolarfrequencies}
\end{equation}
The curve uses 101 collocation points and 50-digit working precision. Checks with 151 and 201 collocation points at $\alpha=0.99,1,1.25,1.5,2$ agree to the quoted precision. The charged scalar belongs to the consistent $SO(9)$ truncation, whose general uplift applies to time-dependent fields~\cite{Bossard:2023jid}. Its growing solution therefore uplifts to an eleven-dimensional gravitational perturbation satisfying the BFSS circle identification. The covariant uplift in Eq.~\eqref{eq:uplift} applies directly, with the linear circle perturbation given in Eq.~\eqref{eq:timefibre}. We thus establish instability in the BFSS geometry directly, rather than importing a growing mode from a Myers--Perry string with a different compact identification.

The order parameter is particularly simple on the matrix side: it is the expectation value of $\mathcal O_2$ in Eq.~\eqref{eq:quadrupole}. The bulk scalar carries charge two under rotations in the $(12)$-plane, and its phase fixes the orientation of the deformation. The solutions with $\Phi\ne0$ preserve rotations by $\pi$. In eleven dimensions they are stationary with respect to the helical Killing fields in Eq.~\eqref{eq:helical}.

\subsection{The nonlinear quadrupolar branch}
To construct the nonlinear solutions, we use the compact coordinate and functions in Eq.~\eqref{eq:compactansatz}, and continue away from the zero mode by fixing
\begin{equation}
 b=q_5(1)=\frac{\phi_{(2)}}{r_+^2}.
\end{equation}
The sign of $b$ can be changed by rotating the deformation through $\pi/2$, so we take $b>0$. Both scalar sources are set to zero throughout. The branch is discretised on a Chebyshev grid with 201 collocation points. For a review of spectral collocation and Newton methods for gravitational boundary-value problems, see~\cite{Dias:2015nua}. We construct solutions from $b=0.01$ to $b=5$. Up to $b=3.71$ we use steps of $0.01$; beyond this we use a quadratic predictor from the preceding three solutions, followed by Newton corrections, with steps between $0.09$ and $0.30$. The extension uses 80-digit working precision and a maximum equation-residual tolerance of $10^{-22}$.

We extract the energy and angular momentum at infinity, and the temperature and entropy at the horizon. In terms of the functions used in the numerical discretisation, choosing the gauge-field sign to agree with Eq.~\eqref{eq:background},
\begin{equation}
 q_1=1+(1-y)^7r_1,\quad q_2=1+(1-y)^8r_2,
 \quad(q_3,q_4,q_5)=(r_3,r_4,r_5),
\end{equation}
the coefficients entering Eq.~\eqref{eq:asymptoticcharges} are $\varepsilon_7=r_1(1)$, $\mu_0=r_4(1)$, and $\eta_2=r_4''(1)/2$. With these assignments, the extracted charges satisfy the scaling relation in Eq.~\eqref{eq:firstlaw}. A local polynomial differentiation along the densely sampled part of the branch, $b\leq3.71$, gives a relative first-law residual below $6\times10^{-11}$. We check the extension by differentiating the boundary problem itself, as described in Section~\ref{sec:convergence}. At the common amplitudes $b=0.01,0.25,0.5,1$, the thermodynamic quantities on the grids with 101 and 201 collocation points agree to better than $10^{-12}$. We also check selected solutions with $b>1$ using 301 and 401 collocation points.

The numerical construction fixes $r_+$, so the angular momentum varies along the saved branch. To present the thermodynamics at fixed $J$, we remove this scale using the source-free scaling relation in Eq.~\eqref{eq:firstlaw}. At fixed $g$ and $G_2$, changing the horizon scale gives $E,F\propto r_+^7$, $S,J\propto r_+^{9/2}$ and $T\propto r_+^{5/2}$. We therefore introduce
\begin{align}
 j&=G_2|J|=\frac{G_2|Q|}{g},\qquad
 \widehat E=\frac{G_2E}{g\,j^{14/9}},\quad
 \widehat S=\frac{G_2S}{j},\nonumber\\
 \widehat T&=\frac{T}{g\,j^{5/9}},\qquad
 \widehat F=\frac{G_2F}{g\,j^{14/9}}.
 \label{eq:fixedJvariables}
\end{align}
These quantities are dimensionless and invariant under changing the horizon scale. They are equivalent to rescaling each solution to a common non-zero angular momentum, and satisfy $\widehat F=\widehat E-\widehat T\widehat S$.

\begin{figure}[htbp]
\centering
\includegraphics[width=0.55\textwidth]{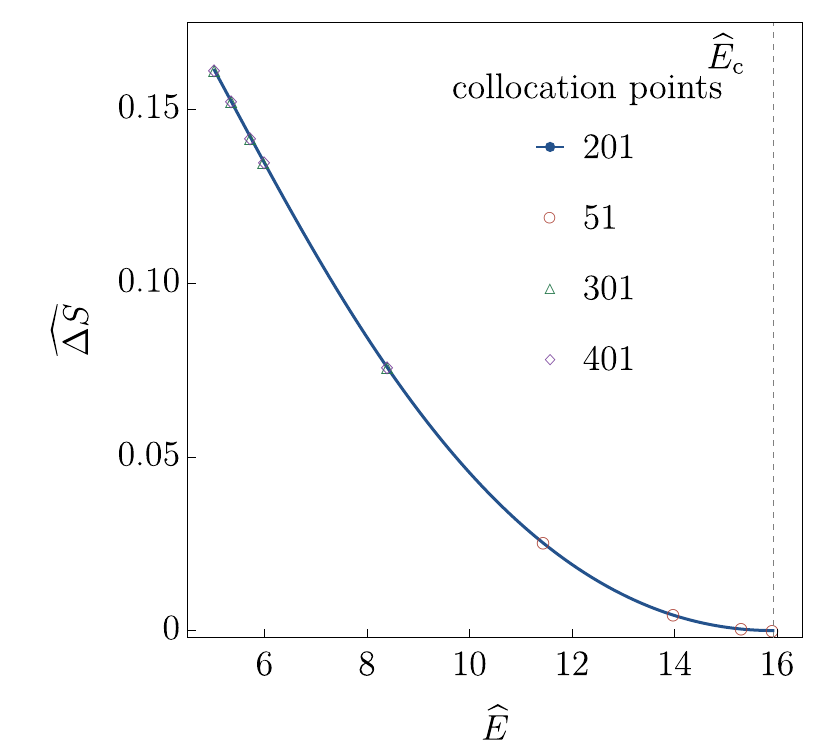}
\caption{Entropy difference $\widehat{\Delta S}=G_2(S-S_{\rm MP})/j$ against the scaled energy $\widehat E=G_2E/(g\,j^{14/9})$, where $j=G_2|J|$. Each axisymmetric reference has the same energy and angular momentum as the corresponding quadrupolar solution. These scale-invariant variables give a comparison at fixed $J$. The blue curve joins all 378 solutions with 201 collocation points, extending to $b=5$. Open circles show the available checks with 51 collocation points for $b\leq1$. Open triangles and diamonds show the checks with 301 and 401 collocation points at $b=2,3.71,4,4.5,5$. The dashed vertical line marks the linear onset. The entropy difference is positive throughout the computed branch.}
\label{fig:nonlinear}
\end{figure}

For an isolated system, the relevant comparison is the entropy at fixed energy and angular momentum. For each quadrupolar solution we therefore choose an axisymmetric solution with the same $(E,J)$, allowing both its horizon radius and spin to vary. Comparing solutions at the same $r_+$ would not keep the energy fixed. We denote the spin of the matched axisymmetric solution by $\alpha_{\rm MP}$ and define
\begin{equation}
 \widehat{\Delta S}=\frac{G_2[S-S_{\rm MP}(E,J)]}{j}.
 \label{eq:entropycomparison}
\end{equation}
We find $\widehat{\Delta S}>0$ along the computed branch. Figure~\ref{fig:nonlinear} plots this entropy difference against $\widehat E$. At fixed $J$, the branch extends towards lower energy from the onset at $\widehat E_{\rm c}\simeq15.922217$, reaching $\widehat E\simeq5.026778$ and $\widehat{\Delta S}\simeq0.161232$ at $b=5$. For comparison, Table~\ref{tab:nonlinear} gives representative values of the fractional gain $\delta_S=(S-S_{\rm MP})/S_{\rm MP}$. Close to the onset the entropy difference is very small. The data are consistent with $\delta_S\propto b^4$ and $\alpha_{\rm MP}-\alpha_{\rm c}\propto b^2$. At $b=1$ the entropy gain is approximately $0.51\%$, increasing to approximately $6.43\%$ at $b=5$, where $\alpha_{\rm MP}\simeq2.504779$. The scaled energy decreases and the entropy difference increases over the computed range, with no observed turning point. We stop at $b=5$ as a computational cutoff; it is not an endpoint of the branch. The nonlinear branch consequently provides an entropically favoured continuation of the quadrupolar zero mode within this family of solutions. Its dynamical stability requires a perturbation calculation about the deformed geometry.

\begin{table}[htbp]
\centering
\begin{tabular}{c r r}
\toprule
$b$ & $\alpha_{\rm MP}$ & $\delta_S$\\
\midrule
0.25 & 1.021452 & $7.1252\times10^{-5}$\\
0.50 & 1.093754 & $7.7295\times10^{-4}$\\
1.00 & 1.276211 & $5.1083\times10^{-3}$\\
2.00 & 1.633051 & $1.9683\times10^{-2}$\\
3.71 & 2.158155 & $4.6302\times10^{-2}$\\
4.00 & 2.239100 & $5.0523\times10^{-2}$\\
4.50 & 2.374369 & $5.7553\times10^{-2}$\\
5.00 & 2.504779 & $6.4275\times10^{-2}$\\
\bottomrule
\end{tabular}
\caption{Entropy of the quadrupolar branch relative to the axisymmetric solution with the same energy and angular momentum. The spin $\alpha_{\rm MP}$ belongs to the matched axisymmetric solution. The entries use the grid with 201 collocation points. Higher-resolution checks at five amplitudes with $b>1$ are given in Table~\ref{tab:convergence}.}
\label{tab:nonlinear}
\end{table}

The ensemble matters for this comparison. The local thermodynamic stability of the axisymmetric spinning-brane family, including its dependence on the ensemble, was analysed in~\cite{Harmark:1999xt}. At fixed $(T,\mu)$ there are two axisymmetric reference solutions whenever the discriminant of
\begin{equation}
 5\nu\alpha_{\rm MP}^2-4\pi\alpha_{\rm MP}+7\nu=0,
 \qquad \nu=\frac{g|\mu|}{T},
 \label{eq:grandcanonicalmatching}
\end{equation}
is positive. The lower-spin solution has $\alpha_{\rm MP}<\sqrt{7/5}$. Evaluating $\mathcal G=E-TS-\mu Q$ at the same $(T,\mu)$, we find that the quadrupolar solutions have higher $\mathcal G$ than this lower-spin reference along the computed branch. There is no contradiction with the entropy comparison: the quantities held fixed are different. In particular, the microcanonical entropy gain should not be read as grand-canonical dominance.

\paragraph{The canonical ensemble.}
We can also compare the two branches at fixed temperature and angular momentum. Since $J=Q/g$, this amounts to fixing $(T,Q)$ at fixed $g$. The relevant thermodynamic potential is the Helmholtz free energy $F=E-TS$. To make this comparison, we must adjust both the horizon radius and the spin of the axisymmetric solution. Denoting the matched spin by $\alpha_{\rm can}\geq0$, we find
\begin{equation}
 \mathfrak q\equiv\frac{G_2g^{4/5}|Q|}{T^{9/5}}
 =\frac{(4\pi)^{9/5}}{8\pi}
 \frac{\alpha_{\rm can}(1+\alpha_{\rm can}^2)^{7/5}}
 {(7+5\alpha_{\rm can}^2)^{9/5}}.
 \label{eq:canonicalmatching}
\end{equation}
The right-hand side is strictly increasing from zero to infinity, so there is a unique axisymmetric reference solution for each $(T,|Q|)$. Its horizon radius and free energy are
\begin{align}
 r_{+,\rm can}&=\left[\frac{4\pi T\sqrt{1+\alpha_{\rm can}^2}}
 {g^{7/2}(7+5\alpha_{\rm can}^2)}\right]^{2/5},\nonumber\\
 F_{\rm can}&=-\frac{g^8r_{+,\rm can}^7}{32\pi G_2}
 (5+\alpha_{\rm can}^2).
 \label{eq:canonicalreference}
\end{align}
Here $F_{\rm can}$ is evaluated at the temperature and charge of the quadrupolar solution. In particular, the matched spin differs from $\alpha_{\rm MP}$ in Table~\ref{tab:nonlinear}, where the energy and angular momentum were held fixed.

\begin{figure}[htbp]
\centering
\includegraphics[width=0.55\textwidth]{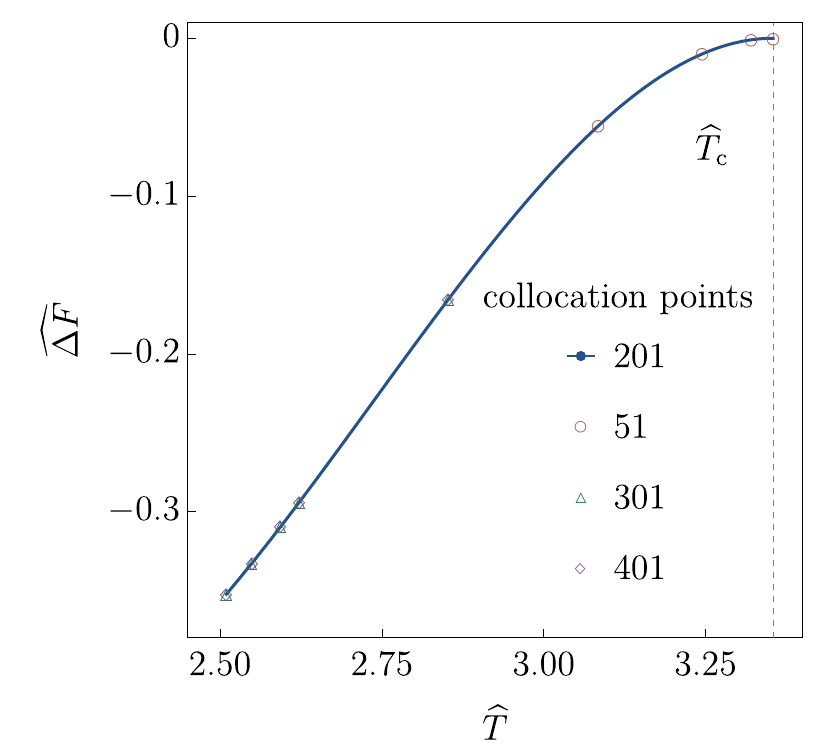}
\caption{Helmholtz free-energy difference $\widehat{\Delta F}=G_2(F-F_{\rm can})/(g\,j^{14/9})$ against the scaled temperature $\widehat T=T/(g\,j^{5/9})$, where $j=G_2|J|$. The axisymmetric reference free energy $F_{\rm can}$ is evaluated at the same temperature and angular momentum as each quadrupolar solution. These scale-invariant variables give a comparison at fixed $J$. The blue curve joins all 378 solutions with 201 collocation points, extending to $b=5$. Open circles show the available checks with 51 collocation points for $b\leq1$. Open triangles and diamonds show the checks with 301 and 401 collocation points at $b=2,3.71,4,4.5,5$. The dashed vertical line marks the linear onset. The free-energy difference is negative throughout the computed branch.}
\label{fig:canonicalfreeenergy}
\end{figure}

Using the charges and horizon quantities in Eqs.~\eqref{eq:asymptoticcharges} and~\eqref{eq:nonlinearTS}, we evaluate
\begin{equation}
 \widehat{\Delta F}\equiv\frac{G_2(F-F_{\rm can})}{g\,j^{14/9}},
 \qquad \delta F\equiv\frac{F-F_{\rm can}}{|F_{\rm can}|}.
 \label{eq:canonicaldifference}
\end{equation}
Figure~\ref{fig:canonicalfreeenergy} plots $\widehat{\Delta F}$ against $\widehat T$, using the same fixed-$J$ normalisation as in Fig.~\ref{fig:nonlinear}. The branch extends towards lower temperature from the onset at $\widehat T_{\rm c}\simeq3.354518$, reaching $\widehat T\simeq2.510044$ and $\widehat{\Delta F}\simeq-0.352503$ at $b=5$. We find $\widehat{\Delta F}<0$, and hence $\delta F<0$, for all 378 computed solutions with $0.01\leq b\leq5$. Thus the quadrupolar branch has lower free energy than the axisymmetric branch in the canonical ensemble. At $b=1,2,5$, the free-energy reduction is approximately $1.48\%$, $6.27\%$ and $26.67\%$, respectively. At the last point, the matched axisymmetric spin is $\alpha_{\rm can}\simeq1.992815$. Close to the bifurcation, the data are consistent with $F-F_{\rm can}\propto-b^4$. The checks with 301 and 401 collocation points at $b=2,3.71,4,4.5,5$ change this free-energy difference by less than $10^{-20}$ in relative terms. These results establish the thermodynamic preference of the quadrupolar branch at fixed $(T,J)$ within the two families considered here; they do not by themselves determine its dynamical stability.

\subsection{Horizon geometry}
\label{sec:horizongeometry}
The uplift also allows us to follow the shape of the horizon. At $r=r_+$, take the quotient of a spatial horizon section by the Kaluza--Klein circle and write $x=\cos\theta$. The quotient metric is the induced metric on an eight-dimensional ellipsoid with two axes in the rotation plane and seven equal transverse axes. Defining
\begin{equation}
 t_+=q_2(0),\qquad H_+=\frac{1}{1+q_3(0)},\qquad
 u_+=\operatorname{arsinh}\Phi_+,\qquad \Phi_+=q_5(0),
\end{equation}
the axes are
\begin{equation}
 L_\perp=r_+t_+,\qquad
 L_\pm=\frac{r_+t_+}{\sqrt{H_+}}e^{\pm u_+},\qquad
 \frac{L_+}{L_-}=e^{2u_+}.
 \label{eq:horizonaxes}
\end{equation}
Here we choose the orientation with $\Phi_+>0$. For example, an embedding of the quotient is
\begin{equation}
 X_1=L_+\sin\theta\cos\phi,\qquad
 X_2=L_-\sin\theta\sin\phi,\qquad
 \boldsymbol X_\perp=L_\perp\cos\theta\,\boldsymbol n_6,
\end{equation}
where $\boldsymbol n_6$ is a unit vector on $S^6$. Its induced metric reproduces the angular terms in Eq.~\eqref{eq:uplift} after removing the circle fibre. The axes describe this quotient geometry; the full horizon section includes the non-trivial circle fibration.

\begin{figure}[htbp]
\centering
\includegraphics[width=\textwidth]{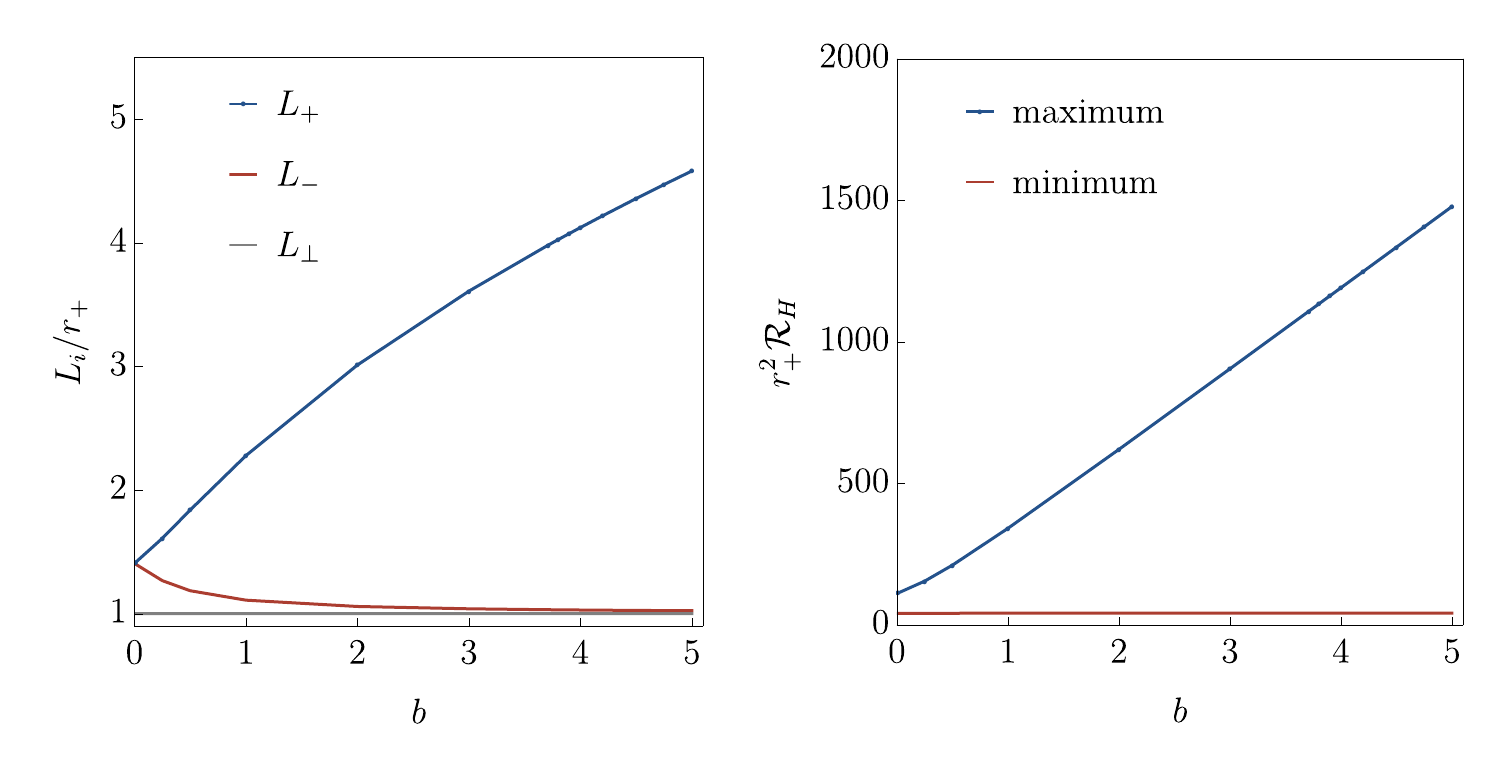}
\caption{Horizon geometry along the quadrupolar branch. Left: principal axes of the eight-dimensional quotient by the Kaluza--Klein circle, in units of $r_+$. Right: sampled maximum and minimum of the intrinsic scalar curvature of the full nine-dimensional horizon section. Lines join the sampled branch amplitudes. The radial solutions use 201 collocation points. The curvature is evaluated on a $41\times41$ angular grid. These are intrinsic horizon diagnostics; they do not measure the spacetime Kretschmann scalar.}
\label{fig:horizongeometry}
\end{figure}

For the full nine-dimensional horizon section, write its metric as $h_H=\gamma+\ell^2({\rm d}z+\mathcal A)^2$, where $\gamma$ is the quotient metric and $\ell^2=H_+^{2/9}/(\Delta_+\rho_+^{14/9})$. Its intrinsic scalar curvature is
\begin{equation}
 \mathcal R_H=\mathcal R_\gamma-2\ell^{-1}\nabla_\gamma^2\ell
 -\frac{\ell^2}{4}\mathcal F_{ij}\mathcal F^{ij},\qquad
 \mathcal F={\rm d}\mathcal A.
 \label{eq:horizoncurvature}
\end{equation}
All contractions in this expression use $\gamma$. We evaluate it using the ellipsoid embedding and the connection in Eq.~\eqref{eq:Ez}. In the axisymmetric limit it agrees with the intrinsic curvature of the Myers--Perry horizon times a line. We sample the angular dependence on a $41\times41$ grid covering $0\leq\theta,\phi\leq\pi/2$, including the axes. The extrema plotted in Fig.~\ref{fig:horizongeometry} are extrema on this angular grid.

The extension mainly increases $L_+$. The axes at two representative amplitudes are
\begin{equation}
 \frac{(L_+,L_-,L_\perp)}{r_+}\simeq
 \begin{cases}
 (3.982,1.032,1.000),&b=3.71,\\
 (4.582,1.024,1.000),&b=5.
 \end{cases}
\end{equation}
At $b=5$ the rotation-plane axis ratio is $L_+/L_-\simeq4.474$. The sampled maximum of $r_+^2\mathcal R_H$ increases from approximately $1108$ to $1478$ between these amplitudes, while its minimum remains close to $41.8$. The quotient axes and the local circle radius remain non-zero, and we find no loss of horizon regularity over this range. This family restricts the quotient to ellipsoidal shapes. The absence of an observed singularity or turning point within it does not exclude further instabilities involving modes outside the truncation.

\subsection{Numerical convergence}
\label{sec:convergence}
To check the part of the branch with $b>1$, we repeat the calculation at $b=2,3.71,4,4.5,5$ with 301 and 401 collocation points. We interpolate the 201-point solutions onto each new Chebyshev grid and apply Newton corrections to the bulk equations and all boundary conditions, including the two source-free conditions. The refinements at $b=2,3.71$ use 70-digit working precision; those on the extension use 90 digits. The final maximum equation residual is below $10^{-30}$ in the numerical normalisation. We count both endpoints: the notebook parameter $n_y$ gives $n_{\rm c}=n_y+1$ collocation points.

We extract the thermodynamic quantities again at each resolution and repeat the axisymmetric matching at fixed $(E,J)$. Table~\ref{tab:convergence} gives the changes relative to the 201-point solution at the same $b$ and horizon radius. We define
\begin{equation}
 \epsilon_{\rm grid}=\max_{X\in\{E,Q,\mu,T,S\}}
 \left|\frac{X_{n_{\rm c}}-X_{201}}{X_{201}}\right|.
\end{equation}
All relative thermodynamic changes and absolute changes in $\delta_S$ are below $10^{-20}$ at these amplitudes. This is sufficient for the precision quoted in Table~\ref{tab:nonlinear}, including the $6.43\%$ entropy gain at $b=5$. The horizon axes and sampled curvature extrema at $b=4,4.5,5$ also agree at the precision shown in Fig.~\ref{fig:horizongeometry}. At this level the thermodynamic comparison is limited by numerical precision and the conditioning of the equations; it does not resolve the asymptotic rate of spectral convergence. The entries in Table~\ref{tab:convergence} are measured changes between resolutions, rather than bounds on the absolute error.

For the extension we also check the first law without differencing widely separated solutions. If $\boldsymbol R(\boldsymbol U,b)=0$ denotes the discretised equations, including the boundary conditions, we solve
\begin{equation}
 \frac{\partial\boldsymbol R}{\partial\boldsymbol U}
 \frac{{\rm d}\boldsymbol U}{{\rm d}b}
 =-\frac{\partial\boldsymbol R}{\partial b}.
\end{equation}
Here $\boldsymbol U$ includes the radial fields and the two parameters fixed by the source-free conditions. Differentiating the boundary and horizon charge expressions then gives $E'$, $Q'$ and $S'$, with primes in this paragraph denoting $b$ derivatives. At $b=4,4.5,5$ on the grid with 301 collocation points, $|E'-TS'-\mu Q'|/|E'|$ is below $10^{-20}$. The same tangent calculation gives positive derivatives of $\alpha_{\rm MP}$ and $\delta_S$ at each of these amplitudes.

\begin{table}[htbp]
\centering
\begin{tabular}{c c r r}
\toprule
$b$ & Collocation points & $\epsilon_{\rm grid}$ & $|\delta_{S,n_{\rm c}}-\delta_{S,201}|$\\
\midrule
2.00 & 301 & $1.97\times10^{-43}$ & $1.63\times10^{-43}$\\
2.00 & 401 & $3.84\times10^{-41}$ & $3.18\times10^{-41}$\\
3.71 & 301 & $1.74\times10^{-23}$ & $1.53\times10^{-23}$\\
3.71 & 401 & $3.00\times10^{-22}$ & $2.64\times10^{-22}$\\
4.00 & 301 & $3.49\times10^{-26}$ & $3.02\times10^{-26}$\\
4.00 & 401 & $3.49\times10^{-26}$ & $3.02\times10^{-26}$\\
4.50 & 301 & $9.66\times10^{-24}$ & $8.47\times10^{-24}$\\
4.50 & 401 & $9.66\times10^{-24}$ & $8.47\times10^{-24}$\\
5.00 & 301 & $3.31\times10^{-24}$ & $2.93\times10^{-24}$\\
5.00 & 401 & $1.10\times10^{-23}$ & $9.97\times10^{-24}$\\
\bottomrule
\end{tabular}
\caption{Resolution checks beyond $b=1$. The reference solutions have 201 collocation points. At each resolution the entropy difference is evaluated against the axisymmetric solution with the same energy and angular momentum.}
\label{tab:convergence}
\end{table}

\FloatBarrier
\section{The matrix side}\label{sec:matside}
Here we simulate the classical physics of bosonic matrices numerically. We work with finite $N$ and try to take $N$ large. This is strictly speaking not the 't Hooft limit described in Section~\ref{sec:review}, but with $\tau \gg 1$ first taken at fixed $N$ and $N \to \infty$ afterwards. These two limits need not commute. In this classical limit, it appears that angular momentum assists spontaneous emission of a single D$0$-brane in the plane of rotation. The one emitted D$0$-brane appears to orbit the central clump of $N-1$ branes at a large radius in the rotation plane. We study this effect in two ways. First, using a hybrid Monte Carlo (HMC) algorithm, we sample configurations in the canonical ensemble (fixed temperature and angular momentum). As we show in Section~\ref{sec:matrixresults}, both single-clump and separated configurations are observed. Among the separated configurations, one brane appears to reside at a large radius in the rotation plane with relatively small radial fluctuations and carries angular momentum. The radius of the separated brane and its angular momentum share both grow with $N$. Second, we numerically evolve single-clump configurations drawn from the HMC sampler in real time and measure the emission lifetime against varying angular momentum. Using the real time trajectories, we confirm the orbital motion of the emitted brane. As we show in Section~\ref{sec:matrixresults}, many orbital periods are observed while the orbital radius only fluctuates by a small amount. We have, however, not established indefinite stability of this orbital motion. If the orbits are stable, this would be a very different behaviour from the ten-dimensional singly-spinning black brane background where no stable orbits exist for a probe D$0$-brane. Previous studies of classical matrix dynamics at zero angular momentum include~\cite{Asplund:2012tg}; D$0$-brane emission along the flat directions of the matrix potential was investigated in~\cite{Berkowitz:2016znt}. We describe our numerical implementation in this section and present results in Section~\ref{sec:matrixresults}.

A single emitted brane with radius increasing with $N$ raises concerns about the regime of validity of the classical approximation. We explain now that for such a separated configuration to be approximately classical, $\tau \gg 1$ is not enough; instead we need $\tau \gg (J/N)^{2/3}$, or in terms of a rotation variable kept fixed in the large-$N$ limit, $\sigma \equiv \frac{J}{N^2 \tau^{3/4}}$, we need $\tau \gg (N \sigma)^{4/3}$. The key requirement is that the off-diagonal matrix components connecting the separated D$0$ to the central clump need to have large thermal occupation for a valid classical approximation, i.e. $\frac{\omega_{\rm off}}{T} \ll  1$, where $\omega_{\rm off}$ is the typical frequency of such components. From the commutator-squared potential, $\omega_{\rm off} \sim r_{\rm orb}$, where $r_{\rm orb}$ is the radius of the separated D$0$. We now find $r_{\rm orb}$ in the classical system and bound the temperature by requiring \be\label{eq:rcondition}
r_{\rm orb}/T \ll 1.
\ee
The orbital radius can be estimated by balancing the centrifugal repulsion due to the angular momentum $J_{\rm emit}$ carried by the separated D$0$ and the effective potential from integrating out the off-diagonal modes, 
\be
F_{\rm cen}(r) = \frac{\l J^2_{\rm emit}}{2 N r^2}, \quad F_{\rm off}(r) \approx c N T \ln r,
\ee
where $c$ is some order one constant irrelevant for this scale estimate but is positive. These potentials give a stable radius at
\be\label{eq:rorb}
r_{\rm orb} = \frac{\sqrt{\lambda} J_{\rm emit}}{N \sqrt{c T}}.
\ee
Then \eqref{eq:rcondition} requires
\be
\frac{J_{\rm emit}}{N \sqrt{c \tau^{3}}} \ll 1.
\ee
Assuming the emitted D$0$ shares a order one fraction of the total angular momentum (this is supported by our results in Section~\ref{sec:matrixresults}), we can replace $J_{\rm emit}$ by $J$; ignoring order one constants, we then have
\be
\frac{N \sigma}{\tau^{3/4}} \ll 1, \quad \Rightarrow \quad \tau \gg (N \sigma)^{4/3}.
\ee

\subsection{The classical phase space}
After fixing the $A = 0$ gauge, the hamiltonian of the classical theory is 
\be
H = \frac{\lambda}{2N} \Tr \left(P^i P_i\right) - \frac{N}{4\lambda} \Tr \left([X_i, X_j]^2 \right), \quad P_i = \frac{N}{\lambda} \dot{X}_i, 
\ee
and angular momenta
\be
J^{ij} = \Tr \left(X^i P^j - X^j P^i\right). 
\ee
The gauge fixing condition leaves a $SU(N)$ gauge constraint
\be
G \equiv [X^i, P_i] = 0,
\ee
which is preserved by the hamiltonian evolution hence just restricts the physical phase space (i.e. only a constraint on initial conditions). 

The classical theory has two scaling symmetries. One is just dimensional analysis. We can use it to absorb the dimensionful constant $\lambda$ into the dynamical time $t$ and degrees of freedom $X^i$ to make them dimensionless --- we define
\be
Y^i \equiv \left(\frac{N}{\lambda}\right)^{1/3} X^i, \quad \hat{t} \equiv \left(\frac{\lambda}{N}\right)^{1/3} t.
\ee
With respect to $\hat{t}$, we have the conjugate momentum to $Y^i$ being $Q^i = \partial_{\hat{t}} Y^i = \left(\frac{\lambda}{N}\right)^{1/3} P^i $. In these new variables, we can define a more convenient set of fictitious hamiltonians and angular momenta
\be
\hat{H} \equiv \frac{1}{2} \Tr \left(Q^i Q_i \right) - \frac{1}{4} \Tr \left( [Y_i, Y_j]^2 \right), \quad \hat{J}^{ij} \equiv \Tr \left(Y^i Q^j - Y^j Q^i \right), 
\ee
which are related to the physical hamiltonian and angular momenta as
\be
H = \left(\frac{\lambda}{N}\right)^{1/3} \hat{H}, \quad J = \hat{J}.
\ee
This is just a convenient rescaling of variables such that all quantities are dimensionless, and that the hamiltonian $\hat{H}$ has no free parameters nor explicit $N$. Another scaling is
\be\label{eq:sclmap}
Y^i \to \alpha^{1/4} Y^i, \quad Q_i \to \alpha^{1/2} Q_i,
\ee
under which 
\be
\hat{H} \to \alpha \hat{H}, \quad \hat{J} \to \alpha^{3/4} \hat{J}.
\ee
This means that classical dynamics of $\hat{H}$ at different non-zero energy scales (or equivalently different temperatures) can be mapped between each other by a simple scaling, at the cost of rescaling the angular momentum (if there is any). In other words, nothing distinguishes the system at different temperatures, apart from a ratio between the angular momenta and temperature 
\be
\sigma = \frac{\hat{J}}{N^{7/4} \hat{T}^{3/4}},
\ee
where $\hat{T}$ is the temperature associated with $\hat{H}$. The extra $N^{7/4}$ ensures this is the same definition of $\sigma$ as above once transformed back into the physical frame. The upshot is that using these two scaling symmetries, the full classical phase space labelled by $(N, \lambda, T, J)$ can be obtained by scaling from the subspace spanned by $(N, \sigma)$. In what follows, we will work with the system governed by $\hat{H}$, and scale back to the physical system only at the end. 

The classical function with fixed angular momenta $\hat{J}^{ij}$ and temperature $\hat{T}$ is given by
\be \label{eq:part}
Z_N(\hat{T}, \hat{J}^{ij}) = \int {\rm d}Y\,{\rm d}Q \, e^{- \hat{H}/\hat{T}} \, \delta^{(N^2-1)}\left(G\right) \prod_{i<j} \delta \left(\hat{J}^{ij} - 2\Tr Y^{[i}Q^{j]} \right), 
\ee
where ${\rm d}Y\,{\rm d}Q$ are products of standard Haar measure on $N$-by-$N$ hermitian traceless matrices. We focus on the case where the only non-zero components of $\hat{J}^{ij}$ are $\hat{J}^{12} = - \hat{J}^{21} = \hat{J}$. An equivalent way of seeing the equivalence map between states at different temperatures is to make the change of variables in \eqref{eq:sclmap} inside the partition function, taking $\alpha = \hat{T}$: 
\be \label{eq:partrescl}
Z_N\left(\hat{T}, \hat{J}^{ij}\right) = \hat{T}^{\frac{3}{4}[8(N^2-1) - 36]} Z_N \left(1, \hat{J}^{ij} \hat{T}^{-3/4}\right).
\ee

From \eqref{eq:part} we can deduce two relations of the system at equilibrium. Both of them are used to verify our numerical algorithm. One is the virial theorem relating the average kinetic and potential energies:
\be \label{eq:virial}
\langle \hat{K} \rangle = 2 \langle \hat{V} \rangle,
\ee
where $\hat{K} \equiv \frac{1}{2} \Tr Q^i Q_i$ and $\hat{V} \equiv -\frac{1}{4} \Tr [Y_i,Y_j]^2$. Here $\langle \cdot \rangle$ denotes average in the ensemble \eqref{eq:part}. This relation can be derived by considering a change of integration variables $Y \to \lambda Y$ and $Q \to \lambda^{-1} Q$, and requiring the partition function to be invariant. Another one is the equation of state, relating the average total energy to temperature and angular momenta: 
\be \label{eq:eos}
\langle \hat{H} \rangle = \frac{3}{4} \left[ \left(8 (N^2-1) - 36 \right) \hat{T} + \left \langle \frac{\hat{J}^2}{I_{\text{eff}}(Y)} \right \rangle \right], 
\ee
where $I_{\text{eff}}(Y)$ only depends on $Y^i$ and can be interpreted as an effective moment of inertia. It can be computed as follows. It is useful to think of an array of hermitian traceless matrices, like $Y^i$ or $Q^i$ in our problem, as an element of the vector space $\mathcal{V} = \mathfrak{su}(N) \otimes \mathbb{R}^9$. An inner product can be defined on $\mathcal{V}$ as
\be \label{eq:inner}
(U, V) \equiv \Tr (U^i V_i), \quad \text{for} \ U, V \in \mathcal{V}. 
\ee
The constraints are all linear in $Q^i$ at fixed $Y^i$, so together they can be thought of as a projection onto a constraint surface in $\mathcal{V}$. Now we proceed to compute the effective moment of inertia. First, we define the vector
\be\label{eq:wform}
w_i \equiv i [Y_i, Z] + K_{ij} Y^j,
\ee
where $Z$ is hermitian traceless and $K^{ij} = - K^{ji}$ have real entries. Then we demand
\be\label{eq:condw}
[w^i, Y_i] = 0, \quad \text{and} \ \Tr\left(Y^i w^j - Y^j w^i \right) = \epsilon^{ij},
\ee
where $\epsilon^{ij}$ is just the anti-symmetric polarisation tensor in the (12)-plane, i.e. $\epsilon^{12} = -\epsilon^{21} = 1,$ and all other components vanishing. For generic $Y^i$,  \eqref{eq:condw} are independent linear equations for $Z$ and $K^{ij}$; they can be numerically solved to find $w^i(Y)$. This $w^i(Y)$ is the vector satisfying \eqref{eq:condw} with minimal norm. The effective moment of inertia is given by
\be
I_\text{eff} (Y) = \left(w(Y), w(Y) \right)^{-1}. 
\ee
Indeed, it would be exactly $\Tr (Y_1^2 + Y_2^2)$, the natural candidate moment of inertia, had we not imposed the gauge condition nor fixed all other components of $\hat{J}^{ij}$ to zero. The derivation of \eqref{eq:eos} is in Appendix \ref{app:eos}. Using the virial theorem to turn \eqref{eq:eos} into an expression for the average kinetic energy, we have
\be
\langle \hat{K} \rangle = \frac{2}{3} \langle \hat{H} \rangle = \frac{8 (N^2-1) - 36 }{2} \hat{T} + \left \langle \frac{\hat{J}^2}{2I_{\text{eff}}(Y)} \right \rangle.
\ee
Each term on the right-hand-side has a clear interpretation, i.e. the equipartition of thermal energy and the rotational kinetic energy, respectively. 

\subsection{Hybrid Monte Carlo}\label{sec:hmcmethod}
We start by sampling the classical partition function of the bosonic matrix model using a hybrid Monte Carlo (HMC) algorithm \cite{Duane:1987de} (see also \cite{Jha:2021exo, Martina:2026qxg} for applications to matrix models).  The algorithm repeats the update procedure below and reproduces \eqref{eq:part} as its stationary distribution after equilibrium without finite-step bias:
\begin{enumerate}
    \item Draw a random noise $\xi_i$ from a Gaussian hermitian traceless ensemble of width $\sqrt{\hat{T}}$. 
    
    \item Deform $\xi_i$ to 
        \be\label{eq:deformQ}
        Q_i = \xi_i + i [Y_i,Z] + K_{ij} Y^j,
        \ee
        where $Z$, a hermitian traceless matrix, and  $K_{ij}$, real antisymmetric in $(i,j)$, are unknowns. Solve for $Z$ and $K_{ij}$ such that
        \be
        [Q^i, Y_i] = 0, \quad 2 \Tr Y^{[i}Q^{j]} = \hat{\jmath}^{ij}.
        \ee
    
    \item Choose $h = \pm \delta t$ with equal probabilities, and then evolve the system along the hamiltonian flow with a leap-frog integrator which is the following updates in sequence:
    \be
    \begin{aligned}
        Q_i &\to Q_i + \frac{h}{2} F_i(Y),\\
        Y^i &\to Y^i + h Q^i,\\
        Q_i &\to Q_i + \frac{h}{2} F_i(Y),
    \end{aligned}
    \ee
    where 
    \be
    F_i(Y) = - \frac{\partial \hat{H}}{\partial Y^i} = -[Y^j, [Y_j, Y_i]]. 
    \ee

    \item Compute the change in total energy, $\Delta \hat{H}$, between the configuration after $n$ leap-frog updates and starting point. Accept the endpoint with probability $\min \left(1, e^{-\Delta \hat{H}/\hat{T}} \right)$. If rejected, it retains the state after step 2 and before step 3.
\end{enumerate}
A full HMC sampling repeats the steps above $\mathcal{N}$ times starting from a random initial $Y^i$. The procedure in step 2 is equivalent to orthogonally projecting $\xi_i$ onto its image $Q_i$ in the constraint hypersurface. This projection is an affine transformation so $Q_i$ is again a Gaussian random noise of spread $\sqrt{\hat{T}}$ but confined to the constraint hypersurface. Fixing non-zero $\hat{\jmath}^{ij}$ breaks time-reversal symmetry, so the random sign choice $h = \pm \delta t$ in step 3 is required to guarantee reversibility of the leap-frog proposals. Each leap-frog update preserves both the Gauss law and angular momentum constraints, and also preserves the phase space volume. Combining these with the Metropolis selection in step 4, the standard detailed-balance argument in \cite{Duane:1987de} applies, and therefore the procedure samples \eqref{eq:part} exactly without bias even at finite $\delta t$; $\Delta \hat{H}$ is only non-zero due to finite-step error which vanishes as $\delta t \to 0$. A proof of the finite-step exactness of this algorithm is in Appendix~\ref{app:HMCproof}. Because the HMC chain starts from a random initial $Y^i$, a number of burn-in steps are required before equilibrium is reached. The precise number is only determined a posteriori, see Section~\ref{sec:matrixresults}. As noted above, the only physical parameter at fixed $N$ is $\sigma$; our numerical runs scan values of $\sigma$ by fixing $\hat{J}$ and varying $\hat{T}$. This is just a choice -- equivalently one could fix $\hat{T}$ and vary  $\hat{J}$. 

As we show in Section~\ref{sec:matrixresults}, the output samples can be approximately grouped into single-clump and emitted-D$0$ configurations. We take the criterion for single-clump to be
\be\label{eq:blobcondition}
\frac{\lambda_1}{\lambda_2} < 4,
\ee
where $\lambda_{1,2}$ are the largest and second largest eigenvalues of $R^2 \equiv Y_1^2 + Y_2^2$. This is a proxy for defining a D$0$ brane as emitted if its in-plane radius is at least twice the radius of the central core. This definition is a choice; one might as well choose a different threshold. As shown by Figure~\ref{fig:emitted} in Section~\ref{sec:matrixresults}, the orbital radius of the emitted D$0$-brane increases with $N$ and is usually much larger than this criterion, so the two groups of configurations are cleanly separated in the large $N$ limit\footnote{The increase of orbital radius $r$ with $N$ is expected heuristically from a balance between the centrifugal barrier, which scales like $\hat{J}_{\rm emit}^2/r^2$, and the attractive potential from integrating out the order $N$ off-diagonal matrix elements, which scales like $N \hat{T} \log r$. Other details such as the numerical coefficients in front of these potentials, the angular momentum distribution, and the eccentricity of the orbit would affect the precise relation.}. We also measure the fraction of angular momentum carried by the emitted D$0$-brane, defined by a projection onto the subspace spanned by the largest eigenvector of $R^2$: 
\be\label{eq:Jsplit}
\begin{aligned}
    \hat{J}^{12}_{\rm emit} &\equiv 2\Tr \left( \Pi Y^{[1} \Pi Q^{2]} \right), \\
    \hat{J}^{12}_{\rm string} &\equiv 2 \Tr \left(\Pi Y^{[1} \Bar{\Pi} Q^{2]}\right) + 2 \Tr \left(\Bar{\Pi} Y^{[1} \Pi Q^{2]}\right), \\ 
    \hat{J}^{12}_{\rm clump} &\equiv 2\Tr \left(\Bar{\Pi} Y^{[1} \Bar{\Pi} Q^{2]} \right), \\
    \hat{J^{12}} &= \hat{J}^{12}_{\rm emit} + \hat{J}^{12}_{\rm string} + \hat{J}^{12}_{\rm clump}
\end{aligned}
\ee
where $\Pi$ is the projector built from the eigenvector corresponding to the largest $R^2$ eigenvalue and $\Bar{\Pi} = \mathbb{1} - \Pi$. Both projectors transform covariantly under $SU(N)$ so definitions in \eqref{eq:Jsplit} are invariant. The angular momentum carried by the orbiting D$0$-brane is $\hat{J}^{12}_{\rm emit}$, while $\hat{J}^{12}_{\rm string}$ and $\hat{J}^{12}_{\rm clump}$ are respectively interpreted as that carried by strings connecting the orbiter to the central clump and that carried by the clump itself. 

Expectation values can be estimated by averaging over the HMC samples. If the samples are not completely independent, the uncertainty on such estimates is widened. The amount of widening depends on the correlation between HMC samples and on the observable whose expectation value is calculated, quantified by the autocorrelation function
\be
\rho_{\mathcal{O}} (\Delta s) = \frac{\langle \delta \mathcal{O}(s + \Delta s) \delta \mathcal{O}(s) \rangle}{\langle \delta \mathcal{O}^2 \rangle}, \quad \text{where} \ \delta \mathcal{O}(s) \equiv \mathcal{O}(s) - \langle \mathcal{O} \rangle,
\ee
where $s$ labels the sample by its step number on the HMC chain and $\Delta s$ the lag between samples in units of steps. All autocorrelation functions computed later in this work are restricted to small enough lags such that their value comes from averaging at least $50$ pairs of samples. To measure how quickly the correlation falls, we also compute the integrated autocorrelation length $\lambda_\mathcal{O}$, defined as
\be
\l_{\mathcal O} \equiv \frac{1}{2}+\sum_{\Delta s=1}^{\infty}\rho_{\mathcal O}(\Delta s).
\ee
Completely independent samples would have $\lambda_{\mathcal{O}} = \frac{1}{2}$. To see its relevance, compare the variance on the sample mean $\langle \mathcal O \rangle $ against the variance of $\mathcal{O}$ itself in the sample
\be\label{eq:var}
\begin{aligned}
\operatorname{Var}(\langle \mathcal{O} \rangle)
&=
\frac{\operatorname{{Var}(\mathcal{O})}}{\mathcal N^2}
\sum_{s,s'=1}^{\mathcal N}
\rho_{\mathcal O}(|s-s'|) \\
&=
\frac{\operatorname{{Var}(\mathcal{O})}}{\mathcal N}
\left[
1+2\sum_{\Delta s = 1}^{\mathcal N-1}
\left(1-\frac{\Delta s}{\mathcal N}\right)
\rho_{\mathcal O}(\Delta s)
\right].
\end{aligned}
\ee
In the large sample size limit, assuming
$\mathcal N\gg\lambda_{\mathcal O}$ and a summable autocorrelation
function, this becomes
\be\label{eq:widening}
\lim_{\mathcal{N} \to \infty} \operatorname{Var}(\langle \mathcal O \rangle) = \frac{2\lambda_{\mathcal O}}{\mathcal N} \operatorname{{Var}(\mathcal{O})}.
\ee
Thus $\lambda_{\mathcal{O}}$ measures the effective number of independent samples as
\be
\mathcal{N}_{\rm eff} =
\frac{\mathcal N}{2\lambda_{\mathcal O}}.
\ee
In the case of completely independent samples, $\mathcal{N}_{\rm eff} = \mathcal{N}$, and $\lim_{\mathcal{N} \to \infty} \operatorname{Var}(\langle \mathcal O \rangle) = \mathcal{N}^{-1} \operatorname{{Var}(\mathcal{O})}$ as usual. In practice, the large-$\Delta s$ tail of $\rho_{\mathcal O}(\Delta s)$ is noisy, so we use the Madras-Sokal self-consistent windowing prescription \cite{Sokal} to estimate $\lambda_{\mathcal{O}}$. We first define a windowed length 
\be
\lambda_{\mathcal O}(W) \equiv  \frac{1}{2} + \sum_{\Delta s = 1}^{W} \hat\rho_{\mathcal O}(\Delta s).
\ee
where $W < \mathcal{N}-1$. Then we find $W$ as the smallest value such that
$W\geq c\,\lambda_{\mathcal O}(W)$, where $c$ is a chosen constant. We choose $c=5$, within the range used in \cite{Sokal}; the estimated $\lambda_{|\mathcal{O}_2|}$ we report later in this manuscript are stable to within $2 \%$ over $c \in [4,6]$. The associated uncertainty for this estimate is given by \cite{Sokal}
\be\label{eq:lambdaerr}
\operatorname{Var}(\lambda_\mathcal{O}) \approx \frac{2 (2 W +1)}{\mathcal{N}} \lambda_\mathcal{O}^2(W).
\ee
When computing ratios of averages using the same set of samples, the uncertainty on the ratio should not be built from combining uncertainties on the numerator and denominator in quadrature. This is because using the same set of samples correlates the two averages. For a ratio like $\frac{\langle a \rangle}{\langle b \rangle}$, we instead use the effective sample size derived from the autocorrelation function of 
\be
\frac{a}{\langle b \rangle} - \frac{\langle a \rangle}{\langle b \rangle^2} b.
\ee
Also, because the autocorrelation function is fluctuation dominated at long lag, we set it to zero for lags beyond the Madras-Sokal self-consistent window when computing the effective sample size.

\subsection{Real time dynamics}
Configurations drawn by the HMC algorithm are classical microstates of a thermal macrostate. To measure the decay rate, we select single-clump configurations at fixed $N$ and $\sigma$ generated by HMC, evolve them classically with $\hat{H}$ in real time using a leap-frog algorithm, and measure the rate of single-D$0$ emission. The emission criterion is the complement of \eqref{eq:blobcondition}. We also require this criterion to be satisfied for at least one sustained time unit measured with $\hat{t}$. The emission time is then the beginning of this confirmation interval. If a threshold crossing happens within one time unit of the maximum observation time, the trajectory is counted as not emitted. Due to storage limitations, we only record $\lambda_1, \lambda_2, \mathcal{O}_{2,3,4}, \Tr[(Y_3 + \rm{i} Y_4)^2], \hat{H}$ every $0.1$ time units, the full spectra of $R^2$ and $Y_{1,2,3}$ every $1.0$ time units, and the full $(Y^i, Q_i)$ only every $10$ time units.

\subsubsection{Decay time estimate and uncertainties}\label{sec:decaytime4}
We assume that the decay follows a Poisson process with constant decay rate, i.e. the survival probability follows a pure exponential decay with time. This is an approximation and it is checked a posteriori with real-time data in Section~\ref{sec:matrixresults}. Drawing $n_{\rm IC}$ HMC samples and labelling their corresponding trajectories by $I$, time elapsed before emission for the $I$th trajectory by $t_{\rm em}^I$, and the total time observed $t^I_{\rm obs}$ for each trajectory, the maximum-likelihood estimate (MLE) for the decay time (inverse of the decay rate) $t_{\rm decay}$ is given by 
\be\label{eq:texp}
\hat{t}_{\rm exp} \equiv \sum_I \min(\hat{t}_{\rm em}^I, \hat{t}^I_{\rm obs}), \quad \hat{t}_{\rm decay} = \frac{\hat{t}_{\rm exp}}{n_{\rm em}} ,
\ee
where $n_{\rm em}$ is the number of trajectories with emission observed, and $\hat{t}_{\rm exp}$ is called the exposure time. Because the real-time emission rate is not an HMC observable, its autocorrelation cannot be measured directly. We use the autocorrelation of $|O_2|$ as a proxy for the correlation between selected initial conditions. The resulting error inflation should be interpreted as a sensitivity estimate rather than an exact sampling variance. The error widening is then captured approximately by the effective sample size defined as
\be\label{eq:deff}
n_{\rm eff} \equiv \frac{n_{\rm IC}^2}{\sum_{IJ} \rho_{|\mathcal{O}_2|}(|s_I - s_J|)}, \quad f \equiv \frac{n_{\rm eff}}{n_{\rm IC}},
\ee
where the sum runs over pairs drawn from the same chain; pairs from different chains contribute zero. The auto-correlation function is noise dominated beyond the Madras-Sokal self-consistent window, so we set it to zero for lags beyond that; this means that for chains in Table~\ref{tab:hmc-metadata} that are too short for a self-consistent window estimate, this $n_{\rm eff}$ is less reliable. The error widening from correlation is captured by an effective-sample-size prescription, i.e. replacing usual formulae for uncorrelated samples by $(\hat{t}_{\rm exp}, n_{\rm em}) \to (f \hat{t}_{\rm exp}, f n_{\rm em})$. Under this replacement, the maximum-likelihood estimate does not change. Using the same prescription, we approximate the lower bound $\hat{t}_{-}$ and upper bound $\hat{t}_+$ of the $1-\alpha$ confidence level interval using the Garwood interval~\cite{Garwood:1936kqp}
\be\label{eq:Garwood}
\hat{t}_- = \frac{ 2 f \hat{t}_{\rm exp}}{Q(1 -\alpha/2; 2 f n_{\rm em} + 2)}, \quad \hat{t}_+ = \frac{2 f \hat{t}_{\rm exp}}{Q(\alpha/2;2 f n_{\rm em})},
\ee
where $Q(p; \nu)$ denotes the inverse cumulative $\chi^2$-distribution function, also called the quantile function; $Q(p; \nu)$ returns the value up to which the $\chi^2$-distribution function with $\nu$ degrees of freedom integrates to $p$. The parameter $1-\alpha$ specifies the nominal coverage of the interval. For independent samples ($f = 1$) with fixed exposure, the Garwood interval would be exact\footnote{Exactness here means that the interval has an infimum coverage $ 1-\alpha$ under varying the ratio between the exposure and lifetime. }. Applied to our data, it is only nominal for two reasons. First, the effective-sample-size prescription is only approximate. Second, even at $f=1$, the Garwood coverage applies if the observations were done at fixed exposure whereas our observation strategy described in Section~\ref{sec:obstime} results in random and outcome-dependent exposure. The impact of the second effect on the lifetime estimate and coverage is tested in Section~\ref{sec:obstime}. For the real data, we report the Garwood interval at $68.3\%$ nominal confidence level ($\alpha = 0.317$) if $n_{\rm em} \geq 3$. For those with $n_{\rm em} \leq 2$, we instead report the $95\%$ nominal confidence level lower bound, 
\be\label{eq:lowerbound}
\hat{t}_{\rm decay} > \frac{2 f \hat{t}_{\rm exp}}{Q(0.95; 2 f n_{\rm em} +2)}.
\ee

An alternative way of estimating the error is performing a moving-block bootstrap on the observed decay lifetimes \cite{Kunsch1989}. Within each run, the trajectories are ordered by the HMC step of their initial conditions. We call $l$ consecutive trajectories in this order a \textit{block} with length $l$. Then we randomly draw $\lceil n_{\rm IC}/l\rceil$ such blocks from this ordered list of trajectories, allowing overlaps of arbitrary lengths including complete repetitions. We concatenate these blocks and truncate the tail to form a new set of trajectories with the same length $n_{\rm IC}$, but now some trajectories may appear more than once while others may not appear at all. Each such new set is called a \textit{resample}, and we compute the MLE lifetime from each resample. Each resample keeps together whether a trajectory has emitted and its contribution to $t_{\rm exp}$, i.e. $\hat{t}_{\rm exp}$ and $n_{\rm em}$ are computed from the same resample and their ratio formed. Resamples with $n_{\rm em} = 0$ are assigned $\hat{t}_{\rm decay} = \infty$ \footnote{For runs with $n_{\rm em} \geq 10$, among the $4000$ resamples none returned zero emission. Enlarging to $40000$ resamples, up to $0.005\%$ resamples are assigned $\hat{t}_{\rm decay} = \infty$ which occurs at $N = 20, \ \sigma = 1.243, \ l = 2$.}. The $16$th to $84$th percentile of the $\hat{t}_{\rm decay}$ estimate over $4000$ such resamples is the error estimate. At the end, we vary $l$ to find the lowest lower bound and highest upper bound as the error estimate. If $n_{\rm IC}$ was large enough to allow picking $1 \ll l \ll n_{\rm IC}$, the error estimate would plateau at large $l$. However, our data sets do not have large enough $n_{\rm IC}$, so $l$ becomes comparable to $n_{\rm IC}$ before the plateau is reached. When $l$ and $n_{\rm IC}$ are comparable, the error estimate falls again simply because there are only a few different blocks. Therefore, we take the maximum error instead. Of the $35$ points with at least ten emissions (listed in Table~\ref{tab:decay-metadata}), the block bootstrap gives a narrower interval than the quoted one at $32$ points, the quoted half-width being larger by a median factor of $1.61$. At the remaining three the block interval is wider: by $35\%$ at $N=20$, $\sigma=1.338$, by $22\%$ at $N=10$, $\sigma=1.465$, and by $9\%$ at $N=17$, $\sigma=1.385$. We nevertheless retain the $|\mathcal{O}_2|$ widening in the quoted intervals, because the $n_{\rm IC}$ we have are not enough to give a reliable block estimate. The correlations relevant to the uncertainty in $\hat{t}_{\rm decay}$ need not be the same as those of $|\mathcal{O}_2|$. A genuinely accurate error estimate would come from an independent-chain sampling. That would require completely new HMC chains and real time evolution of independent initial conditions. We have not done that due to the computational cost.

\subsubsection{Adaptive observation time}\label{sec:obstime}
To balance the precision of measurements and the computational cost, we decide the observation time $\hat{t}_{\rm obs}^{I}$ for the real time trajectories adaptively as follows: We call the set of all trajectories with the same $(N, \sigma)$ a \textit{run}. Trajectories are evolved in chunks of 100 time units and check-pointed at each boundary. We stop a run once all of its trajectories had emitted (according to the criterion in \eqref{eq:blobcondition}). The initial scheduled ceiling is $\hat{t}=3000$ and then adaptively extended to $\hat{t}=9000$; these are runs in Table~\ref{tab:obs-windows} with a single $\hat{t}_{\rm obs}$ being either $3000$ or $9000$. Two runs at $N=20$ with $\sigma = 1.243$ and $1.304$ have pre-set observation ceilings $19000$ and $30000$. For runs with two registered $\hat{t}_{\rm obs}$, two batches of trajectories are evolved to different pre-set maximal observation ceilings and results pooled; for example $(N, \sigma) = (20, 1.374)$ has one batch stopped at $\hat{t}_{\rm obs} = 3000$ and another at $9000$. No run at $N=40$ has two batches. To save computation, whenever the survivors fell to half or fewer of the currently active trajectories, emitted trajectories were removed from the integration. Some runs have all trajectories emitted before their pre-set observation ceiling, so their maximum observation times $\hat{t}_{\rm obs}^{\rm max}$ as recorded in Table~\ref{tab:obs-windows} are shorter. The emitted trajectories have their emission times recorded before being removed, so the lifetime estimates are unaffected, but their configuration records end at that point.

The decay time estimate through \eqref{eq:texp} is subtly different for runs with all trajectories emitted and those with some survived trajectories. When some trajectories are censored, the lifetime estimate relies more strongly on the constant-rate assumption and is more of an extrapolation than measurement. To be precise about how much of the estimate for each run should be interpreted as extrapolation, we report the share of the exposure time contributed by trajectories that never emitted within the maximal observation time:
\begin{equation}
  f_{\rm cens}=\frac{1}{\hat t_{\rm exp}}
  \sum_{I\,\in\,\mathrm{censored}}\hat t_{{\rm obs}}^I,
\end{equation}
where we call such un-emitted trajectories \textit{censored}.
Table~\ref{tab:obs-windows} lists, for every $(N,\sigma)$, the longest
window among the runs pooled there, the windows at which trajectories were
censored with the number censored at each, and $f_{\rm cens}$. Each
censored trajectory enters with its own observation time, so at $N=10$,
$\sigma=1.198$, the censored exposure is $21\times 9000+90\times 27000$. 

Our observation protocol does not fix either the exposure or the emission count, but is instead a fixed-observation-time design with adaptive extension. The formula for the MLE remains the same, but its sampling distribution is different. This would affect the coverages of the confidence intervals \eqref{eq:Garwood} and \eqref{eq:lowerbound} which are adapted to the fixed-exposure protocol. We estimate the scale of this bias by simulating our adaptive protocols on a fictitious decay process with a known decay rate. Here we are simulating only the bias caused by the observation protocol, excluding effects from correlation between trajectories (i.e. setting $f = 1$). Synthetic emission times were drawn as independent exponential variates with a prescribed mean $t_{\rm true}$, representing what would be seen with infinite observation time;
an observation window was then imposed, so that a trajectory whose drawn time exceeded its window was recorded only as censored; and the exposure, the emission count, and the uncertainty interval were formed as in Section~\ref{sec:decaytime4}. The problem is scale-independent, meaning results only depend on time scales measured in units of $t_{\rm true}$ and on the batch size. To mimic the procedure used on real data, we take a first-stage window $t_1/t_{\rm true}=10$, $10/3$, $1$, $1/3$, $1/10$ and
$1/30$, an extended window $t_2=\frac{10}{3}t_1$. We take three different run sizes with $n_{\rm IC}=32$, $65$ and $120$ trajectories. In total, we have eighteen configurations. We generate $2 \times 10^6$ synthetic emission histories for each configuration, and then filter them under each of the six protocols to produce measurements mimicking observation. The first protocol keeps the window at $t_1$ and never extends it. The next three carry the survivors on to $t_2$ whenever the fraction that has emitted by $t_1$ falls below $0.1$, $0.5$ or $0.8$. The fifth extends exactly when fewer than three trajectories have emitted. The last extends whenever the interval computed at $t_1$ would have missed $t_{\rm true}$; this rule is not realisable since it requires knowing the ground-truth answer, and is included as an adversarial test.

Table~\ref{tab:cov-cv} shows the mean of the MLE of the decay lifetime among the $2 \times 10^6$ experiments with Monte Carlo error bars for each configuration under each observation protocol. For experiments with zero emissions, the MLE is infinite and the Garwood interval is one-sided with only a lower boundary. The reported mean is computed by excluding zero-emission experiments. Table~\ref{tab:cov-frac} shows the fraction of experiments with zero emissions and those with $\leq 2$ emissions (the latter includes the former). Only configurations with $t_1/t_{\rm true}\leq 1/10$ have a significant fraction of such experiments. Table~\ref{tab:cov-cov} shows the coverage of the Garwood interval and Table~\ref{tab:cov-lcov} the $95 \%$ lower bound, measured as the fraction of those $2 \times 10^6$ experiments in which $t_{\rm true}$ lies within the intervals. Table~\ref{tab:cov-cal} lists the calibration factor of the $95 \%$ lower bound; calling the lower bound $L$, the calibration factor is defined as $t_{\rm true}$ divided by the $95$th percentile of the distribution of $L$ formed from all experiments. This factor should be unity for an exactly calibrated bound, and a large calibration factor corresponds to a higher confidence level (i.e. the nominal bound is below the actual $95\%$ bound). Across all $108$ combinations of configuration and protocol, the mean MLE ranges between $0.793$ to $1.343$ of the true lifetime; the Garwood interval has coverage between $70.3\%$ and $96.2\%$, and the $95\%$ lower bound covers between $95.4\%$ and $100\%$. The statistical uncertainties on these coverages are at most $0.03 \%$ and $0.02 \%$, respectively.

There is a clear bias on the decay time MLE as indicated by the mean ratio to the ground truth. However, the bias may be less severe than the extremal values suggest. First, experiments with no emission are dropped from computing the mean; this would pull down the mean when the zero-emission fraction is large. This effect would be present even if the observational protocol fixes the exposure. The two cells with highest zero-emission fraction ($n_{\rm IC} = 32$, $t_1 = t_{\rm true}/30$, fixed and adv.) account for the two lowest recorded means\footnote{Working with the decay rate (inverse of the lifetime) avoids the problem of infinite MLE for zero-emission experiments. The mean rate estimator should be unbiased under fixed exposure. Results are shown in Table~\ref{tab:cov-rate}. The mean rate bias for these two cells become $1.017$ and $0.885$. However, bias of the rate is not the same as bias of the lifetime because of their reciprocal relation.}. Also, these two cells have $91\%$ of experiments yielding two emissions or fewer, so our reporting rule would likely quote the $95\%$ lower bound instead. Excluding these two cells would raise the lower end of the range to $0.967$. Second, the bias range shrinks steeply with the expected number of emissions. For cells with average $n_{\rm em}$ between $10$ and $40$ (including $10$) the mean lies in $[0.967,1.103]$, and those at least $40$ in $[1.000, 1.015]$. Among the real data points plotted in Figure~\ref{fig:lifetime}, $12$ out of $41$ have $10 \leq n_{\rm em} < 40$, and $23$ more have $n_{\rm em} \geq 40$.

This synthetic test is only a test on the scale of bias rather than an exact calibration. It examines representative adaptive-extension rules, but does not reproduce the full combination of batch pooling and observation windows used in our data. It reveals a bias in the MLE estimate of the lifetime under observation protocols mimicking our procedure. However, the amount is unlikely to be severe enough to change the qualitative feature in Figure~\ref{fig:lifetime} where the lifetime varies across orders of magnitude. Among the configurations and extension protocols we have tested, the coverages of the Garwood interval and the lower bound are always above their nominal level, the smallest margins being $61$ and $24$ standard deviations, respectively.

An additional subtlety is that the constant-rate assumption is violated in the short observation time limit: the emission criterion requires the outer-most eigenvalue to climb to the threshold first, so a very short observation window means that trajectories spend a larger share of their exposure in the process of climbing to the threshold. Our observation times are long enough that this has negligible effect on the decay time estimates. We measure this effect directly by re-censoring every trajectory at a common cut $\hat{t}_{\rm obs} = 3000$. This truncation changes the measured lifetime by a factor whose median is $1.008$ over the twenty-seven points at which the cut removes data, with a $16$th--$84$th percentile range of $0.78$--$1.16$. Thus this check finds little median shift across the scan while individual estimates change. The threshold sensitivity check in Section~\ref{sec:validation} shows that the qualitative trend of faster separation with increasing angular momentum persists, although the numerical lifetimes depend on the emission criterion.

\section{Results in the classical matrix model}\label{sec:matrixresults}

\subsection{Hybrid Monte Carlo}
\begin{figure}[htbp]
\centering
\begin{subfigure}{0.48\textwidth}
\includegraphics[width=\linewidth]{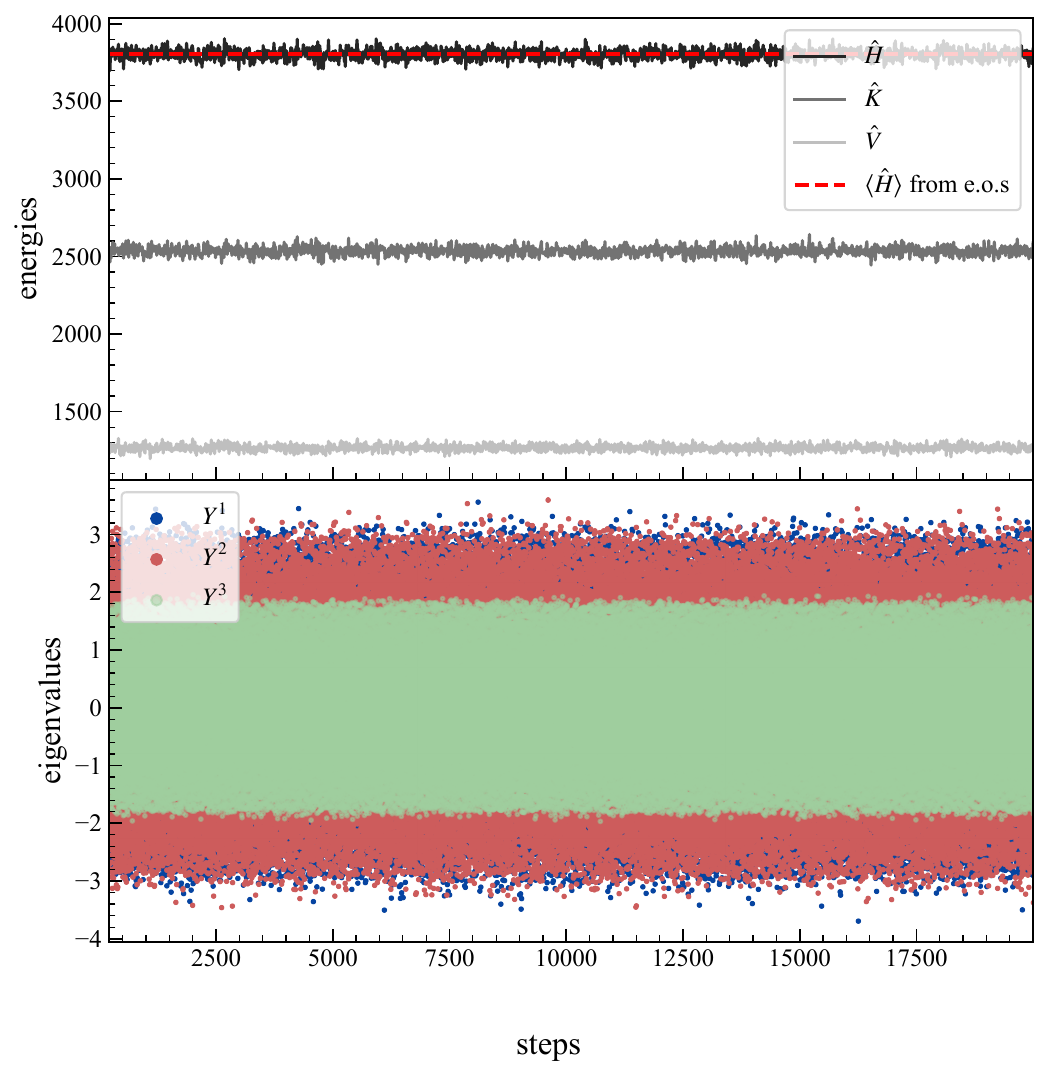}
\caption{$\sigma=1.217$.}
\end{subfigure}\hfill
\begin{subfigure}{0.48\textwidth}
\includegraphics[width=\linewidth]{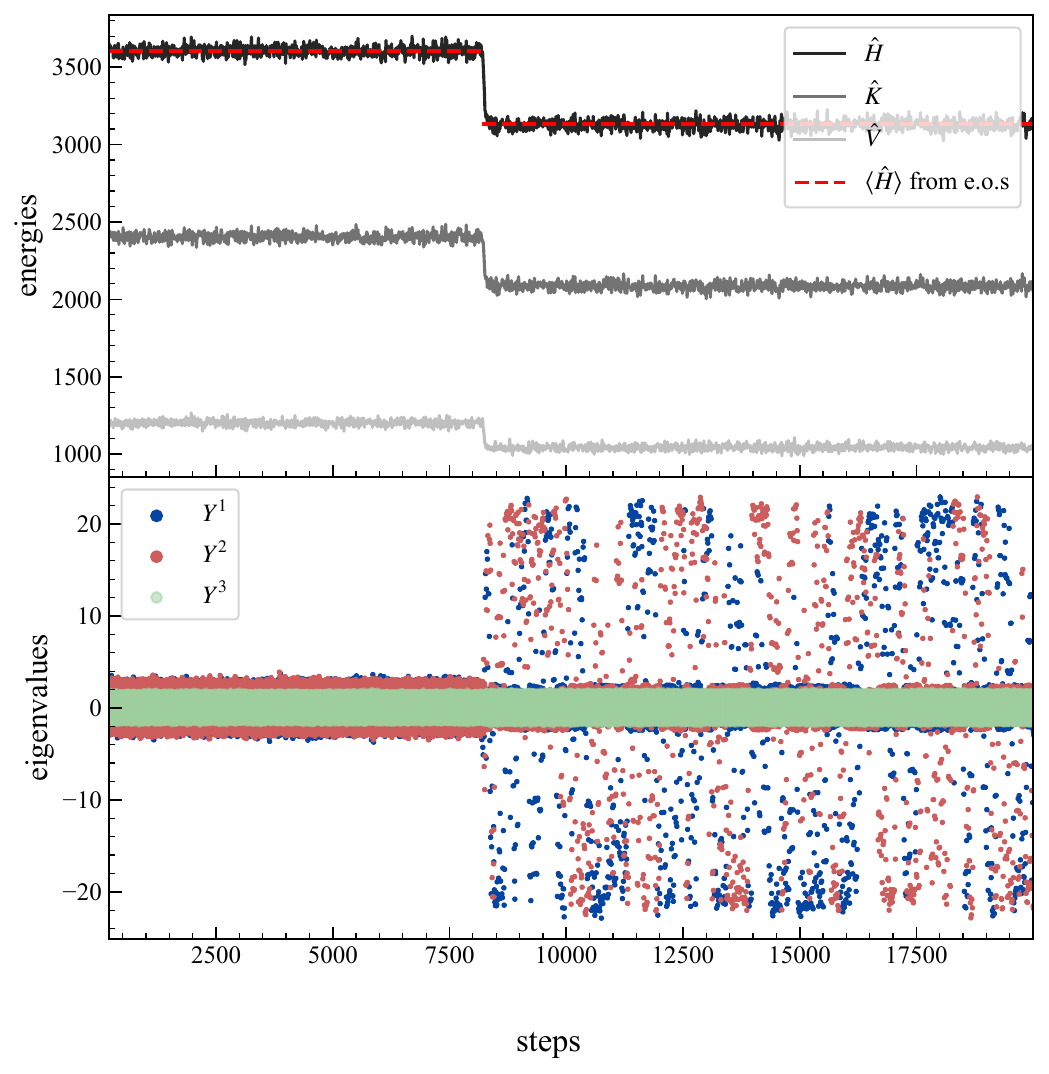}
\caption{$\sigma=1.275$.}
\end{subfigure}
\caption{Representative HMC histories at $N=40$. The upper panels show the total, kinetic and potential energies, together with the equation-of-state prediction for the mean total energy. The lower panels show eigenvalues of $Y^1$, $Y^2$ and $Y^3$. The larger-spin history exhibits separation in the rotation plane.}
\label{fig:matrixhistories}
\end{figure}

\begin{figure}[htbp]
\centering
\includegraphics[width=0.9\textwidth]{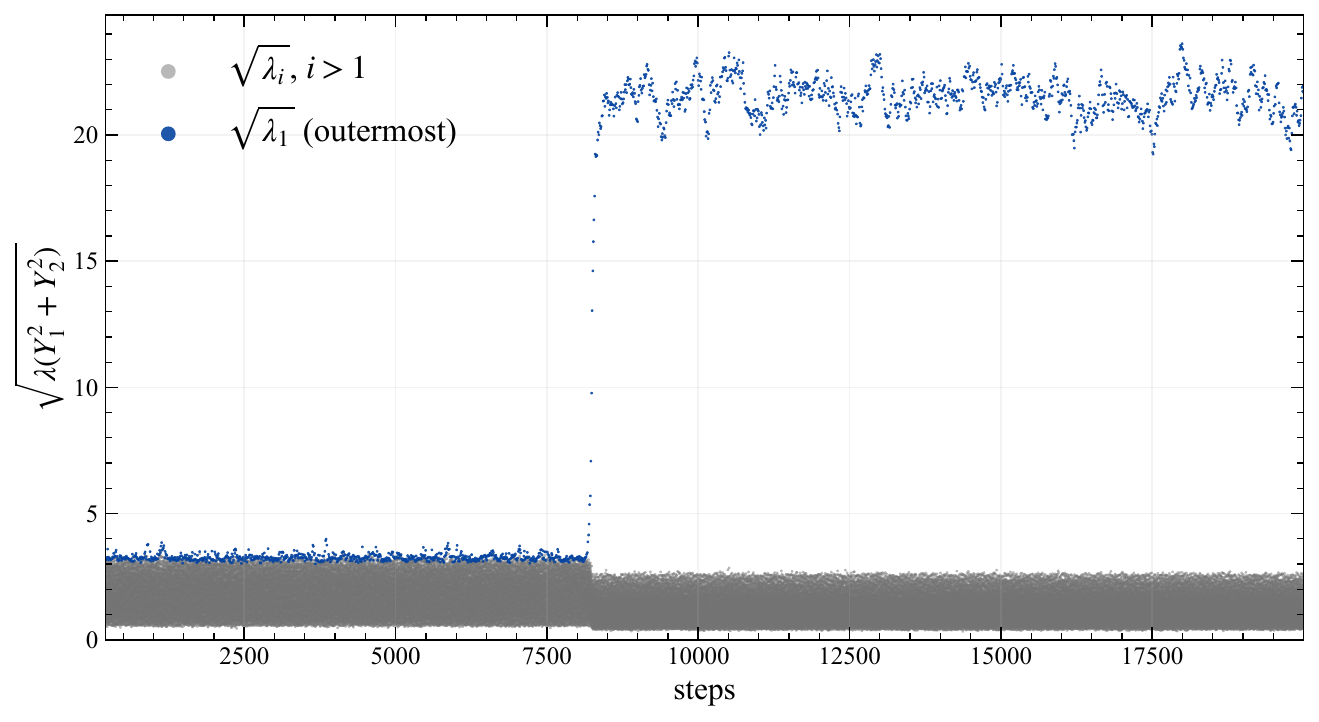}
\caption{Eigenvalue history of $R = \sqrt{Y_1^2 + Y_2^2}$ for the HMC chain with $N = 40$ and $\sigma = 1.275$. The separated eigenvalue only explores a small range of values compared to the mean separation.}
\label{fig:radii}
\end{figure}

\begin{figure}[htbp]
\centering
\begin{subfigure}{0.48\textwidth}
\includegraphics[width=\linewidth]{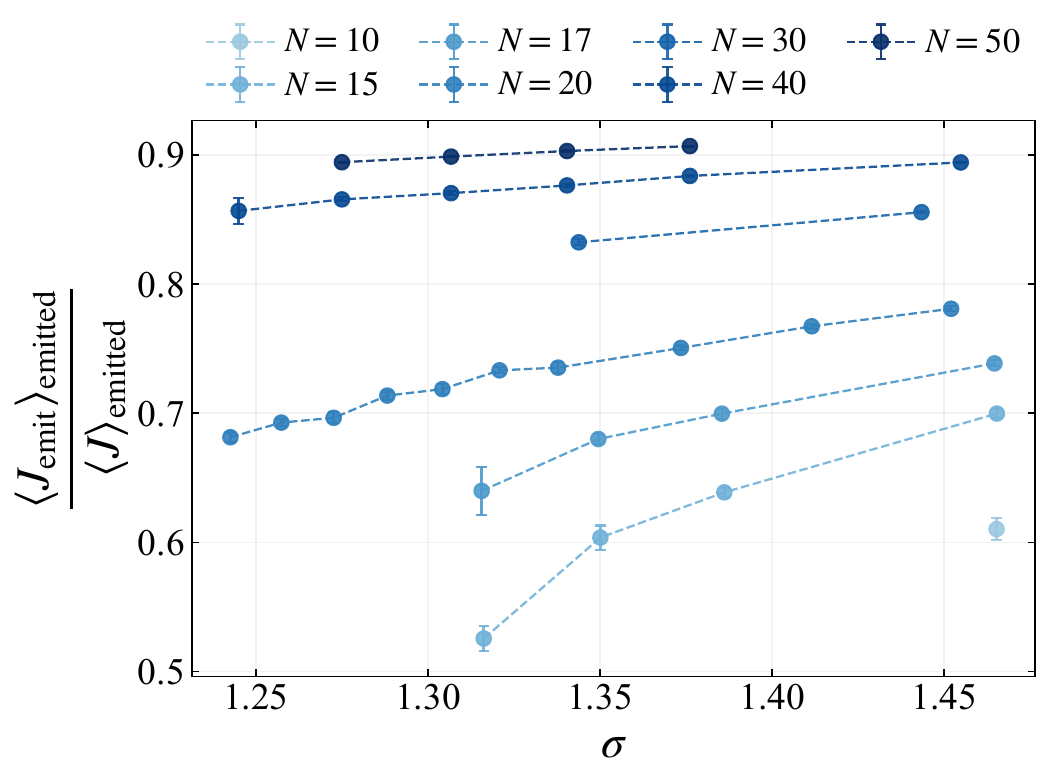}
\caption{Angular momentum share.}
\end{subfigure}\hfill
\begin{subfigure}{0.48\textwidth}
\includegraphics[width=\linewidth]{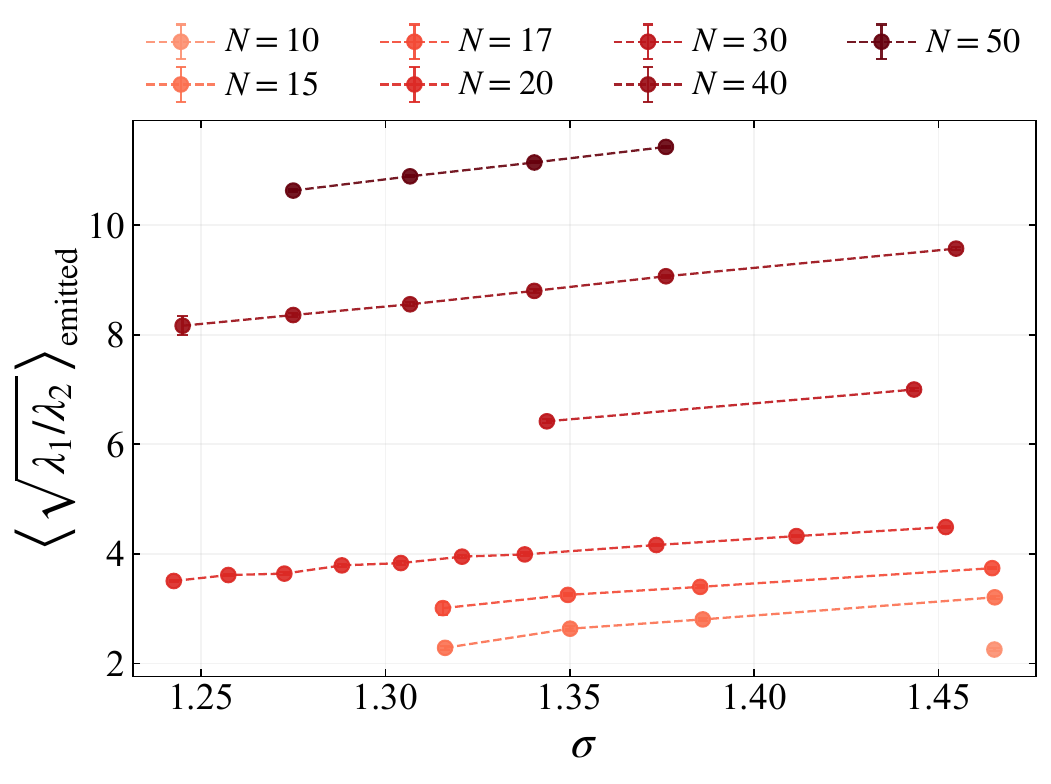}
\caption{Orbital radius.}
\end{subfigure}
\caption{Average angular momentum share and in-plane orbital radius of the emitted brane. Both are ratios invariant under rescaling of variables, hence are the same regardless of measured with $(X, P)$ or $(Y, Q)$. Both grow with increasing $N$, qualitatively agreeing with the interpretation of a single emitted D$0$-brane carrying most of the angular momentum through large orbital radius.}
\label{fig:emitted}
\end{figure}

Metadata of the full HMC scan is recorded in Table~\ref{tab:hmc-metadata}. We choose the burn-in length to be the last HMC step with energy above twice the median of the whole chain. The median burn-in length among our chains is $35$ steps with a maximum of $48$ steps, and they do not vary much if the criterion was changed to between $1.5$ to $10$ times the median energy. Figure~\ref{fig:matrixhistories} shows two representative HMC histories at $N=40$ after burn-in. Equilibrium is checked by both the virial theorem and the equation of state. At $\sigma=1.217$ the matrices remain in a single clump over the displayed history. At $\sigma=1.275$ an eigenvalue separates in the $(12)$-plane after approximately $8\times10^3$ sampling steps. The corresponding eigenvalues of $Y^1$ and $Y^2$ then explore radii much larger than the core, whereas $Y^3$ remains concentrated around the origin. The change is accompanied by a decrease in the sampled energy. The mean energies shown by the dashed lines agree with the equation-of-state prediction used in the simulations. At low $\sigma$, the sampler almost always outputs single-clump matrix configurations. As $\sigma$ is increased, we find that the sampler explores more quickly configurations with one almost-simultaneous eigenvalue of $X_{1,2}$ emitted. Figure~\ref{fig:radii} shows the radial eigenvalues in the $(12)$-plane for the $\sigma = 1.275$ chain. The horizontal axes here count HMC updates not physical time. The large $Y^1$ and $Y^2$ eigenvalues are correlated such that the separated radial eigenvalue only explores a small range of values compared to the overall separation. Combined with Figure~\ref{fig:emitted} showing the angular momentum share of the emitted brane, our results are consistent with the interpretation of these as configurations of a single emitted D$0$-brane orbiting the central clump.

\begin{figure}[htbp]
\centering
\includegraphics[width=0.7\textwidth]{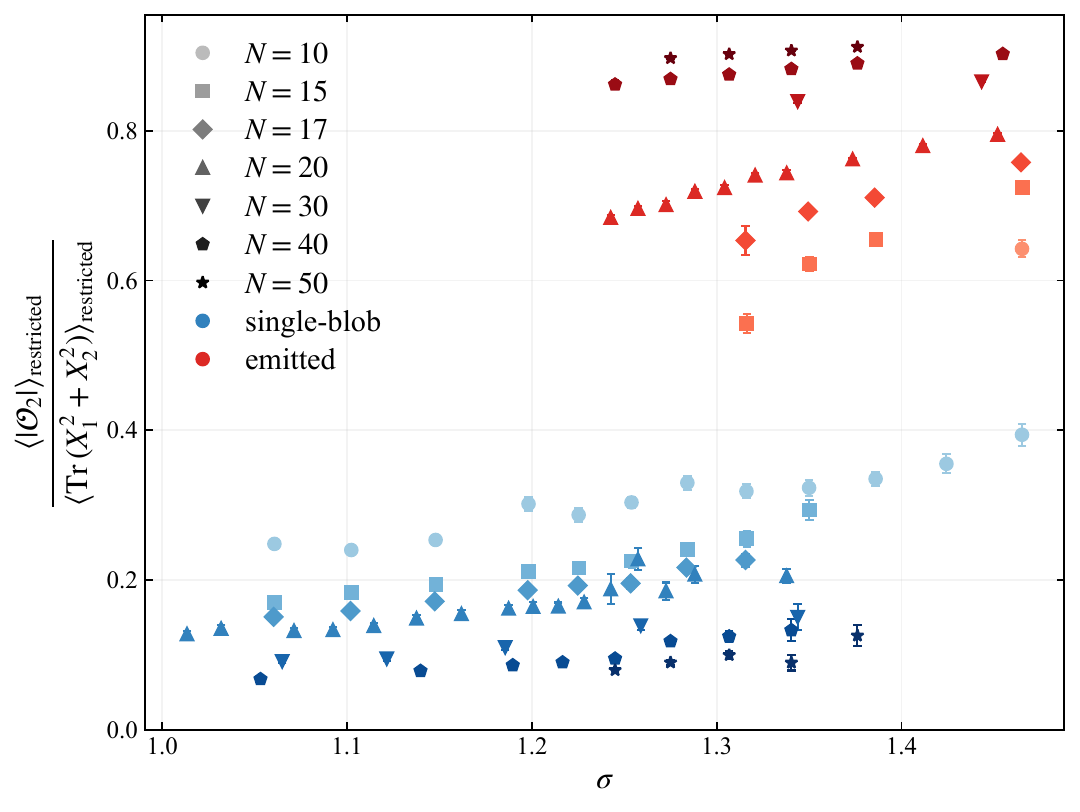}
\caption{Expectation values of the normalised quadrupole magnitude separately from the single-clump and emitted samples. Results from single-clump samples (coloured blue) decrease with $N$, indicating no appreciable quadrupole moment; those from emitted samples (coloured red) are significant and increase with $N$, but are due to the single outlier eigenvalue not a collective mode.}
\label{fig:matrixo2}
\end{figure}

We also compute the expectation value of the magnitude of the quadrupole normalised by the spread in $(12)$-plane, $  \frac{\langle|\mathcal{O}_2|\rangle}{\langle \text{Tr} (X_1^2 + X_2^2) \rangle} $, for each $(N, \sigma)$ separately when restricted to single-clump and emitted samples. The results are shown in Figure~\ref{fig:matrixo2}. We use the magnitude here because $\mathcal{O}_2$ would average to zero due to the rotational symmetry. Since this ratio is invariant under rescaling of $X^i$, it is equivalent to the result measured from $Y^i$. Only chains with a pre-fixed maximum step are used\footnote{This means no measurement is available at $(N, \sigma) = (15,1.424)$, $(17, 1.424)$, $(30, 1.391)$, $(40, 1.414)$, $(50, 1.414)$, and ($50, 1.455)$}. The uncertainties are computed through the effective sample size as described in Section~\ref{sec:hmcmethod}; the autocorrelation functions are restricted to segments of the chain in the same type of configuration. Results from segments in which the Madras-Sokal self-consistency condition is never met are not included in the plot. Among the single-clump data points, the minimum effective sample size is $18.6$, median $242.4$, and maximum $1220.2$; among the emitted points, the minimum is $13.9$, median $138.5$, and maximum $313.4$. There is no appreciable expectation value among the single-clump samples and the value decreases with $N$. This is consistent with having no collective quadrupole in the single-clump region. In contrast, there is significant quadrupole in the emitted region, increasing with $N$; however, this is just due to the single large-radius eigenvalue rather than a collective mode similar to the gravity side. This is in stark contrast to the gravity results where the quadrupole arises from a collective deformation rather than a single outlying component. We verify that the outlier dominates the quadrupole by computing the quadrupole carried by the remaining central clump for the emitted samples. We project $Y^{1,2}$ to the clump subspace, remove its centre of mass, and define the quadrupole for the leftover clump as 
\be
\tilde{Y}^{1,2} \equiv \Bar{\Pi} Y^{1,2} \Bar{\Pi} - \frac{\Tr(\Bar{\Pi} Y^{1,2})}{N-1}\Bar{\Pi},\quad \tilde{\mathcal{O}}_2 \equiv \Tr((\tilde{Y}^1 + \mathrm{i} \tilde{Y}^2)^2).
\ee
Using the same selection of chains and methods for computing uncertainties, we arrive at results in Table~\ref{tab:clump-quad}, ratios between the clump and the full quadrupole. The ratios are generally small and again decrease with $N$.

\begin{table}[htbp]
\centering
\small
\setlength{\tabcolsep}{4pt}
\begin{tabular}{rrrl|rrrl}
\toprule
$N$ & $\sigma$ & $\mathcal{N}^{\rm emitted}_{\rm eff}$ & \multicolumn{1}{c|}{$\frac{\langle|\tilde{\mathcal{O}}_2|\rangle_{\rm emitted}}{\langle|\mathcal{O}_2|\rangle_{\rm emitted}}$} & $N$ & $\sigma$ & $\mathcal{N}^{\rm emitted}_{\rm eff}$ & \multicolumn{1}{c}{$\frac{\langle|\tilde{\mathcal{O}}_2|\rangle_{\rm emitted}}{\langle|\mathcal{O}_2|\rangle_{\rm emitted}}$} \\[4pt]
\hline
  15 & 1.350 & 84.2 & $0.0826 \pm 0.0057$ & 20 & 1.411 & 792.9 & $0.02231 \pm 0.00044$ \\
  15 & 1.386 & 406.7 & $0.0642 \pm 0.0019$ & 20 & 1.452 & 873.1 & $0.02044 \pm 0.00038$ \\\cline{5-8}
  15 & 1.465 & 466.6 & $0.0451 \pm 0.0011$ & 30 & 1.344 & 645.2 & $0.00951 \pm 0.00020$ \\\cline{1-4}
  17 & 1.316 & 30.2 & $0.0555 \pm 0.0070$ & 30 & 1.443 & 791.6 & $0.00785 \pm 0.00015$ \\\cline{5-8}
  17 & 1.349 & 228.9 & $0.0449 \pm 0.0017$ & 40 & 1.245 & 113.3 & $0.00621 \pm 0.00037$ \\
  17 & 1.385 & 423.6 & $0.0422 \pm 0.0012$ & 40 & 1.275 & 1104.6 & $0.005521 \pm 0.000093$ \\
  17 & 1.465 & 658.3 & $0.03121 \pm 0.00064$ & 40 & 1.307 & 1181.7 & $0.005320 \pm 0.000083$ \\\cline{1-4}
  20 & 1.243 & 500.9 & $0.03932 \pm 0.00096$ & 40 & 1.340 & 1447.9 & $0.005047 \pm 0.000071$ \\
  20 & 1.257 & 357.6 & $0.03768 \pm 0.00099$ & 40 & 1.376 & 1878.6 & $0.004660 \pm 0.000055$ \\
  20 & 1.273 & 236.4 & $0.0363 \pm 0.0013$ & 40 & 1.455 & 1744.2 & $0.004041 \pm 0.000051$ \\\cline{5-8}
  20 & 1.288 & 192.7 & $0.0305 \pm 0.0013$ & 50 & 1.275 & 1510.9 & $0.003453 \pm 0.000046$ \\
  20 & 1.304 & 360.4 & $0.03145 \pm 0.00099$ & 50 & 1.307 & 1792.9 & $0.003163 \pm 0.000039$ \\
  20 & 1.321 & 415.9 & $0.02972 \pm 0.00078$ & 50 & 1.340 & 1752.8 & $0.003024 \pm 0.000038$ \\
  20 & 1.338 & 436.7 & $0.02753 \pm 0.00067$ & 50 & 1.376 & 2021.1 & $0.002868 \pm 0.000033$ \\
  20 & 1.374 & 691.7 & $0.02400 \pm 0.00048$ &  &  &  \\
\bottomrule
\end{tabular}
\caption{The clump quadrupole among emitted samples. The effective sample size rounded to one decimal place is listed under $\mathcal{N}^{\rm emitted}_{\rm eff}$. The error bars are kept to two significant figures with the precision levels of the averages matched.}
\label{tab:clump-quad}
\end{table}

\begin{figure}[htbp]
\centering
\begin{minipage}[c]{0.50\textwidth}
  \centering
  \includegraphics[width=\linewidth]{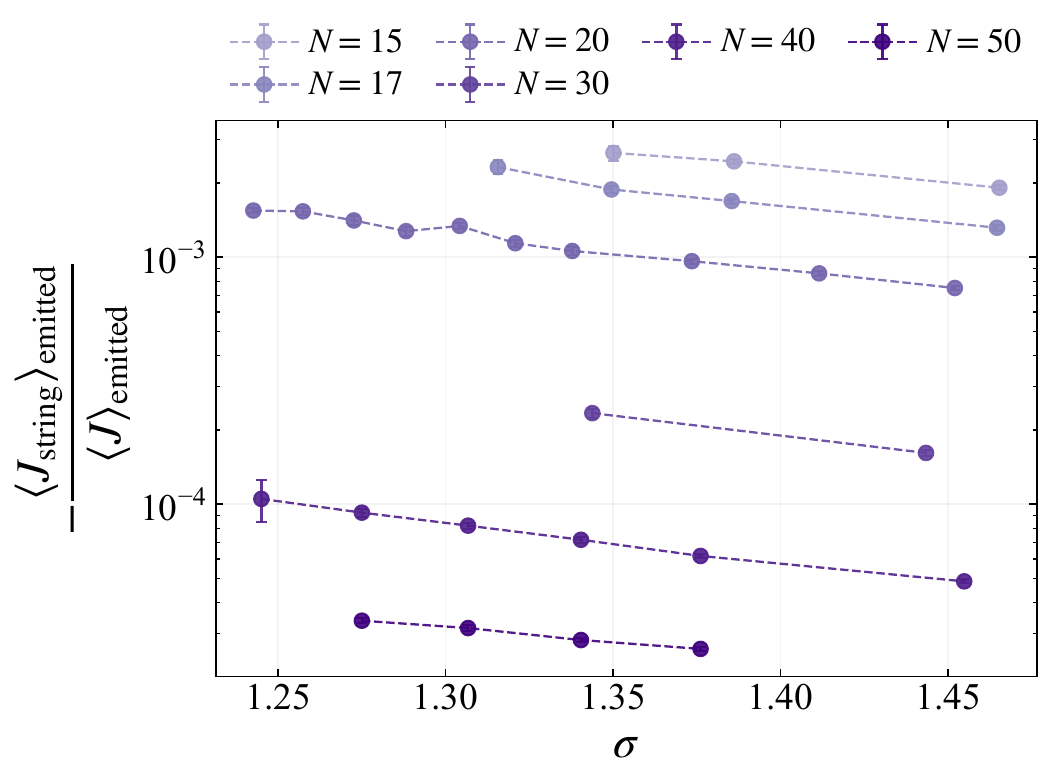}
\end{minipage}
\hfill
\begin{minipage}[c]{0.47\textwidth}
  \centering
  \footnotesize
  \setlength{\tabcolsep}{3pt}
  \begin{tabular}{rrr|rrr}
    \toprule
    $N$ & $\sigma$ & $f_{-}$ (\%) & $N$ & $\sigma$ & $f_{-}$ (\%) \\
    \hline
    15 & 1.350 & 92.1 & 20 & 1.411 & 95.7 \\
    15 & 1.386 & 91.2 & 20 & 1.452 & 96.4 \\\cline{4-6}
    15 & 1.465 & 93.9 & 30 & 1.344 & 94.4 \\\cline{1-3}
    17 & 1.316 & 93.9 & 30 & 1.443 & 94.3 \\\cline{4-6}
    17 & 1.349 & 96.2 & 40 & 1.245 & 93.5 \\
    17 & 1.385 & 94.9 & 40 & 1.275 & 94.0 \\
    17 & 1.465 & 95.0 & 40 & 1.307 & 93.7 \\\cline{1-3}
    20 & 1.243 & 95.4 & 40 & 1.340 & 93.4 \\
    20 & 1.257 & 97.0 & 40 & 1.376 & 92.4 \\
    20 & 1.273 & 92.7 & 40 & 1.455 & 90.8 \\\cline{4-6}
    20 & 1.288 & 96.2 & 50 & 1.275 & 91.6 \\
    20 & 1.304 & 96.7 & 50 & 1.307 & 91.8 \\
    20 & 1.321 & 96.2 & 50 & 1.340 & 90.9 \\
    20 & 1.338 & 94.7 & 50 & 1.376 & 91.5 \\
    20 & 1.374 & 95.3 &  &  &  \\
    \bottomrule
  \end{tabular}
\end{minipage}
\caption{The string contribution to the angular momentum of emitted
configurations. \emph{Left}: negative of the string share of the angular momentum averaged over the emitted configurations on log scale. \emph{Right}: A table showing the percentage of
individual samples with $J_{\rm string} < 0$, labelled by $f_-$, for
every point shown at left.}
\label{fig:Jstring}
\end{figure}

Restricting to the emitted configurations, we measure how the angular momentum share and the average orbital radius change with $(N, \sigma)$. Results are shown in Figure~\ref{fig:emitted}. As in Figure~\ref{fig:matrixo2}, we only use chains stopped at a pre-fixed maximum HMC step; the same prescription for computing the uncertainties is used; and results from short segments not enough to support a self-consistent window are not included. For the angular momentum share, the minimum effective sample size is $13.8$, median $196.2$, and maximum $1127.3$; for the radius ratio, the minimum is $11.0$, median $153.0$, and maximum $509.3$. With increasing $N$, the single emitted D$0$-brane carries larger fractions of the total angular momentum (reaching approximately $90 \%$ for $N=50$ within the range of $\sigma$ studied). We also computed the average $J_{\rm string}$ normalised by the total $J$, and results are shown in the left panel of Figure~\ref{fig:Jstring}. The magnitudes of this share are very small and get smaller with increasing $N$. It is interesting to note that all data points included have a negative average $J_{\rm string}$; moreover, as listed in the right panel of Figure~\ref{fig:Jstring}, an overwhelming portion of emitted samples have a consistent negative $J_{\rm string}$\footnote{The total angular momentum $J_{\rm emit} + J_{\rm string} + J_{\rm clump}$ remains agreeing with the fixed $J$ sample by sample to $\mathcal{O}(10^{-14})$, so this sign is not caused by numerical rounding error. All $31$ independent chains return negative $J_{\rm string}$, so it is unlikely due to correlations within each chain.}. We did not manage to find an explanation for this phenomenon. Back to Figure~\ref{fig:emitted}, the relative orbital radius, the squared-rooted ratio between the largest and second largest eigenvalues of $R^2$, also increases with $N$. This justifies a posteriori the claim in Section~4 that the single-clump and emitted configurations are cleanly separated at large $N$. 

We note that results in Figures~\ref{fig:matrixo2}, \ref{fig:emitted}, and \ref{fig:Jstring} are conditional averages, not an equilibrium average for either subset of samples. To find an equilibrium average and the correct weighting for each type of configuration, there need to be several transitions both from single-clump to emitted configurations and the reverse. The great majority of our HMC chains only contain one transition without the reverse. Also, the chains used have pre-determined stopping points, so the proportion of these two configurations in each chain is determined by the stopping point not equilibrium. Therefore, equilibrium weights for these two types of configurations cannot be computed with the samples available.

\subsection{Real time dynamics}
\begin{figure}[htbp]
\centering
\includegraphics[width=0.9\textwidth]{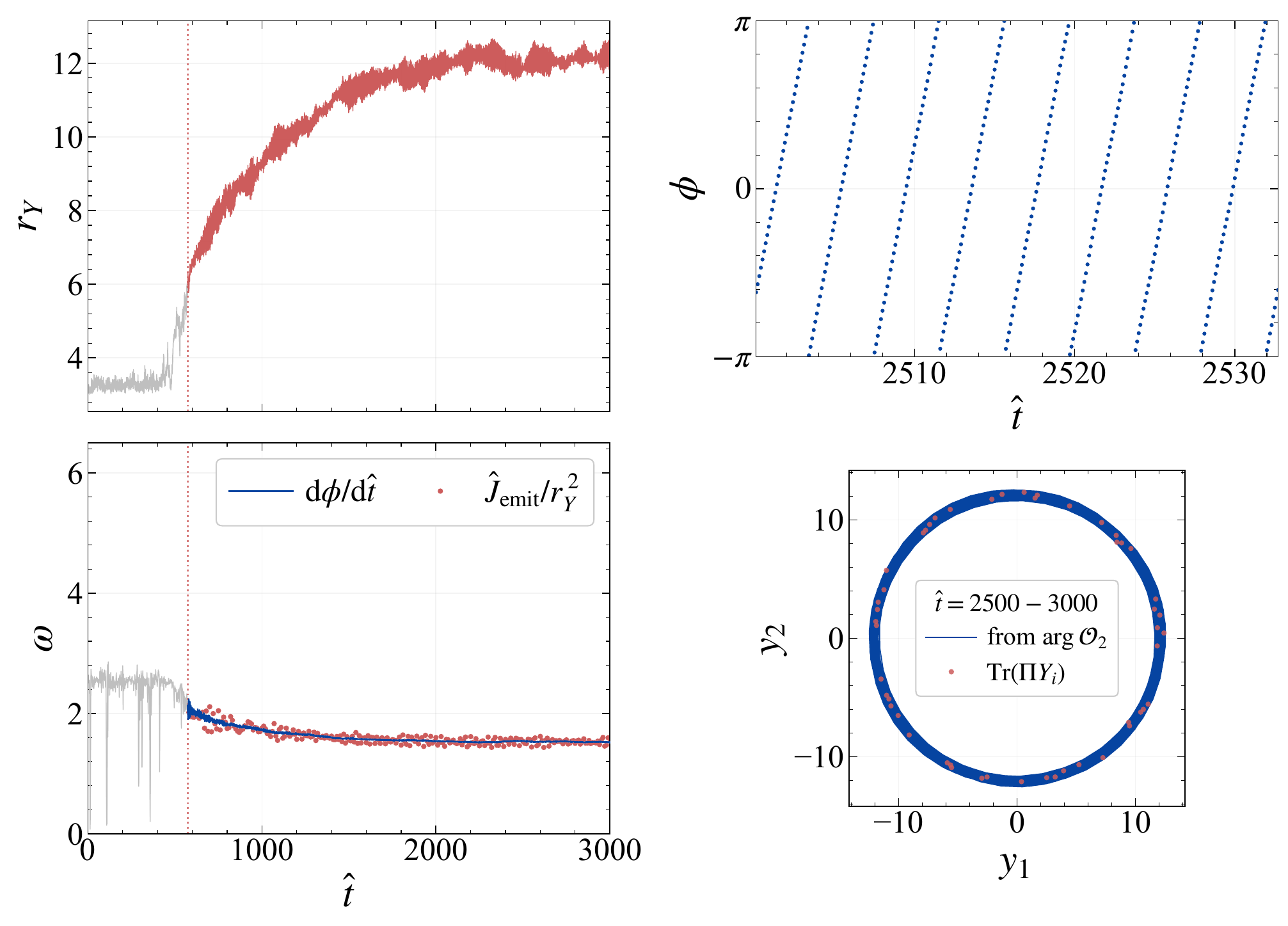}
\caption{Real time trajectory with $N = 40, \sigma = 1.340$. The top-left panel shows the largest eigenvalue $r_Y$ of $R = \sqrt{Y_1^2 + Y_2^2}$, also denoted $\sqrt{\l_1}$ elsewhere. The top-right panel shows the phase of the emitted brane, $\phi$, reconstructed from the phase of $\mathcal{O}_2$, taking value between $\pm \pi$. The bottom-left panel shows angular velocity measured in two ways: the time-derivative of $\phi$ (blue), and the value derived from angular momentum share and radius (red). The bottom-right panel shows the trajectory of the emitted brane, in blue that reconstructed from $r$ and $\phi$ as in the top panels, and in red direct measurements from projecting $Y_{1,2}$ to the subspace of the largest $R^2$ eigenvalue. The two left panels plot the entire trajectory; the top-right panel zooms in to $2500 < \hat{t} < 2530$ to show just a number of periods, and the bottom-right panel shows $2500 < \hat{t} < 3000$. All data coloured red are direct measurements from the real-time trajectory with a projection, and those in blue involve reconstruction using $\mathcal{O}_2$. }
\label{fig:rt_orbit}
\end{figure}

The claim above that the separated eigenvalue represents an orbiting D$0$-brane is supported by real-time evolution after emission. An example trajectory at $N = 40$ and $\sigma = 1.340$ is shown in Figure \ref{fig:rt_orbit}. The in-plane radius of the emitted brane oscillates in a small range compared to the average value, and the angular velocity remains approximately constant. Because we only store the full $(Y^i, Q_i)$ every $10$ time units, it is not enough to resolve the orbiting angle of the emitted brane. We instead use an angle $\phi$ reconstructed from the phase of $\mathcal{O}_2$ since it is dominated by the emitted brane. We first calculate the phase of $\mathcal{O}_2$ as defined in the range $[-\pi, \pi)$; then unwrap the phase by adding $2\pi$ to the current and subsequent values whenever the phase drops by more than $\pi$, and subtracting $2\pi$ whenever it increases by more than $\pi$; then divide the unwrapped phase by $2$ to get $\phi_{\rm unwrap}$; finally we map each $\phi_{\rm unwrap}$ back into $[-\pi, \pi)$ using $\phi = (\phi_{\rm unwrap} + \pi)\  \mathrm{mod} \ 2\pi - \pi$. There is a remaining $\phi \to \phi + \pi$ ambiguity after this procedure which we fix using $Y^{1,2}$ projected to the subspace of the emitted brane. We check that this is a good representation by finding good agreement between the angular velocity derived from this $\phi$ to that estimated from $\hat{J}_{\rm emit}/r_Y^2$, where $r = \sqrt{\l_1}$ is the radius of the brane, also shown in Figure~\ref{fig:rt_orbit}. 

\begin{figure}[htbp]
\centering
\includegraphics[width=0.78\textwidth]{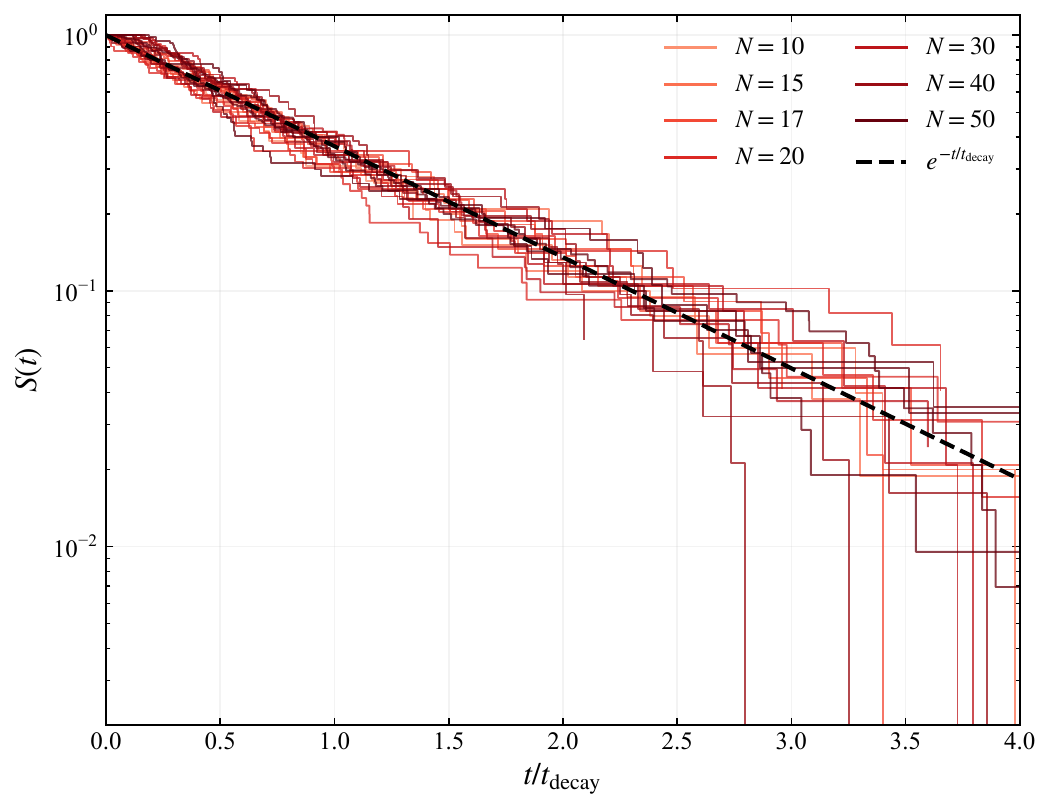}
\caption{Survival probability for the classical real-time trajectories, with time measured in units of the estimated decay time. The dashed line is $e^{-t/t_{\rm decay}}$. The fluctuations at late times reflect the smaller number of surviving trajectories.}
\label{fig:survival}
\end{figure}

To measure the lifetime, we draw a number of single-clump configurations from each HMC chain, evolve them in real time, and apply the emission criterion described in Section~4. We use step size $\delta \hat{t} = 0.005$ for the leap-frog evolution. The survival probability is the fraction of trajectories that have not yet emitted a D0-brane, estimated by a Kaplan-Meier curve
\be
S(\hat{t}) = \prod_{\hat{t}_i \leq \hat{t}}\left(1 - \frac{n_{\rm em}(\hat{t}_i)}{n_{\rm risk}(\hat{t}_i)}\right),
\ee
where $n_{\rm em}(\hat{t}_i)$ is the number of emissions at $\hat{t}_i$ and $n_{\rm risk}(\hat{t}_i)$ the number of trajectories still being observed and not emitted before $\hat{t}_i$. Figure~\ref{fig:survival} plots this probability against $t/t_{\rm decay}$. The data for the different matrix sizes follow the exponential $e^{-t/t_{\rm decay}}$ over the measured range, supporting the constant-rate description used to estimate the lifetime. The spread in energy and lifetime within a canonical run is measured in Section~\ref{sec:energyspread}. The rescaling by the fitted lifetime removes its dependence on $N$ and $\sigma$; the collapse tests the shape of the survival law, rather than establishing an $N$-independent decay rate.

\begin{figure}[htbp]
\centering
\includegraphics[width=0.78\textwidth]{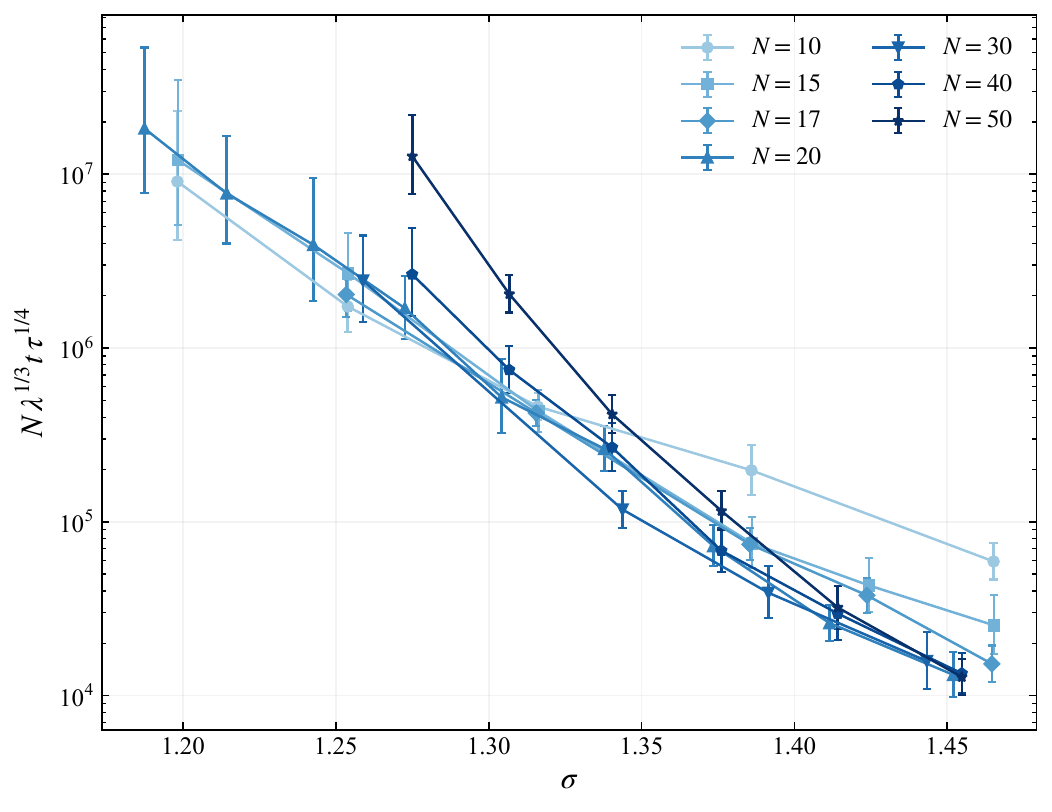}
\caption{Scaled decay time as a function of $\sigma$ for $N=10,15,17,20,30,40,50$. Increasing angular momentum accelerates single-D0 emission from the classical clump. The ordinate is the lifetime estimated from the real-time trajectories.}
\label{fig:lifetime}
\end{figure}

The dependence of the lifetime on angular momentum is listed in Table~\ref{tab:decay-metadata} and plotted in Fig.~\ref{fig:lifetime}. For all matrix sizes studied, increasing $\sigma$ strongly reduces the decay time. For example, the $N=20$ data span more than three orders of magnitude over the displayed interval. For $N\gtrsim20$, the combination $N\lambda^{1/3}t_{\rm decay}\tau^{1/4}$ (the dimensionless and scale invariant decay time scale multiplied by $N$) approximately collapses at the larger values of $\sigma$; the smaller matrix sizes and the lower-spin data show more pronounced differences. The finite range of $N$ does not by itself determine the asymptotic large-$N$ lifetime.

These results exhibit two effects of rotation: a collective quadrupolar bifurcation in the gravitational regime and enhanced single-D$0$ emission in the classical regime. The eigenvalue histories establish separation in the rotation plane. Over the matrix sizes and spins studied, the single-clump data show no evidence for a collective quadrupolar condensate. The emitted samples exhibit large quadrupole but that is due to the single emitted D$0$ brane, not a collective mode. On the gravity side, the scalar probe in Appendix~\ref{app:greybody} can extract energy from the rotating black hole in a range of frequencies, while the radial potential can strongly suppress transmission to infinity at large $N$. This calculation does not determine the emission timescale.

\subsection{Validation}\label{sec:validation}
\begin{figure}[htbp]
\centering
\begin{subfigure}{0.48\textwidth}
\includegraphics[width=\linewidth]{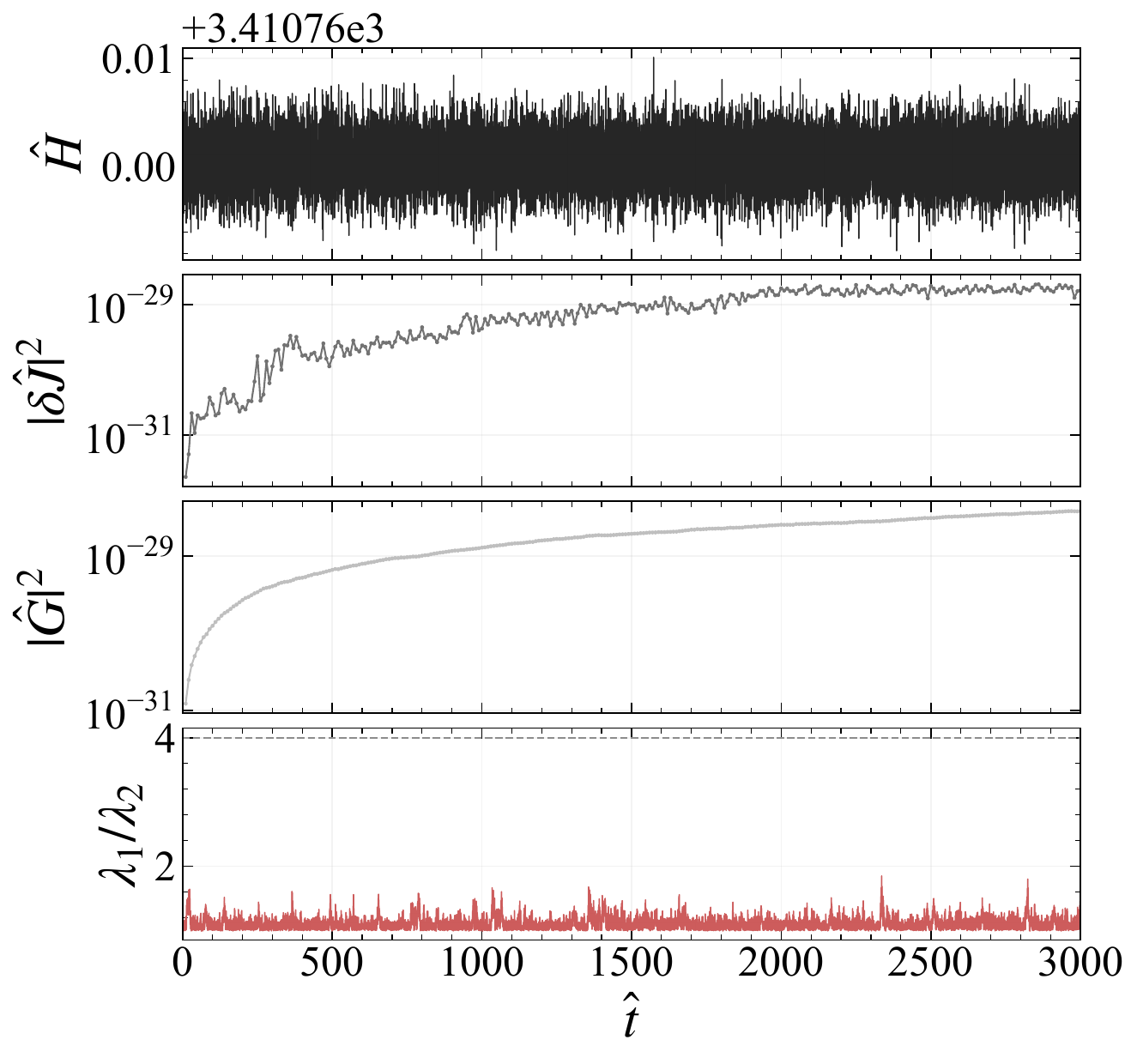}
\caption{Without emission}
\end{subfigure}\hfill
\begin{subfigure}{0.48\textwidth}
\includegraphics[width=\linewidth]{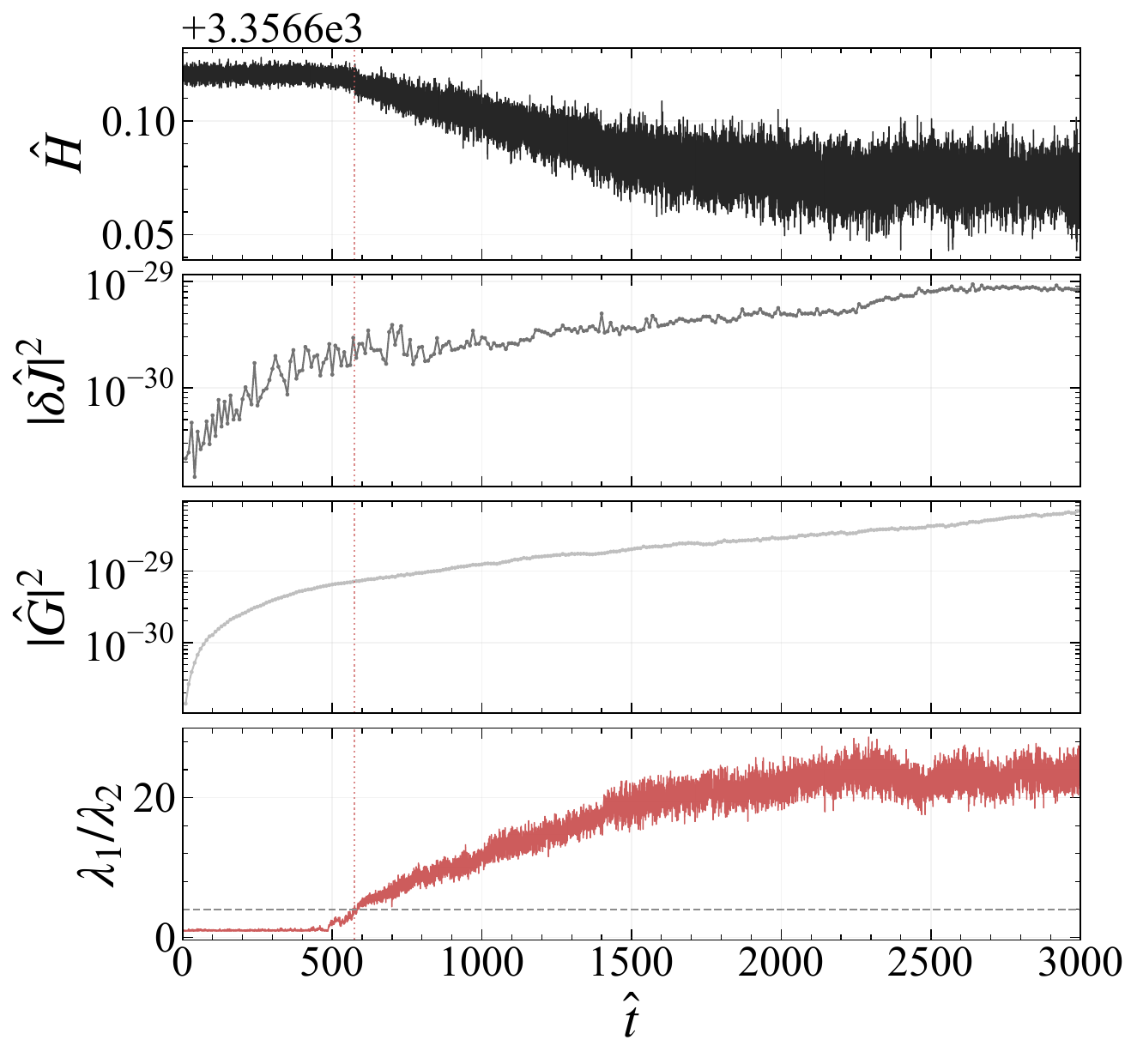}
\caption{Emission at $\hat t \approx 574$.}
\end{subfigure}
\caption{Representative real-time histories at $N=40, \sigma = 1.340$ integrated with $\delta t = 0.005$. Each panel from top to bottom respectively shows the total energy, angular-momentum constraint violation, Gauss constraint violation, and the ratio of the largest and second largest eigenvalues of $R^2$. The red dashed line in (b) labels the point where the emission criterion is met. The trajectory in (b) is the same one as that in Figure~\ref{fig:rt_orbit}.}
\label{fig:rt}
\end{figure}

The numerical stability of real-time trajectories is verified by monitoring the fluctuation of total energy, angular momenta, and the Gauss constraint. Drifts of angular momenta and Gauss constraint are respectively measured by 
\be
|\delta \hat{J}(t)|^2 \equiv  \frac{\sum_{i,j}\left(\hat{J}^{ij}(t) - \hat{J}^{ij}(0)\right)^2}{\sum_i \Tr Y_i^2 \sum_j \Tr Q_i^2}, \quad  |\delta \hat{G}(t)|^2 \equiv  - \frac{\Tr [Y^i, Q_i]^2}{\sum_i \Tr Y_i^2 \sum_j \Tr Q_i^2}. 
\ee
For all of our trajectories, both the energy fluctuation and violation of either constraint remain small. Two example trajectories at $N=40$ and $\sigma = 1.340$ are shown in Figure~\ref{fig:rt}. In the emitted trajectory, the Hamiltonian undergoes a transient relative change of $\Delta\hat{H}/\hat{H}(0)\sim10^{-5}$ during the separation immediately following emission, after which it again exhibits bounded oscillations without further drift. This change is of the same order as $\delta\hat t^{\,2}=2.5\times10^{-5}$, the expected scale of the leading leap-frog discretisation error. Since the change begins only after the emission, and the angular momenta and Gauss constraint remain preserved to high precision throughout, we find no indication that accumulated numerical error triggers the emission. To further strengthen this point, a time-step test for a few representative runs is performed and results are shown below. 

\subsubsection{Time-step sensitivity}
\begin{figure}[ht]
    \centering

    \begin{subfigure}{0.32\textwidth}
        \centering
        \includegraphics[width=\linewidth]{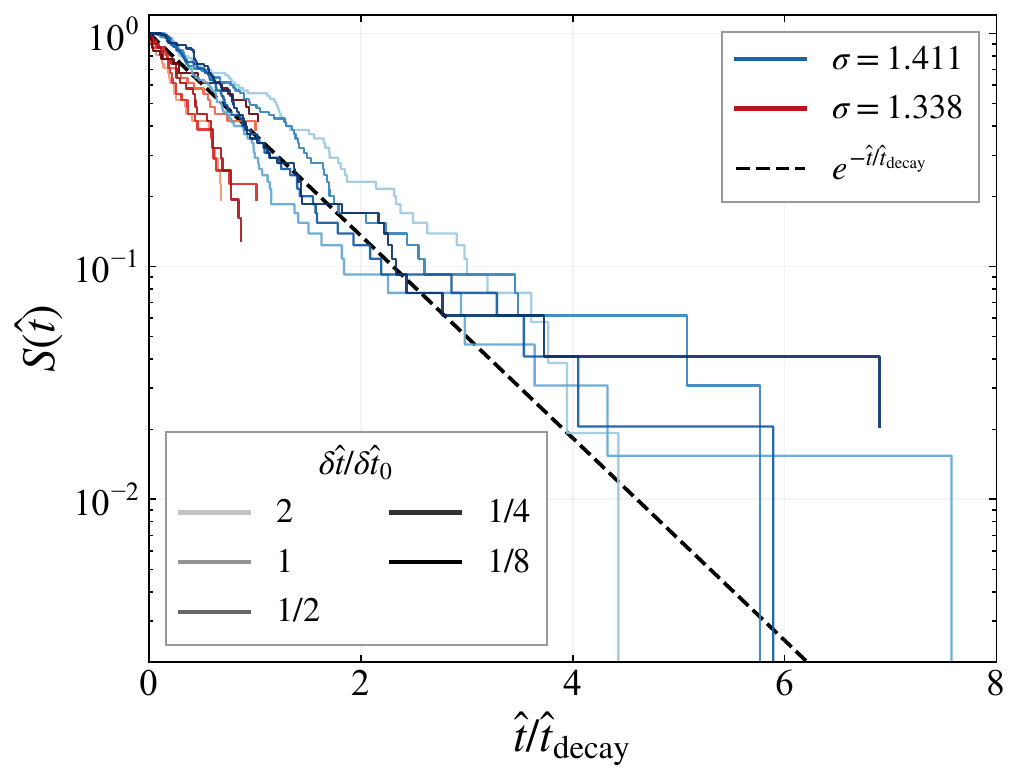}
        \caption{$N = 20$}
    \end{subfigure}
    \hfill
    \begin{subfigure}{0.32\textwidth}
        \centering
        \includegraphics[width=\linewidth]{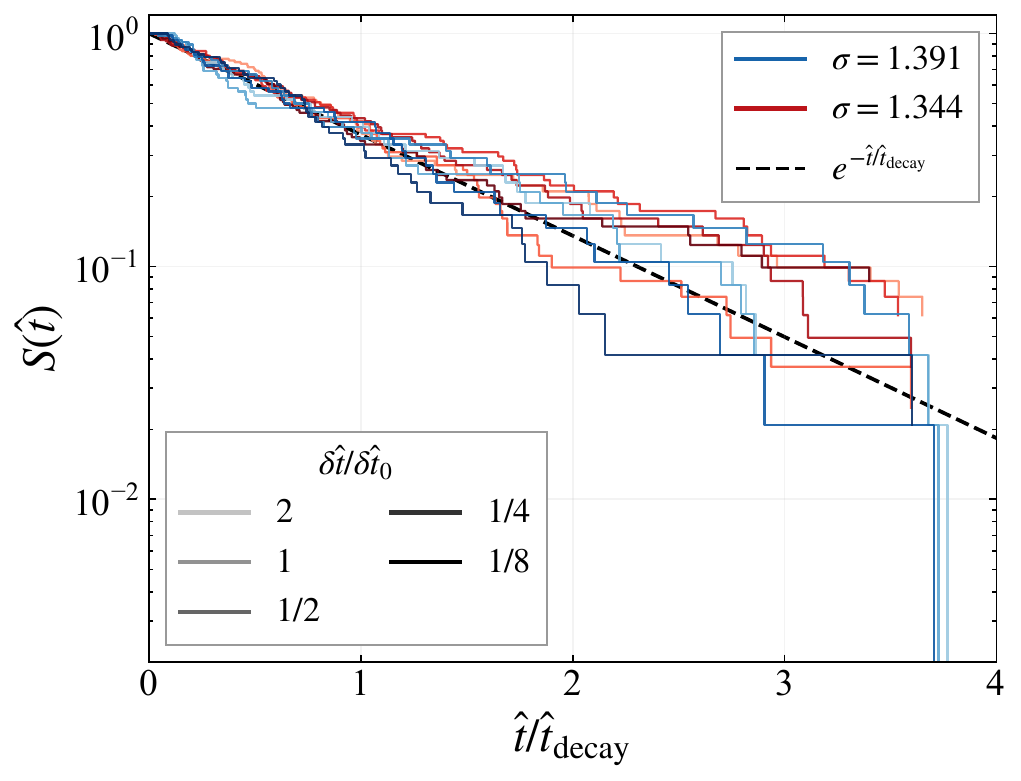}
        \caption{$N = 30$}
    \end{subfigure}
    \hfill
    \begin{subfigure}{0.32\textwidth}
        \centering
        \includegraphics[width=\linewidth]{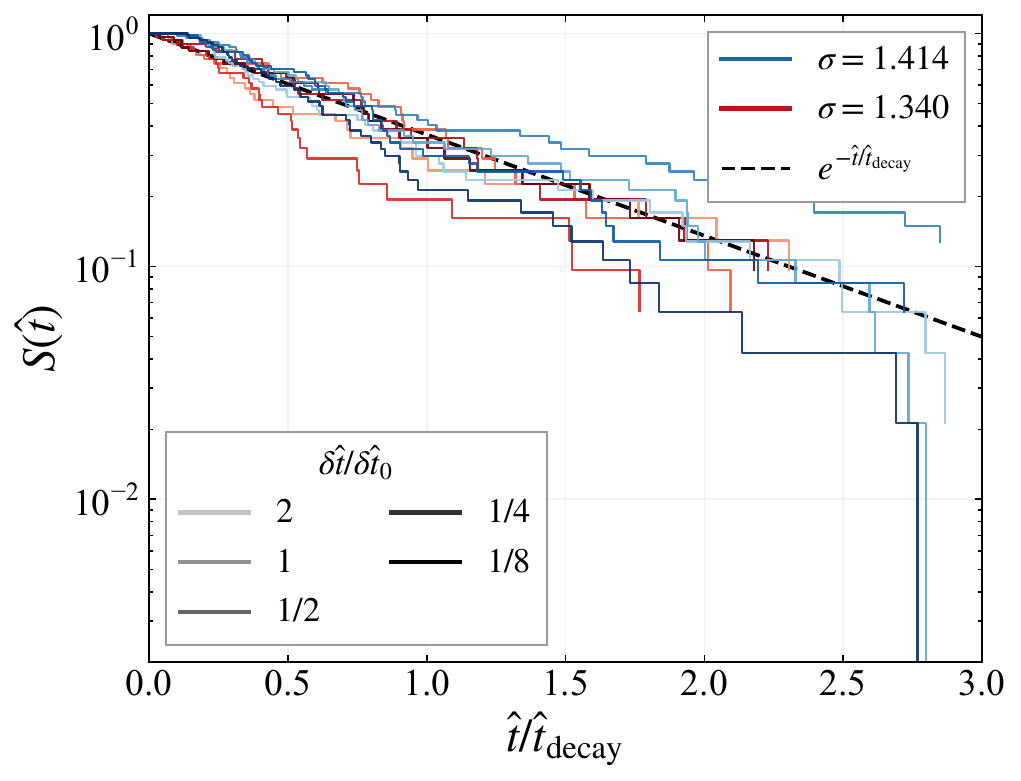}
        \caption{$N = 40$}
    \end{subfigure}
    \caption{Survival curve with varying leap-frog step size. All curves are plotted in time units rescaled by their respective baseline lifetime $\hat{t}_{\rm decay}$, i.e. that with $\delta \hat{t} = 0.005$. The dashed line is the reference curve with baseline lifetime $\hat{t}_{\rm decay}$. Each subplot shows the two runs with the same $N$. The lower $\sigma$ run is plotted red and the higher one plotted blue. Darker shades denote shorter $\delta \hat{t}$, with their respective ratio to the baseline labelled $\delta \hat{t}/\delta \hat{t}_0$. The curves disperse over time due to low surviving counts. }
    \label{fig:timestep}
\end{figure}

The classical matrix dynamics is chaotic, so the same initial condition under a different leap-frog step size evolves to a very different configuration over time. Here we examine whether varying the leap-frog step size changes the measured lifetime. The emission time of individual trajectories need not agree, but the MLE lifetime, being a statistical average, should approach the same value. Re-evolving the whole $41$ runs at varying $\delta \hat{t}$ would be computationally costly, so we pick two values of $\sigma$ near the middle and upper end of our scan for each $N = 20, 30, 40$ for this test. We vary $\delta \hat{t}$ to $2 \times$, $1/2 \times$, $1/4 \times$ and $1/8 \times$ the baseline ($\delta \hat{t} = 0.005$) and measure the lifetime in each case following the same stopping rules and emission criteria. Results are shown in Table~\ref{tab:timestep}. We also plot the survival curves as shown in Figure~\ref{fig:timestep}, where curves from variant $\delta \hat{t}$ are normalised with the baseline $\hat{t}_{\rm decay}$. From both the table and the plots, it appears that the $(N, \sigma) = (20, 1.338)$ run has three measurements ($2 \times, 1/2 \times, 1/4\times$) visibly lower than the other two (baseline and $1/8 \times)$; however, there appears to be no systematic order on how the lifetime varies with $\delta \hat{t}$.

\begin{table}[htbp]
\small
\centering
\begin{tabular}{rrr|rrrrr}
\toprule
\multicolumn{1}{c}{$N$} & \multicolumn{1}{c}{$\sigma$} & \multicolumn{1}{c|}{$n_{\rm IC}$} & \multicolumn{5}{c}{$\delta \hat{t}/\delta \hat{t}_0$ ratio} \\
 &  &  & \multicolumn{1}{c}{$2$} & \multicolumn{1}{c}{$1$} & \multicolumn{1}{c}{$1/2$} & \multicolumn{1}{c}{$1/4$} & \multicolumn{1}{c}{$1/8$} \\
\hline
20 & 1.338 & 31 & $4600^{+1400}_{-1000}$ & $8500^{+3000}_{-2200}$ & $5000^{+1500}_{-1100}$ & $4700^{+1300}_{-1000}$ & $9900^{+3600}_{-2600}$ \\
 &  & {\scriptsize $n_{\rm em}$} & \multicolumn{1}{c}{\scriptsize $(25)$} & \multicolumn{1}{c}{\scriptsize $(19)$} & \multicolumn{1}{c}{\scriptsize $(25)$} & \multicolumn{1}{c}{\scriptsize $(27)$} & \multicolumn{1}{c}{\scriptsize $(18)$} \\[2pt]
20 & 1.411 & 65 & $1200^{+330}_{-250}$ & $860^{+230}_{-180}$ & $1120^{+310}_{-240}$ & $940^{+260}_{-200}$ & $1060^{+290}_{-220}$ \\
 &  & {\scriptsize $n_{\rm em}$} & \multicolumn{1}{c}{\scriptsize $(64)$} & \multicolumn{1}{c}{\scriptsize $(65)$} & \multicolumn{1}{c}{\scriptsize $(63)$} & \multicolumn{1}{c}{\scriptsize $(64)$} & \multicolumn{1}{c}{\scriptsize $(63)$} \\[2pt]
\midrule
30 & 1.344 & 81 & $2990^{+860}_{-660}$ & $2450^{+690}_{-530}$ & $3160^{+910}_{-690}$ & $2880^{+820}_{-620}$ & $2880^{+850}_{-640}$ \\
 &  & {\scriptsize $n_{\rm em}$} & \multicolumn{1}{c}{\scriptsize $(76)$} & \multicolumn{1}{c}{\scriptsize $(79)$} & \multicolumn{1}{c}{\scriptsize $(76)$} & \multicolumn{1}{c}{\scriptsize $(78)$} & \multicolumn{1}{c}{\scriptsize $(74)$} \\[2pt]
30 & 1.391 & 48 & $870^{+370}_{-250}$ & $820^{+350}_{-240}$ & $1010^{+440}_{-300}$ & $800^{+340}_{-230}$ & $760^{+330}_{-220}$ \\
 &  & {\scriptsize $n_{\rm em}$} & \multicolumn{1}{c}{\scriptsize $(48)$} & \multicolumn{1}{c}{\scriptsize $(48)$} & \multicolumn{1}{c}{\scriptsize $(46)$} & \multicolumn{1}{c}{\scriptsize $(48)$} & \multicolumn{1}{c}{\scriptsize $(47)$} \\[2pt]
\midrule
40 & 1.340 & 31 & $3320^{+1310}_{-910}$ & $3630^{+1400}_{-980}$ & $2530^{+980}_{-680}$ & $3800^{+1500}_{-1000}$ & $3700^{+1500}_{-1000}$ \\
 &  & {\scriptsize $n_{\rm em}$} & \multicolumn{1}{c}{\scriptsize $(28)$} & \multicolumn{1}{c}{\scriptsize $(29)$} & \multicolumn{1}{c}{\scriptsize $(29)$} & \multicolumn{1}{c}{\scriptsize $(28)$} & \multicolumn{1}{c}{\scriptsize $(28)$} \\[2pt]
40 & 1.414 & 47 & $380^{+170}_{-110}$ & $410^{+180}_{-120}$ & $560^{+270}_{-170}$ & $420^{+190}_{-130}$ & $323^{+140}_{-94}$ \\
 &  & {\scriptsize $n_{\rm em}$} & \multicolumn{1}{c}{\scriptsize $(46)$} & \multicolumn{1}{c}{\scriptsize $(47)$} & \multicolumn{1}{c}{\scriptsize $(41)$} & \multicolumn{1}{c}{\scriptsize $(44)$} & \multicolumn{1}{c}{\scriptsize $(47)$} \\[2pt]
\bottomrule
\end{tabular}
\caption{Lifetime MLE from different leap-frog time steps. Both the MLE and the uncertainty intervals are computed with the same methods as for the reported values. Uncertainties are rounded to two significant figures and the MLE reported with precision level matching the smaller error bar. The emission count $n_{\rm em}$ are listed in brackets. }
\label{tab:timestep}
\end{table}

\subsubsection{Varying emission criteria}
We vary the emission threshold $\lambda_1/\lambda_2 > 4$ to nearby values $5, 7, 10$ and the confirmation period to $0.5, 1, 2, 5$, and $10$ time units to check their effect on the lifetime estimate. Results are tabulated in Tables~\ref{tab:threshold-ratio} and \ref{tab:hold-ratio} in Appendix~\ref{app:sens}, respectively. For the threshold test, we restrict to runs with $n_{\rm em} \geq 3$ under the original criterion (these are the ones shown in Figure~\ref{fig:lifetime}). Also, for $N = 10, 15, 17$ the typical radius of the outlier is too low, as shown in Figure~\ref{fig:emitted}, that increasing the threshold may completely miss the emission; we thus restrict to runs with $N \geq 20$. For each run, we re-compute the MLE lifetime under the new threshold $\mathfrak{r}$ and take its ratio with the original estimate; the result is shown in Table~\ref{tab:threshold-ratio} in the column labelled $B$, 
\be
B \equiv \frac{\hat{t}_{\rm decay}(\mathfrak{r})}{\hat{t}_{\rm decay}(4)}.
\ee
The uncertainties in $B$ are estimated by a moving-block bootstrap, but with paired trajectories: we conduct the same block resampling as described in Section~\ref{sec:decaytime4} with the block length $l \in [1, \lfloor n_{\rm IC}/5 \rfloor]$; for each $l$, we compute the ratio $B$ from each resample, i.e. both the numerator and the denominator are computed from the same resample; then we record the $16$th and $84$th percentile from $4000$ resamples as the lower and upper uncertainty bound, respectively; finally we take the lowest lower bound and highest upper bound among all $l$ to be the quoted error bars. Resamples with zero emission for both the denominator and numerator are discarded; those with zero emission only for the numerator is assigned $B = \infty$ \footnote{Among rows with a quoted $B$, only one row $(N, \sigma) = (50, 1.275)$ has a maximum of $7.25\%$ resamples discarded across the $l$-scan; this is not surprising since the run has $n_{\rm em}/n_{\rm IC} = 8/127$. All other rows have less than $0.60 \%$ discarded resamples. A maximum of $5.10 \%$ of resamples are assigned $B=\infty$, and this happens at $(N, \sigma) = (20, 1.214)$ and $\mathfrak{r} = 10$. Crucially, the fraction of resamples with $B = \infty$ is smaller than $16\%$ so that the $84$th percentile remains finite. }. As described above, the uncertainty interval is formed from the paired distribution of $B$ rather than separate distributions of the numerator and the denominator; in this way, we have accounted for the two lifetime estimates being computed from the same set of trajectories. We took the maximal interval among all $l$ because typical values of our $n_{\rm IC}$ are not large enough for a reliable block uncertainty estimate. Apart from the bias introduced by our extension rules discussed in Section~\ref{sec:decaytime4}, the $\mathfrak{r}>4$ estimates would suffer additional bias from the procedure of stopping the emitted trajectories early based on the $\mathfrak{r}=4$ criterion. In other words, additional bias comes from misaligning the $\mathfrak{r}=4$ observation protocol with $\mathfrak{r}>4$ criteria. This additional effect is only absent if all emitted trajectories in a run also reach the higher threshold before their evolutions are cut off. In Table~\ref{tab:threshold-ratio}, we also list the percentage of trajectories that emitted under the $\mathfrak{r}=4$ threshold but did not cross the higher threshold before being cut off, under the column labelled $q$. Note that $q$ does not include trajectories that have not reached the higher threshold before a pre-set observation ceiling. A different measure of the change in lifetime is $1/A$ with
\be
A \equiv \frac{1}{n_{\rm em}} \sum_{I \in \mathrm{emitted}} \frac{\hat{t}_{\rm em}^I (4)}{\hat{t}_{\rm em}^I(\mathfrak{r})},
\ee
where the sum and the emission count $n_{\rm em}$ are both based on the $\mathfrak{r} = 4$ threshold. If a trajectory did not cross the higher threshold before its evolution stopped, $\hat{t}^I_{\rm em}$ is strictly bounded between $\hat{t}^I_{\rm obs}$ and $\infty$. Therefore, we also report an interval bound for $1/A$ in Table~\ref{tab:threshold-ratio}. The upper bound on $1/A$ is finite unless all emitted trajectories were stopped before reaching the higher threshold, whereas a similar upper bound on $B$ would be infinite just because one trajectory was cut off. However, $1/A$ is not measuring how the MLE lifetime would change under raising the threshold; instead, it measures on average how each trajectory's decay time would change.  

We also vary the confirmation interval from $1$ time unit to $2, 5, 10$ time units while keeping the threshold unchanged. This test is restricted to runs where all trajectories have at least $10$ time units on record after emission. The ratio between the new MLE lifetime and the one with $1$ time unit of confirmation interval is again labelled $B$ and recorded in Table~\ref{tab:hold-ratio}. The uncertainty is calculated using the same block estimate as above \footnote{Among cells with a reported $B$, a maximum of $37.80 \%$ of resamples are discarded for $(N, \sigma) = (20, 1.187)$ with $0.5$ time units of confirmation; this run has $n_{\rm em}/n_{\rm IC} = 3/117$. The next highest discard fraction is from $(N, \sigma) = (50, 1.275)$ with $6.75 \%$, the same row as in the threshold test. All others have less than $0.73 \%$ resamples discarded. The largest fraction of resamples assigned $B = \infty$ is $1.30 \%$, which occurs in all cells for $(N, \sigma) = (20, 1.214)$.}. The MLE lifetime increases with increasing confirmation interval due to some emitted trajectories falling below the threshold again within the longer confirmation interval. This either reduces $n_{\rm em}$ if the variant emission criterion was never met before evolution stopped, or increases $\hat{t}_{\rm exp}$ if it sustained longer at a later threshold crossing; both increase the MLE. 

To summarise, based on the restricted set of runs, while the qualitative behaviour in Figure~\ref{fig:lifetime} remains, the precise lifetime depends on how we define emission. Among runs with $q=0$, raising the threshold to $\mathfrak{r} = 5$ can increase the MLE lifetime by up to about $45 \%$ and to $\mathfrak{r} = 10$ by up to about $120\%$. In some runs, such as $N = 50, \ \sigma = 1.340$, the effect of raising the threshold is small in magnitude but still increases the measured lifetime with statistical significance. Among runs with at least $10$ more time units of recording after emission for every trajectory, the MLE lifetime increases up to $26 \%$ under doubling the confirmation interval and up to $46 \%$ under increasing tenfold. In both tests, lifetimes measured from runs with larger $N$ appear to be more stable against varying criteria, which would be consistent with the orbital radius increasing with $N$. This behaviour is, however, not conclusive from the tests at hand -- since they are limited to a subset of runs, this may be a selection effect. 

\FloatBarrier
\subsubsection{Energy spread}\label{sec:energyspread}
Each run of fixed $(N, \sigma)$ consists of trajectories with a spread of energies. If the emission rate depends on energy, the pooled survival function is a mixture of exponentials. Here we quantify this spread by binning trajectories in the same run by their energies and computing the MLE lifetime in each energy bin. We restrict to runs with $n_{\rm em} > 20$ and less than $5 \%$ censored trajectories, and group all trajectories into $\lfloor n_{\rm em}/10 \rfloor$ bins. All bins happen to have at least one emitted trajectory. This selection bias towards the high $\sigma$ end where the lifetimes are shorter. The spread in energy and MLE lifetime are shown in Table~\ref{tab:energy-spread}. The standard deviation of the lifetime is calculated from the MLE lifetime of each energy bin, so it also includes statistical fluctuations that are also present even if the true lifetimes are the same across bins.
\begin{table}[htbp]
\centering
\small
\setlength{\tabcolsep}{8pt}
\newcommand{\headstrut}{\rule[-1.5ex]{0pt}{4.6ex}}
\begin{tabular}{rrrrr}
\toprule
\headstrut $N$ & $\sigma$ & bins & $\dfrac{\mathrm{std}(E)}{\mathrm{avg}(E)}$ (\%) & $\dfrac{\mathrm{std}(\hat{t}_{\rm bins})}{\hat{t}_{\rm decay}}$ \\
\midrule
15 & 1.424 & 5 & 1.81 & 0.38 \\
15 & 1.465 & 5 & 1.97 & 0.52 \\
\cmidrule(lr){1-5}
17 & 1.385 & 4 & 2.18 & 0.47 \\
17 & 1.424 & 4 & 2.31 & 0.65 \\
17 & 1.465 & 4 & 2.27 & 0.40 \\
\cmidrule(lr){1-5}
20 & 1.411 & 6 & 1.49 & 0.52 \\
20 & 1.452 & 6 & 1.70 & 0.36 \\
\cmidrule(lr){1-5}
30 & 1.344 & 7 & 1.17 & 0.58 \\
30 & 1.391 & 4 & 1.15 & 0.49 \\
30 & 1.443 & 4 & 1.16 & 0.19 \\
\cmidrule(lr){1-5}
40 & 1.376 & 6 & 0.70 & 0.30 \\
40 & 1.414 & 4 & 0.78 & 0.23 \\
40 & 1.455 & 4 & 0.81 & 0.37 \\
\cmidrule(lr){1-5}
50 & 1.340 & 10 & 0.64 & 0.40 \\
50 & 1.376 & 14 & 0.72 & 0.35 \\
50 & 1.414 & 6 & 0.70 & 0.46 \\
50 & 1.455 & 5 & 0.77 & 0.28 \\
\midrule
\multicolumn{3}{l}{median} & 1.16 & 0.40 \\
\bottomrule
\end{tabular}
\caption{Energy spread within each canonical run. The ratio of the standard deviation to the average energy for trajectories within each run is listed under the column $\mathrm{std}(E)/\mathrm{avg}(E)$. The standard deviation of MLE lifetimes across energy bins is denoted $\mathrm{std}(\hat{t}_{\rm bins})$, and its ratio to the reported MLE lifetime is shown in the last column. }
\label{tab:energy-spread}
\end{table}

\FloatBarrier
\section{Discussion}
We have studied the effect of angular momentum on BFSS states in two coupling regimes. At strong coupling, the axisymmetric black hole develops a collective quadrupolar instability beyond a critical angular momentum. A new family of non-axisymmetric rotating black holes branches off above the critical rotation and is thermodynamically favoured in the microcanonical and canonical ensembles. At weak coupling, the matrix dynamics is approximately classical and we find no evidence for such a collective quadrupolar condensate over the sizes and spins studied. Rotation instead assists the separation of a single brane carrying a substantial fraction of the total angular momentum. 

On the gravity side, we start with the ten-dimensional singly-spinning D$0$ black hole uplifted to eleven dimensions. This uplift is a solution to the eleven-dimensional vacuum Einstein equations and locally equivalent to the singly-spinning Myers-Perry string. Around this background, an 11D metric perturbation satisfying BFSS boundary conditions becomes unstable at $a/r_+\simeq0.991384$, where $a$ and $r_+$ are respectively the rotation parameter and horizon radius in Myers-Perry convention. This mode breaks both time-translation symmetry and rotational symmetry in the plane of rotation, but preserves a linear combination of the two. To follow this mode non-linearly, we work within a two-dimensional consistent truncation of 11D Einstein gravity. From the two-dimensional perspective this unstable mode is a charged scalar resembling that in holographic superconductors~\cite{Gubser:2008px,Hartnoll:2008vx,Hartnoll:2008kx}; here the charge is angular momentum and the charged condensate breaks the $U(1)$ rotational symmetry in one plane to $\mathbb Z_2$. Uplifted back to eleven dimensions, it produces a family of non-axisymmetric black holes, thermodynamically favoured than the symmetric family. 

Near the bifurcation, the axisymmetric solution with the same energy and angular momentum as a quadrupolar solution lies on the unstable side of the threshold. The fixed-$J$ variables in Eq.~\eqref{eq:fixedJvariables} allow us to follow the branch in energy and temperature without choosing a horizon scale. The non-axisymmetric branch emerges at $\widehat E_{\rm c}\simeq15.922217$ and $\widehat T_{\rm c}\simeq3.354518$, and extends towards lower energy and temperature over the computed range. Figures~\ref{fig:nonlinear} and~\ref{fig:canonicalfreeenergy} show that its entropy difference at fixed $(E,J)$ is positive and its Helmholtz free-energy difference at fixed $(T,J)$ is negative. Close to the onset, the data are consistent with $\widehat{\Delta S}\propto(\widehat E-\widehat E_{\rm c})^2$ and $\widehat{\Delta F}\propto-(\widehat T-\widehat T_{\rm c})^2$. These scalings, together with the growing linear mode, support the interpretation of a continuous symmetry-breaking bifurcation. The branch does not dominate over the lower-spin axisymmetric reference in the grand-canonical ensemble. Its thermodynamic preference at fixed $(E,J)$ and $(T,J)$ makes the deformed solutions candidates for the outcome of the instability. Their dynamical stability and global thermodynamic dominance remain to be established. If they are globally dominant, they would predict a symmetry-breaking large-$N$ phase transition of the matrix quantum mechanics.

On the classical matrix side, we measured the same charged quadrupole directly using a hybrid Monte Carlo sampler. The sampled matrix configurations separate into two types, single-clump and emitted. The former has all eigenvalues of the two matrices in the rotation plane taking values inside a clumped range; the latter has one eigenvalue from each matrix separated from the rest. The normalised magnitude of the quadrupole decreases with $N$ in single-clump samples, which is consistent with a fluctuation contribution rather than a collective condensate. The emitted samples have a large quadrupole dominated by the one distant brane. This brane orbits the rest of the clump and can carry approximately $90\%$ of the total angular momentum in the largest matrices studied. The angular-momentum decomposition and radial eigenvalues therefore support a different interpretation from the collective gravity mode: much of the total angular momentum is carried by a separated constituent. A quantitative large-$N$ extrapolation of the clump observable would sharpen the comparison.

We also measured the real-time lifetime of the first sustained separation from the clump, according to the criterion in Section~\ref{sec:matside}. It does not require the separated brane to escape to infinity, and should not be identified with the growth time of the gravitational quadrupole either. The scalar probe in Appendix~\ref{app:greybody} likewise admits frequencies for which a wave can extract energy from the rotating black hole and propagate to infinity. The intervening radial potential can suppress transmission at large $N$, so identifying these frequencies does not determine how rapidly D0-branes are emitted.

To compare the two coupling regimes, it is useful to work with common physical variables, $\tau=T/\lambda^{1/3}$ and $J/N^2$. On the gravity side, hyperscaling gives $S\propto N^2\tau^{9/5}$ at fixed dimensionless spin. Since the quadrupolar instability sets in at $J/S=\alpha_{\rm c}/(2\pi)$, the critical angular momentum scales as $J_{\rm c}/N^2\propto\tau^{9/5}$. In the classical regime, the natural angular momentum scale is instead $J/N^2\sim \tau^{3/4}$, with fixed $\sigma$ corresponding to $J/N^2=\sigma\tau^{3/4}$. Different temperature scalings are familiar from the non-rotating system: gravity predicts $E/(N^2\lambda^{1/3})\propto\tau^{14/5}$, whereas classical equipartition gives $E/(N^2\lambda^{1/3})\propto\tau$. These describe different limits of the same physical quantity. For angular momentum, however, the classical scaling does not by itself define a critical curve, and our results do not establish whether a collective instability with a critical spin $J_{\rm c}(\tau)$ persists at high temperature. The classical limit is also taken at fixed $N$ before studying increasing matrix size. The growth of the separated-brane radius with $N$ therefore requires care when relating these results to the full quantum theory. For separated-brane configurations carrying an order-one fraction of the total angular momentum, the classical treatment of the off-diagonal modes connecting the brane to the clump requires $\tau\gg(N\sigma)^{4/3}$, in addition to $\tau\gg1$.

Our results distinguish the mechanisms observed in the two regimes, but do not determine how they interpolate or whether a phase boundary connects them at intermediate coupling. Below we discuss some possible future directions.

\paragraph{Higher multipoles and giant gravitons} Our quadrupole is the $k=2$ member of the highest-weight family $\mathcal O_k=\operatorname{Tr}Z^k$, where $Z=X^1+{\rm i}X^2$. These operators are annihilated by half of the supercharges and enter the giant-graviton construction of~\cite{Batra:2025ivy}. This operator property does not make a state with a quadrupole condensate supersymmetric, nor does it protect the thermal expectation values studied here. In particular, BFSS does not have the conformal state--operator correspondence that would identify these operators with BPS energy eigenstates.

There is nevertheless a useful algebraic connection. The coefficients of $\det(1+uZ)$ are the antisymmetric Schur operators used in the spherical giant-graviton description. Since $\operatorname{Tr}Z=0$, the first non-trivial coefficient is
\begin{equation}
 \det(1+uZ)=1-\frac{u^2}{2}\mathcal O_2+O(u^3).
\end{equation}
This suggests studying higher members of the same family to sharpen the distinction between a collective deformation and a distant constituent. For example, the rotation-invariant ratio $R_4=|\operatorname{Tr}Z^4|/|\operatorname{Tr}Z^2|^2$ is of order unity when both moments are dominated by one outlier. For a coherent deformation involving $O(N)$ constituents, with non-zero second and fourth moments, simple trace counting instead gives $R_4=O(N^{-1})$. These are scaling expectations to test, rather than supersymmetric identities; configurations with a small quadrupole require separate treatment. A gravitational comparison would also require the nonlinear Kaluza--Klein dictionary, including mixing with $\mathcal O_2^2$. The consistent truncation does not imply the absence of higher multipole expectation values.

On the gravity side, the spherical giant gravitons proposed for BFSS are D6-branes wrapping an $S^6\subset S^8$, which uplift to Kaluza--Klein monopoles~\cite{Batra:2025ivy}. A probe calculation in our rotating backgrounds could test whether they provide another thermodynamically favourable carrier of angular momentum. At fixed total $J$, this would compare the free-energy cost of the probe with the gain from reducing the black-hole angular momentum, whose conjugate potential is $g\mu$. The monopole geometry also involves a non-trivial degeneration of the M-theory circle, so a backreacted construction would require topology beyond the present ansatz. A single spherical giant has angular momentum at most $N$; for a black hole with $J=O(N^2)$, transferring a macroscopic fraction of its spin would therefore require many such objects and their backreaction. We have not performed this probe calculation, and neither the collective quadrupolar branch nor the emitted D0-brane has been identified with a giant graviton.

\paragraph{Mean field theory} It may be possible to identify the critical spin and the non-axisymmetric phase directly in the strongly coupled matrix quantum mechanics using a mean field approximation developed in \cite{Kabat:1999hp, Kabat:2000zv,Kabat:2001ve}. This method aims to mimic the strongly coupled theory by a Gaussian theory but with a different width for each Matsubara mode; these widths are then determined by a set of self-consistency equations, often chosen to be the loop-truncated Schwinger-Dyson equations. Using this approximation, the power-law scaling of the free energy with temperature in the black-hole limit, $F \sim \tau^{14/5}$, was recovered approximately for $1/4 \lesssim \tau  \lesssim 1$ \cite{Kabat:2000zv}. We attempted but failed to recover the thermodynamics of spinning black holes by deforming the equations in \cite{Kabat:2000zv} with a rotational chemical potential. This failure is consistent with results in \cite{Lin:2013jra}, where a refined analysis of the one-loop Schwinger-Dyson equations fails to recover the non-rotating thermodynamics, and the success in \cite{Kabat:2000zv} appears to be a transient feature that fails at lower $\tau$. However, as mentioned in \cite{Lin:2013jra}, a better chosen set of equations and truncation may lead to matrix results directly comparable to supergravity.

\paragraph{Bootstrap bounds} An alternative would be to adapt the bounds on Euclidean two-point functions developed in~\cite{Cho:2025vws} and applied to complex-time singularities in~\cite{Dodelson2026MatrixScattering} to the charged quadrupole. Bootstrap bounds on BFSS observables have already been obtained at finite energy~\cite{Lin:2023owt} and for ground-state correlators at infinite $N$~\cite{Lin:2024vvg}. In particular, studying $\langle\mathcal O_2(t)\mathcal O_2^\dagger(0)\rangle$ could help identify a collective mode whose frequency approaches zero in the co-rotating frame, complementing our classical measurements of the quadrupole magnitude and single-brane separation. Adapting the finite-time bounds to rotating thermal BFSS would require including the fermions, imposing the Gauss constraint and working at fixed angular momentum. The high-temperature expansion in~\cite{Dodelson2026MatrixScattering} could also provide a way to calculate leading quantum corrections to the classical quadrupole observables. The finite-time methods have so far been demonstrated in simpler models; determining whether they can resolve the rotating BFSS instability remains an open problem.

\paragraph{Large spin limit} We have not studied the dynamical stability of the non-axisymmetric black hole, nor have we determined its global dominance in thermal ensembles. In particular, it would be interesting to investigate whether solutions with residual symmetry different from $\mathbb{Z}_2$ or different horizon topologies branch off at increasing angular momentum. It would also be interesting to determine the behaviour of the dominant gravity solution in the large angular momentum or zero temperature limit, which probes the ground states of the matrix quantum mechanics in fixed R-charge sectors. The zero-temperature limits of holographic superconductors~\cite{Gubser:2009cg,Horowitz:2009ij} provide a useful analogy for this question. On the matrix side, this limit might be solvable by some large-charge expansion similar to those for CFTs in \cite{Hellerman:2015nra, Monin:2016jmo}. 

\paragraph{Gregory-Laflamme instability} We have been working in a temperature regime where the black hole remains uniform along the M-theory circle. At low enough temperature, $\tau \sim N^{-5/9}$, the uniform black hole suffers Gregory-Laflamme instability and the dominant solution is either non-uniform or localised on the M-theory circle. This transition with zero spin in the transverse sphere has been studied to high precision in \cite{DiasLowEnergy2025, Dias:2025jxx}. It would be interesting to investigate the effect of non-zero spin. The bulk solution might be similar to non-uniform rotating black string solutions with helical isometries in lower dimensions \cite{GLSR, resonator, helix}.

\section*{Acknowledgements}
It is a pleasure to thank Sean A. Hartnoll for very helpful discussions and for reading an earlier version of this manuscript. We also thank \'Oscar J. C. Dias for reading an earlier version of this manuscript and for his helpful comments. This work has been partially supported by STFC consolidated grant ST/X000664/1. JES is partially supported by Hughes Hall, Cambridge. JL is supported by a Harding Distinguished Postgraduate Scholarship.

\appendix

\section{Eleven-dimensional gravitational perturbations}\label{app:11d}
The two-dimensional charged scalar selects the quadrupolar sector. To determine whether another azimuthal harmonic bifurcates at a lower spin, we consider gravitational perturbations of the eleven-dimensional uplift directly. This also checks that the radial zero mode is a mode of the higher-dimensional geometry.

Locally, the background can be written as
\begin{multline}
 {\rm d}s^2=-{\rm d}T^2
 +\frac{2M_0}{r^5\Sigma}
 ({\rm d}T-a\sin^2\theta\,{\rm d}\psi)^2
 +\frac{\Sigma}{\Delta_r}{\rm d}r^2+\Sigma{\rm d}\theta^2
 \\
 +(r^2+a^2)\sin^2\theta\,{\rm d}\psi^2
 +r^2\cos^2\theta\,{\rm d}\Omega_6^2+{\rm d}Z^2,
 \label{eq:MPstring}
\end{multline}
where $\Sigma=r^2+a^2\cos^2\theta$ and $\Delta_r=r^2+a^2-2M_0/r^5$. This is an eleven-dimensional black string with a ten-dimensional Myers--Perry factor. The dimensional distinction matters: the transverse sphere at infinity is $S^8$, and the spectator circle supplies the eleventh coordinate.

We preserve the $SO(7)$ symmetry of the $S^6$ and decompose
\begin{equation}
 h_{AB}=e^{-{\rm i}\varpi T+{\rm i}m_\varphi\psi+{\rm i}k_ZZ}
 \widehat h_{AB}(r,\theta).
 \label{eq:11dansatz}
\end{equation}
For the modes descending from stationary fields in the two-dimensional frame, Eq.~\eqref{eq:Carroll} gives
\begin{equation}
 \varpi=m_\varphi\Omega_H,\qquad
 k_Z=-m_\varphi\Omega_H.
 \label{eq:synchronisation}
\end{equation}
These relations specify the helical sector of interest. They do not cover arbitrary Kaluza--Klein perturbations of an asymptotically flat string.

The perturbation has ten components on $(T,r,\theta,\psi)$ and a common trace component on the $S^6$. In transverse-traceless gauge,
\begin{equation}
 h^A{}_A=0,\qquad \nabla^Ah_{AB}=0,
 \qquad -\nabla^2h_{AB}-2R_{ACBD}h^{CD}=0.
 \label{eq:Lichnerowicz}
\end{equation}
The trace condition eliminates the $S^6$ amplitude, leaving ten coupled functions of two coordinates and four divergence constraints. Smoothness at the two angular axes fixes the powers of the polar coordinates factored from the components. Horizon regularity is imposed in a co-rotating ingoing frame. At infinity we select the decaying branch compatible with the source-free D0-brane asymptotics.

We use compact coordinates $u,w\in[0,1]$ defined by $r=r_+/(1-w^2)$ and $\cos\theta=u\sqrt{2-u^2}$. The horizon is $w=0$, infinity is $w=1$, and the angular axes are $u=0,1$. After extracting the endpoint powers, the equations are discretised on a tensor product of Chebyshev grids. An amplitude condition removes the freedom to rescale $h_{AB}$, and the spin $\alpha=a/r_+$ is solved for together with the perturbation. The divergence constraints provide an independent diagnostic of the solutions.

We have solved the coupled eleven-dimensional perturbation equations directly as PDEs in two variables. Given the separation of transverse-sphere-invariant gravitational perturbations of singly spinning Myers--Perry black holes~\cite{Santos:2026pbu} and of Kaluza--Klein perturbations of the Kerr string~\cite{Dias:2026zri}, we expect that these methods can be extended to the Myers--Perry string considered here, reducing the calculation to angular and radial ODEs. We leave this extension, with the BFSS boundary conditions imposed, for future work.

The threshold data are listed in Table~\ref{tab:thresholds}. The two finer grids agree especially well for $m_\varphi=2,3,4$, while the highest harmonics require more resolution. We report the actual grid comparison rather than interpreting the working precision as numerical accuracy.
\begin{table}[htbp]
\centering
\begin{tabular}{c r r r}
\toprule
 $m_\varphi$ & $17\times17$ & $51\times51$ & Absolute difference\\
\midrule
 2 & 0.991383498 & 0.991383962 & $4.64\times10^{-7}$\\
 3 & 1.401367808 & 1.401368800 & $9.91\times10^{-7}$\\
 4 & 1.715960330 & 1.715904692 & $5.56\times10^{-5}$\\
 5 & 1.981777157 & 1.981067028 & $7.10\times10^{-4}$\\
 6 & 2.220796077 & 2.214684655 & $6.11\times10^{-3}$\\
 7 & 2.473237287 & 2.425896942 & $4.73\times10^{-2}$\\
\bottomrule
\end{tabular}
\caption{Onset values of $a/r_+$ in the helical eleven-dimensional gravitational sector. The two columns give the thresholds on grids with $17\times17$ and $51\times51$ collocation points, including the endpoints. The $m_\varphi=2$ radial calculation gives $a/r_+\simeq0.991384$ independently.}
\label{tab:thresholds}
\end{table}

The quadrupole has the smallest threshold among these sectors, with a substantial separation from the $m_\varphi=3$ onset. It therefore controls the first resolved non-axisymmetric bifurcation. The calculation is a comparison of stationary onset modes. It neither orders their growth rates at arbitrary spin nor excludes instabilities in sectors outside the imposed symmetry and helical conditions.

\section{Deriving the equation of state} \label{app:eos}
The mean energy can be computed as
\be\label{eq:E}
\langle \hat{H} \rangle = - \frac{\partial}{\partial \hat{\beta}} \ln Z_N(1/\hat{\beta}, \hat{J}),
\ee
where $\hat{\beta} = 1/\hat{T}$. Using the change of variables $Y =\hat{\beta}^{-1/4} y$ and $Q= \hat{\beta}^{-1/2} q$, we get
\be
Z_N(1/\hat{\beta},\hat{J}) = \hat{\beta}^{-\frac{3}{4}[8(N^2-1) - 36]} Z_N(1, \hat{J} \hat{\beta}^{3/4})
\ee

This is another way to see the scaling symmetry of the classical system and that the ratio $\hat{\jmath} \equiv \hat{J} \hat{\beta}^{3/4}$ is the only non-trivial parameter. Then \eqref{eq:E} becomes
\be \label{eq:E2}
\langle \hat{H} \rangle = \frac{3}{4}\left(\left[8(N^2-1) - 36\right] \hat{\beta}^{-1} - \hat{J}\hat{\beta}^{-1/4} \frac{\partial}{\partial \hat{\jmath}} \ln Z_N (1, \hat{\jmath})\right).
\ee
Now we deal with the second term. The derivative acts on one delta function inside the partition function, and we can exchange it for a derivative on the integration variable $q$ taken in a direction that preserves all other constraints except the angular momentum in (12)-plane:
\be \label{eq:deltasub}
\frac{\partial}{\partial \hat{\jmath}} \, \delta \left( \hat{\jmath} - \Tr(y^1 q^2 - y^2 q^1) \right) = - w(y) \cdot \frac{\partial}{\partial q}\, \delta \left( \hat{\jmath} - \Tr(y^1 q^2 - y^2 q^1) \right),
\ee
where the dot is the inner product defined in \eqref{eq:inner} and $w(y)$ satisfies
\be\label{eq:condwapp}
[y^i, w_i] = 0, \quad \ \text{and} \ \Tr(y^i w^j - y^j w^i) = \epsilon^{ij},
\ee
just like in the main text. These conditions ensure that the $q$-derivative taken along the $w$-direction only changes the (12)-plane angular momentum preserving all other constraints. There are infinitely many solutions  to \eqref{eq:condwapp}, since one can add to $w$ any vector solving the homogeneous equations (i.e. constraints with $\epsilon^{ij} = 0$). For later convenience, we choose the solution with minimum norm; this is the vector normal to the constraint surface (i.e. orthogonal to all solutions to the homogeneous equations). Minimizing the norm of $w$ with \eqref{eq:condwapp} imposed by Lagrange multipliers implies that it must take the form in \eqref{eq:wform}. Making the substitution \eqref{eq:deltasub} in the partition function and integrating by parts, we have
\be
\begin{aligned}
    \frac{\partial}{\partial \hat{\jmath}} \ln Z_N (1, \hat{\jmath}) &= \frac{1}{Z_N(1, \hat{\jmath})}\int dy dq \Delta(y, q; \hat{\jmath}) \, w \cdot \frac{\partial}{\partial q} e^{- \frac{1}{2}\Tr(q^i q_i) + \frac{1}{4}\Tr([y_i, y_j]^2)} \\
    &= -\frac{1}{Z_N(1, \hat{\jmath})}\int dy dq \Delta(y, q; \hat{\jmath}) \, \left(w \cdot q\right) \, e^{- \hat{H}},
\end{aligned}
\ee
where $\Delta(y, q; \hat{\jmath})$ denotes the product of delta functions. The only non-zero contributions to the integral come from $q$ satisfying the same conditions on $w$ except replacing $\epsilon^{ij}$ by $\hat{\jmath} \epsilon^{ij}$ in \eqref{eq:condwapp}. The general solution is $q = \hat{\jmath} w$ up to addition by a solution to the homogeneous equations. Since we have chosen $w(y)$ to be orthogonal to all homogeneous solutions, the only contribution to the integral becomes
\be
\begin{aligned}
    \frac{\partial}{\partial \hat{\jmath}} \ln Z_N (1, \hat{\jmath}) &= -\frac{\hat{\jmath}}{Z_N(1, \hat{\jmath})}\int dy dq \Delta(y, q; \hat{\jmath}) \, \left(w \cdot w \right) \, e^{- \hat{H}}\\
    & = - \hat{\jmath} \langle \left(w(y), w(y)\right) \rangle_{(1,\hat{\jmath})}\\
    & = - \hat{J} \hat{\beta}^{1/4} \left \langle \left(w(Y), w(Y)\right) \right \rangle_{(\hat{T},\hat{J})}.
\end{aligned}
\ee
In the last step, we used $w(Y) = w(\hat{\beta}^{-1/4} y) = \hat{\beta}^{1/4} w(y)$ since \eqref{eq:condwapp} are constraints only on bilinears of $y$ and $w$. Substituting this result into \eqref{eq:E2}, we get 
\be
    \langle \hat{H} \rangle  = \frac{3}{4}\left[\left(8(N^2-1) - 36\right) \hat{T} + \hat{J}^2 \left \langle \left(w(Y), w(Y)\right) \right \rangle_{(\hat{T},\hat{J})} \right],
\ee
which is the formula \eqref{eq:eos} in the main text.

\section{Finite-step exactness of the HMC algorithm}\label{app:HMCproof}
To establish finite-step exactness, it is sufficient to show two results. First, at fixed $Y$, the projected Gaussian refresh samples the exact conditional distribution of $Q$ implied by \eqref{eq:part}. Second, the
leap-frog proposal preserves the constraints and the associated measure, and is reversible under $h\to-h$. The Metropolis step (step 4) then enforces detailed balance with respect to the full Boltzmann-weighted constrained measure in \eqref{eq:part}. The standard HMC detailed-balance argument \cite{Duane:1987de} consequently implies that this measure is an exact stationary distribution at finite $\delta t$. 

To demonstrate the first point, we first show that the conditional momentum distribution in \eqref{eq:part} is equivalent to a Gaussian distribution orthogonally projected at fixed $Y$. Then we show that step 2 of our algorithm in the main text precisely executes this projection. We write the Gauss and angular momentum constraints collectively as 
\be\label{eq:cons}
\mathcal{A}_Y Q = c, 
\ee
where $\mathcal{A}_Y$ acts on $\mathcal{V}$, the vector space in which $Q_i$ lives, and only depends on $Y$. $c$ is only non-zero because of the finite fixed angular momentum in the $(12)$-plane. The fact that all constraints can be written in this form, i.e. only affine constraints on momenta, is important for ensuring the first point above. To demonstrate this, we write an arbitrary momentum as
\be
Q = Q_Y + q + p,
\ee
where
\be
\mathcal{A}_Y Q_Y = c, \quad \mathcal{A}_Y q = 0, \quad (Q_Y, q) = (q, p) = 0.
\ee
We also require $Q_Y$ to have minimal norm. $q$ is tangent to the constraint surface, and $p$ is orthogonal to every vector tangent to that surface. Together, $Q_Y + q$ is the orthogonal projection of $Q$ onto the constraint hypersurface. Note that $Q_Y$ only depends on $Y$, not $Q$ itself before projection. Under the change of integration variables $Q \to (q, p)$, the original conditional momentum distribution in \eqref{eq:part} becomes
\be
\begin{aligned}
\dd\mu_Y(Q) &= \dd Q \delta\left(\mathcal{A}_Y Q - c\right) e^{-\frac{||Q||^2}{2 \hat{T}}}\\
&= \dd q \dd p \delta(\mathcal{A}_Y p) e^{-\frac{||Q_Y + p||^2 + ||q||^2 }{2 \hat{T}}}\\
&= \frac{\dd q}{\sqrt{\det^\prime \mathcal{A}_Y\mathcal{A}_Y^\dagger}} e^{-\frac{||Q_Y||^2 + ||q||^2}{2 \hat{T}}},
\end{aligned}
\ee
where in going to the final line we write $p$ in an orthonormal basis in the image of $\mathcal{A}_Y^\dagger$ and marginalise over it. Here $\det^\prime$ denotes the determinant restricted to the non-zero eigenspace of $\mathcal A_Y\mathcal A_Y^\dagger$, namely the product of its non-zero eigenvalues. The Jacobian from $p$-marginalisation only depends on $Y$, so it is only a normalisation in the conditional $Q$-distribution at fixed $Y$. This factor should be retained when comparing different $Y$ configurations. The calculation above shows that the conditional momentum distribution is equivalent to orthogonally projecting a free Gaussian distribution with width $\sqrt{\hat{T}}$ to the constraint hypersurface. Now we show that step 2 of the HMC algorithm in the main text correctly performs this projection. This step uses the fact that orthogonal projection is equivalent to finding the nearest vector on the constraint hypersurface: 
\be
\min_{\tilde{Q}} \frac{1}{2} || \tilde{Q} - \xi ||^2, \quad \text{subject to} \ [Y^i, \tilde{Q}_i] = 0, \quad 2\Tr(Y^{[i} \tilde{Q}^{j]})  = \hat{J}^{ij},
\ee
where $\xi$ is a random Gaussian vector in $\mathcal{V}$ and $\tilde{Q}$ is its orthogonal projection. This $\tilde{Q}$ would be the $Q$ used in the description of the HMC algorithm in the main text, equivalently $Q_Y + q$ above. To avoid confusion, $Q$ in this paragraph is the integration variable in the partition function, which in general does not satisfy any constraint. This minimisation problem can be solved using Lagrange multipliers; the deformed lagrangian is 
\be
\mathcal{L}\left(\tilde{Q}, K^{ij}, Z \right) = \frac{1}{2} || \tilde{Q} - \xi ||^2 + \frac{1}{2}K^{ij} \left( 2\Tr(Y^{[i} \tilde{Q}^{j]})  - \hat{J}^{ij}\right) + i \Tr\left(Z[Y^i, \tilde{Q}_i]\right).
\ee
The saddle-point equation is 
\be
\tilde{Q}_i = \xi_i + \mathrm{i} [Y_i, Z] + K_{ij} Y^j,
\ee
the same as in \eqref{eq:deformQ}. Here $Z$ and $K_{ij}$ are Lagrange multipliers, which can be solved from requiring $\tilde{Q}$ to satisfy the constraints. This is precisely what our algorithm then executes. Abstractly, writing $\Pi_Y$ for the orthogonal projector onto $\ker\mathcal A_Y$, the minimizer is $\tilde Q=Q_Y+\Pi_Y\xi$. Hence its tangent fluctuation is $q=\Pi_Y\xi$. Since $\xi$ is an isotropic Gaussian of width $\sqrt{\hat T}$, $q$ is a Gaussian of the same width on $\ker\mathcal A_Y$, reproducing the conditional distribution derived
above. 

On the second point, we begin by showing that each update in a leap-frog step preserves all constraints and the measure. Below, we revert to the notation $Q$ for the constrained momentum used in the leap-frog evolution, same as in the main text. Under the position update $Y^i \to Y^i + h Q^i$, the constraints are trivially preserved:
\be
\begin{aligned}
[Y^i, Q_i] &\to [Y^i + h Q^i, Q_i] = [Y^i, Q_i], \\ 2\Tr(Y^{[i} Q^{j]}) &\to 2\Tr((Y^{[i} + h Q^{[i}) Q^{j]}) = 2\Tr(Y^{[i} Q^{j]}).
\end{aligned}
\ee
Under the update $Q_i \to Q_i + \frac{h}{2}F_i(Y)$, 
\be
\begin{aligned}
[Y^i, Q_i] &\to [Y^i, Q_i] + \frac{h}{2} [Y^i, F_i],\\
2\Tr(Y^{[i} Q^{j]}) &\to 2\Tr(Y^{[i} Q^{j]}) + h \Tr(Y^{[i} F^{j]}).
\end{aligned}
\ee
The additional terms multiplied by $h$ are zero by $SU(N)$ and $SO(9)$ invariance of the potential. Consider an infinitesimal $SU(N)$ transformation parametrised by $\Lambda \in \mathfrak{su}(N)$, under which $\delta_{\Lambda} Y^i = \mathrm{i} [\Lambda , Y^i]$. The infinitesimal change of the potential is 
\be
\begin{aligned}
\delta_\Lambda \hat{V} &= \Tr\left( \frac{\partial \hat{V}}{\partial Y^i} \delta_\Lambda Y^i\right) \\
&= - \mathrm{i}\Tr \left(F_i [\Lambda, Y^i] \right) \\
&= - \mathrm{i} \Tr\left(\Lambda [Y^i, F_i] \right).
\end{aligned}
\ee
Setting this to zero for arbitrary $\Lambda$ gives $[Y^i, F_i] =0$. Similarly, the infinitesimal change in potential under an arbitrary $SO(9)$ transformation leads to $\Tr(Y^i F^j - Y^j F^i) = 0$. These show that every leap-frog update preserves all the Gauss and angular momentum constraints. The updates are always triangular in $(Y, Q)$, hence resulting in a unit Jacobian. These combine to preserve the constrained reference measure 
$dY\,dQ\,\delta(\mathcal A_YQ-c)$. Finally, for detailed balance, reversibility of the leap-frog proposals is important. Suppose that a sequence of leap-frog steps with $h=+\delta t$ proposes a configuration $(Y',Q')$ starting from $(Y,Q)$. By symmetry of the integrator, applying the same number of steps with $h=-\delta t$ to $(Y',Q')$ would return $(Y,Q)$. The required reversibility is then guaranteed by selecting the two signs with equal probability. In ordinary HMC, reversing the momentum is equivalent to reversing the sign of the integration step. Since the Gaussian momentum refresh draws $Q$ and $-Q$ with equal probability, no explicit random choice of the sign of $h$ is required. Here, however, $Q\to-Q$ maps the fixed-$\hat J$ sector to the sector with angular momentum $-\hat J$. We therefore enforce the equal probabilities of the forward and reverse trajectories directly by choosing $h=\pm\delta t$ with equal probability. Together with preservation of the constrained reference measure, this makes the proposal reversible. The Metropolis rule in step 4 then enforces detailed balance with respect to \eqref{eq:part}, by the standard HMC argument~\cite{Duane:1987de}.

\section{Bias test for adaptive observation windows}

\begin{table}[htbp]
\centering
\small
\begin{tabular}{rr|rrrrrr}
\toprule
\multicolumn{2}{c|}{} & \multicolumn{6}{c}{mean MLE $t_{\rm decay}/t_{\rm true}$} \\
\multicolumn{1}{c}{$n_{\rm IC}$} & \multicolumn{1}{c|}{$t_1/t_{\rm true}$} & \multicolumn{1}{c}{fixed} & \multicolumn{1}{c}{$0.1$} & \multicolumn{1}{c}{$0.5$} & \multicolumn{1}{c}{$0.8$} & \multicolumn{1}{c}{$n<3$} & \multicolumn{1}{c}{adv.} \\
\hline
32 & $10$ & $1.000(1)$ & $1.000(1)$ & $1.000(1)$ & $1.000(1)$ & $1.000(1)$ & $1.000(1)$ \\
32 & $10/3$ & $1.004(1)$ & $1.004(1)$ & $1.004(1)$ & $1.004(1)$ & $1.004(1)$ & $1.001(1)$ \\
32 & $1$ & $1.031(1)$ & $1.031(1)$ & $1.015(1)$ & $1.003(1)$ & $1.031(1)$ & $1.011(1)$ \\
32 & $1/3$ & $1.119(1)$ & $1.096(1)$ & $1.026(1)$ & $1.027(1)$ & $1.110(1)$ & $1.045(1)$ \\
32 & $1/10$ & $1.302(1)$ & $1.034(1)$ & $1.119(1)$ & $1.119(1)$ & $1.020(1)$ & $1.343(1)$ \\
32 & $1/30$ & $0.793(1)$ & $1.307(1)$ & $1.312(1)$ & $1.312(1)$ & $1.284(1)$ & $0.835(1)$ \\
\cline{1-8}
65 & $10$ & $1.000(1)$ & $1.000(1)$ & $1.000(1)$ & $1.000(1)$ & $1.000(1)$ & $1.000(1)$ \\
65 & $10/3$ & $1.002(1)$ & $1.002(1)$ & $1.002(1)$ & $1.002(1)$ & $1.002(1)$ & $1.001(1)$ \\
65 & $1$ & $1.015(1)$ & $1.015(1)$ & $1.011(1)$ & $1.002(1)$ & $1.015(1)$ & $1.005(1)$ \\
65 & $1/3$ & $1.051(1)$ & $1.050(1)$ & $1.013(1)$ & $1.013(1)$ & $1.051(1)$ & $1.019(1)$ \\
65 & $1/10$ & $1.226(1)$ & $0.979(1)$ & $1.051(1)$ & $1.051(1)$ & $1.112(1)$ & $1.077(1)$ \\
65 & $1/30$ & $1.185(1)$ & $1.197(1)$ & $1.199(1)$ & $1.199(1)$ & $1.099(1)$ & $1.288(1)$ \\
\cline{1-8}
120 & $10$ & $1.000(1)$ & $1.000(1)$ & $1.000(1)$ & $1.000(1)$ & $1.000(1)$ & $1.000(1)$ \\
120 & $10/3$ & $1.001(1)$ & $1.001(1)$ & $1.001(1)$ & $1.001(1)$ & $1.001(1)$ & $1.000(1)$ \\
120 & $1$ & $1.008(1)$ & $1.008(1)$ & $1.007(1)$ & $1.001(1)$ & $1.008(1)$ & $1.003(1)$ \\
120 & $1/3$ & $1.026(1)$ & $1.026(1)$ & $1.007(1)$ & $1.007(1)$ & $1.026(1)$ & $1.011(1)$ \\
120 & $1/10$ & $1.103(1)$ & $0.967(1)$ & $1.026(1)$ & $1.026(1)$ & $1.100(1)$ & $1.045(1)$ \\
120 & $1/30$ & $1.318(1)$ & $1.090(1)$ & $1.090(1)$ & $1.090(1)$ & $1.028(1)$ & $1.162(1)$ \\
\bottomrule
\end{tabular}
\caption{Mean MLE for the lifetime from the synthetic experiments. The number of initial samples is labelled by $n_{\rm IC}$. The six columns on the right each label a stopping rule: from left to right are respectively fixed observation time  $t_1$, extension to $t_2$ if the fraction of emitted samples is below $0.1, 0.5, 0.8$ by $t_1$, extension if the number of emissions is less than or equal to two, and extension if the Garwood interval does not cover the true lifetime. The number in parentheses is the standard error of the tabulated mean over the $2 \times 10^6$ synthetic experiments, expressed in units of the last digit shown, e.g. $1.318(1)$ means $1.318 \pm 0.001$. It measures how precisely the simulation determines the bias, not any confidence interval of the MLE distribution.}
\label{tab:cov-cv}
\end{table}

\begin{table}[htbp]
\centering
\small
\begin{tabular}{rr|rrrrrr}
\toprule
\multicolumn{2}{c|}{} & \multicolumn{6}{c}{mean MLE $\Gamma\,t_{\rm true}$, $\Gamma=n_{\rm em}/t_{\rm exp}$} \\
\multicolumn{1}{c}{$n_{\rm IC}$} & \multicolumn{1}{c|}{$t_1/t_{\rm true}$} & \multicolumn{1}{c}{fixed} & \multicolumn{1}{c}{$0.1$} & \multicolumn{1}{c}{$0.5$} & \multicolumn{1}{c}{$0.8$} & \multicolumn{1}{c}{$n<3$} & \multicolumn{1}{c}{adv.} \\
\hline
32 & $10$ & $1.032(1)$ & $1.032(1)$ & $1.032(1)$ & $1.032(1)$ & $1.032(1)$ & $1.032(1)$ \\
32 & $10/3$ & $1.030(1)$ & $1.030(1)$ & $1.030(1)$ & $1.030(1)$ & $1.030(1)$ & $1.031(1)$ \\
32 & $1$ & $1.021(1)$ & $1.021(1)$ & $1.029(1)$ & $1.033(1)$ & $1.021(1)$ & $1.022(1)$ \\
32 & $1/3$ & $1.018(1)$ & $1.022(1)$ & $1.026(1)$ & $1.022(1)$ & $1.019(1)$ & $1.010(1)$ \\
32 & $1/10$ & $1.016(1)$ & $1.167(1)$ & $1.018(1)$ & $1.018(1)$ & $1.163(1)$ & $0.978(1)$ \\
32 & $1/30$ & $1.017(1)$ & $1.060(1)$ & $1.017(1)$ & $1.017(1)$ & $1.148(1)$ & $0.885(1)$ \\
\cline{1-8}
65 & $10$ & $1.016(1)$ & $1.016(1)$ & $1.016(1)$ & $1.016(1)$ & $1.016(1)$ & $1.016(1)$ \\
65 & $10/3$ & $1.014(1)$ & $1.014(1)$ & $1.014(1)$ & $1.014(1)$ & $1.014(1)$ & $1.015(1)$ \\
65 & $1$ & $1.010(1)$ & $1.010(1)$ & $1.012(1)$ & $1.015(1)$ & $1.010(1)$ & $1.011(1)$ \\
65 & $1/3$ & $1.009(1)$ & $1.009(1)$ & $1.011(1)$ & $1.011(1)$ & $1.009(1)$ & $1.006(1)$ \\
65 & $1/10$ & $1.008(1)$ & $1.115(1)$ & $1.009(1)$ & $1.009(1)$ & $1.031(1)$ & $0.998(1)$ \\
65 & $1/30$ & $1.008(1)$ & $1.018(1)$ & $1.008(1)$ & $1.008(1)$ & $1.195(1)$ & $1.004(1)$ \\
\cline{1-8}
120 & $10$ & $1.008(1)$ & $1.008(1)$ & $1.008(1)$ & $1.008(1)$ & $1.008(1)$ & $1.008(1)$ \\
120 & $10/3$ & $1.008(1)$ & $1.008(1)$ & $1.008(1)$ & $1.008(1)$ & $1.008(1)$ & $1.008(1)$ \\
120 & $1$ & $1.006(1)$ & $1.006(1)$ & $1.006(1)$ & $1.008(1)$ & $1.006(1)$ & $1.006(1)$ \\
120 & $1/3$ & $1.005(1)$ & $1.005(1)$ & $1.006(1)$ & $1.006(1)$ & $1.005(1)$ & $1.003(1)$ \\
120 & $1/10$ & $1.004(1)$ & $1.084(1)$ & $1.005(1)$ & $1.005(1)$ & $1.005(1)$ & $0.996(1)$ \\
120 & $1/30$ & $1.004(1)$ & $1.005(1)$ & $1.004(1)$ & $1.004(1)$ & $1.107(1)$ & $0.988(1)$ \\
\bottomrule
\end{tabular}
\caption{Mean MLE for the decay rate from the synthetic experiments. The column labels and error bar convention are the same as in Table~\ref{tab:cov-cv}.}
\label{tab:cov-rate}
\end{table}

\begin{table}[htbp]
\centering
\small
\begin{tabular}{rr|rrrrrr}
\toprule
\multicolumn{2}{c|}{} & \multicolumn{6}{c}{fraction of experiments: $n_{\rm em}=0$ $|$ $n_{\rm em}\leq 2$} \\
\multicolumn{1}{c}{$n_{\rm IC}$} & \multicolumn{1}{c|}{$t_1/t_{\rm true}$} & \multicolumn{1}{c}{fixed} & \multicolumn{1}{c}{$0.1$} & \multicolumn{1}{c}{$0.5$} & \multicolumn{1}{c}{$0.8$} & \multicolumn{1}{c}{$n<3$} & \multicolumn{1}{c}{adv.} \\
\hline
32 & $10$ & 0.00 $|$ 0.00 & 0.00 $|$ 0.00 & 0.00 $|$ 0.00 & 0.00 $|$ 0.00 & 0.00 $|$ 0.00 & 0.00 $|$ 0.00 \\
32 & $10/3$ & 0.00 $|$ 0.00 & 0.00 $|$ 0.00 & 0.00 $|$ 0.00 & 0.00 $|$ 0.00 & 0.00 $|$ 0.00 & 0.00 $|$ 0.00 \\
32 & $1$ & 0.00 $|$ 0.00 & 0.00 $|$ 0.00 & 0.00 $|$ 0.00 & 0.00 $|$ 0.00 & 0.00 $|$ 0.00 & 0.00 $|$ 0.00 \\
32 & $1/3$ & 0.00 $|$ 0.00 & 0.00 $|$ 0.00 & 0.00 $|$ 0.00 & 0.00 $|$ 0.00 & 0.00 $|$ 0.00 & 0.00 $|$ 0.00 \\
32 & $1/10$ & 0.04 $|$ 0.40 & 0.00 $|$ 0.00 & 0.00 $|$ 0.00 & 0.00 $|$ 0.00 & 0.00 $|$ 0.00 & 0.00 $|$ 0.36 \\
32 & $1/30$ & 0.34 $|$ 0.91 & 0.03 $|$ 0.33 & 0.03 $|$ 0.33 & 0.03 $|$ 0.33 & 0.03 $|$ 0.33 & 0.34 $|$ 0.91 \\
\cline{1-8}
65 & $10$ & 0.00 $|$ 0.00 & 0.00 $|$ 0.00 & 0.00 $|$ 0.00 & 0.00 $|$ 0.00 & 0.00 $|$ 0.00 & 0.00 $|$ 0.00 \\
65 & $10/3$ & 0.00 $|$ 0.00 & 0.00 $|$ 0.00 & 0.00 $|$ 0.00 & 0.00 $|$ 0.00 & 0.00 $|$ 0.00 & 0.00 $|$ 0.00 \\
65 & $1$ & 0.00 $|$ 0.00 & 0.00 $|$ 0.00 & 0.00 $|$ 0.00 & 0.00 $|$ 0.00 & 0.00 $|$ 0.00 & 0.00 $|$ 0.00 \\
65 & $1/3$ & 0.00 $|$ 0.00 & 0.00 $|$ 0.00 & 0.00 $|$ 0.00 & 0.00 $|$ 0.00 & 0.00 $|$ 0.00 & 0.00 $|$ 0.00 \\
65 & $1/10$ & 0.00 $|$ 0.05 & 0.00 $|$ 0.00 & 0.00 $|$ 0.00 & 0.00 $|$ 0.00 & 0.00 $|$ 0.00 & 0.00 $|$ 0.00 \\
65 & $1/30$ & 0.11 $|$ 0.64 & 0.00 $|$ 0.03 & 0.00 $|$ 0.03 & 0.00 $|$ 0.03 & 0.00 $|$ 0.03 & 0.00 $|$ 0.54 \\
\cline{1-8}
120 & $10$ & 0.00 $|$ 0.00 & 0.00 $|$ 0.00 & 0.00 $|$ 0.00 & 0.00 $|$ 0.00 & 0.00 $|$ 0.00 & 0.00 $|$ 0.00 \\
120 & $10/3$ & 0.00 $|$ 0.00 & 0.00 $|$ 0.00 & 0.00 $|$ 0.00 & 0.00 $|$ 0.00 & 0.00 $|$ 0.00 & 0.00 $|$ 0.00 \\
120 & $1$ & 0.00 $|$ 0.00 & 0.00 $|$ 0.00 & 0.00 $|$ 0.00 & 0.00 $|$ 0.00 & 0.00 $|$ 0.00 & 0.00 $|$ 0.00 \\
120 & $1/3$ & 0.00 $|$ 0.00 & 0.00 $|$ 0.00 & 0.00 $|$ 0.00 & 0.00 $|$ 0.00 & 0.00 $|$ 0.00 & 0.00 $|$ 0.00 \\
120 & $1/10$ & 0.00 $|$ 0.00 & 0.00 $|$ 0.00 & 0.00 $|$ 0.00 & 0.00 $|$ 0.00 & 0.00 $|$ 0.00 & 0.00 $|$ 0.00 \\
120 & $1/30$ & 0.02 $|$ 0.24 & 0.00 $|$ 0.00 & 0.00 $|$ 0.00 & 0.00 $|$ 0.00 & 0.00 $|$ 0.00 & 0.00 $|$ 0.15 \\
\bottomrule
\end{tabular}
\caption{Fraction of synthetic experiments with all samples censored and those with $n_{\rm em} \leq 2$. The column labels are the same as in Table~\ref{tab:cov-cv}.}
\label{tab:cov-frac}
\end{table}

\begin{table}[htbp]
\centering
\small
\begin{tabular}{rr|rrrrrr}
\toprule
\multicolumn{2}{c|}{} & \multicolumn{6}{c}{coverage of the $68.3\%$ Garwood interval} \\
\multicolumn{1}{c}{$n_{\rm IC}$} & \multicolumn{1}{c|}{$t_1/t_{\rm true}$} & \multicolumn{1}{c}{fixed} & \multicolumn{1}{c}{$0.1$} & \multicolumn{1}{c}{$0.5$} & \multicolumn{1}{c}{$0.8$} & \multicolumn{1}{c}{$n<3$} & \multicolumn{1}{c}{adv.} \\
\hline
32 & $10$ & $0.7201(3)$ & $0.7201(3)$ & $0.7201(3)$ & $0.7201(3)$ & $0.7201(3)$ & $0.7202(3)$ \\
32 & $10/3$ & $0.7199(3)$ & $0.7199(3)$ & $0.7199(3)$ & $0.7199(3)$ & $0.7199(3)$ & $0.7448(3)$ \\
32 & $1$ & $0.7296(3)$ & $0.7296(3)$ & $0.7416(3)$ & $0.7183(3)$ & $0.7296(3)$ & $0.8313(3)$ \\
32 & $1/3$ & $0.7466(3)$ & $0.7496(3)$ & $0.7274(3)$ & $0.7282(3)$ & $0.7471(3)$ & $0.8774(2)$ \\
32 & $1/10$ & $0.8755(2)$ & $0.7693(3)$ & $0.7468(3)$ & $0.7468(3)$ & $0.8009(3)$ & $0.9361(2)$ \\
32 & $1/30$ & $0.9135(2)$ & $0.7441(3)$ & $0.7506(3)$ & $0.7506(3)$ & $0.7025(3)$ & $0.9616(1)$ \\
\cline{1-8}
65 & $10$ & $0.7106(3)$ & $0.7106(3)$ & $0.7106(3)$ & $0.7106(3)$ & $0.7106(3)$ & $0.7106(3)$ \\
65 & $10/3$ & $0.7101(3)$ & $0.7101(3)$ & $0.7101(3)$ & $0.7101(3)$ & $0.7101(3)$ & $0.7402(3)$ \\
65 & $1$ & $0.7173(3)$ & $0.7173(3)$ & $0.7191(3)$ & $0.7105(3)$ & $0.7173(3)$ & $0.8240(3)$ \\
65 & $1/3$ & $0.7308(3)$ & $0.7308(3)$ & $0.7161(3)$ & $0.7161(3)$ & $0.7308(3)$ & $0.8685(2)$ \\
65 & $1/10$ & $0.7479(3)$ & $0.7205(3)$ & $0.7305(3)$ & $0.7305(3)$ & $0.7720(3)$ & $0.8926(2)$ \\
65 & $1/30$ & $0.7985(3)$ & $0.7794(3)$ & $0.7802(3)$ & $0.7802(3)$ & $0.7952(3)$ & $0.9248(2)$ \\
\cline{1-8}
120 & $10$ & $0.7033(3)$ & $0.7033(3)$ & $0.7033(3)$ & $0.7033(3)$ & $0.7033(3)$ & $0.7034(3)$ \\
120 & $10/3$ & $0.7028(3)$ & $0.7028(3)$ & $0.7028(3)$ & $0.7028(3)$ & $0.7028(3)$ & $0.7358(3)$ \\
120 & $1$ & $0.7091(3)$ & $0.7091(3)$ & $0.7091(3)$ & $0.7031(3)$ & $0.7091(3)$ & $0.8185(3)$ \\
120 & $1/3$ & $0.7221(3)$ & $0.7221(3)$ & $0.7081(3)$ & $0.7081(3)$ & $0.7221(3)$ & $0.8634(2)$ \\
120 & $1/10$ & $0.7704(3)$ & $0.7332(3)$ & $0.7215(3)$ & $0.7215(3)$ & $0.7705(3)$ & $0.8951(2)$ \\
120 & $1/30$ & $0.8068(3)$ & $0.7685(3)$ & $0.7686(3)$ & $0.7686(3)$ & $0.8291(3)$ & $0.9229(2)$ \\
\bottomrule
\end{tabular}
\caption{Coverage of the Garwood interval from the synthetic experiments. The associated uncertainties are binomial, $\sqrt{p(1-p)/N}$, with $p$ denoting the coverage and $N = 2 \times 10^6$. The lowest coverage is $p = 0.7025$ with uncertainty $0.0003$. The errors are rounded to one significant figure; if rounded to two, the $0.683$ nominal level is $61$ standard deviations below the lowest coverage. The column labels are the same as in Table~\ref{tab:cov-cv}.}
\label{tab:cov-cov}
\end{table}

\begin{table}[htbp]
\centering
\small
\begin{tabular}{rr|rrrrrr}
\toprule
\multicolumn{2}{c|}{} & \multicolumn{6}{c}{coverage of the $95\%$ lower bound} \\
\multicolumn{1}{c}{$n_{\rm IC}$} & \multicolumn{1}{c|}{$t_1/t_{\rm true}$} & \multicolumn{1}{c}{fixed} & \multicolumn{1}{c}{$0.1$} & \multicolumn{1}{c}{$0.5$} & \multicolumn{1}{c}{$0.8$} & \multicolumn{1}{c}{$n<3$} & \multicolumn{1}{c}{adv.} \\
\hline
32 & $10$ & $0.9649(1)$ & $0.9649(1)$ & $0.9649(1)$ & $0.9649(1)$ & $0.9649(1)$ & $0.9649(1)$ \\
32 & $10/3$ & $0.9641(1)$ & $0.9641(1)$ & $0.9641(1)$ & $0.9641(1)$ & $0.9641(1)$ & $0.9657(1)$ \\
32 & $1$ & $0.9638(1)$ & $0.9638(1)$ & $0.9804(1)$ & $0.9639(1)$ & $0.9638(1)$ & $0.9724(1)$ \\
32 & $1/3$ & $0.9701(1)$ & $0.9758(1)$ & $0.9634(1)$ & $0.9634(1)$ & $0.9711(1)$ & $0.9788(1)$ \\
32 & $1/10$ & $0.9591(1)$ & $0.9703(1)$ & $0.9700(1)$ & $0.9700(1)$ & $0.9727(1)$ & $0.9929(1)$ \\
32 & $1/30$ & $1.0000$ & $0.9713(1)$ & $0.9713(1)$ & $0.9713(1)$ & $0.9713(1)$ & $1.0000$ \\
\cline{1-8}
65 & $10$ & $0.9612(1)$ & $0.9612(1)$ & $0.9612(1)$ & $0.9612(1)$ & $0.9612(1)$ & $0.9613(1)$ \\
65 & $10/3$ & $0.9605(1)$ & $0.9605(1)$ & $0.9605(1)$ & $0.9605(1)$ & $0.9605(1)$ & $0.9613(1)$ \\
65 & $1$ & $0.9603(1)$ & $0.9603(1)$ & $0.9665(1)$ & $0.9607(1)$ & $0.9603(1)$ & $0.9689(1)$ \\
65 & $1/3$ & $0.9613(1)$ & $0.9614(1)$ & $0.9603(1)$ & $0.9603(1)$ & $0.9613(1)$ & $0.9759(1)$ \\
65 & $1/10$ & $0.9536(1)$ & $0.9619(1)$ & $0.9612(1)$ & $0.9612(1)$ & $0.9890(1)$ & $0.9786(1)$ \\
65 & $1/30$ & $1.0000$ & $0.9728(1)$ & $0.9728(1)$ & $0.9728(1)$ & $0.9728(1)$ & $0.9855(1)$ \\
\cline{1-8}
120 & $10$ & $0.9584(1)$ & $0.9584(1)$ & $0.9584(1)$ & $0.9584(1)$ & $0.9584(1)$ & $0.9584(1)$ \\
120 & $10/3$ & $0.9578(1)$ & $0.9578(1)$ & $0.9578(1)$ & $0.9578(1)$ & $0.9578(1)$ & $0.9585(1)$ \\
120 & $1$ & $0.9580(1)$ & $0.9580(1)$ & $0.9581(1)$ & $0.9581(1)$ & $0.9580(1)$ & $0.9662(1)$ \\
120 & $1/3$ & $0.9599(1)$ & $0.9599(1)$ & $0.9577(1)$ & $0.9577(1)$ & $0.9599(1)$ & $0.9739(1)$ \\
120 & $1/10$ & $0.9756(1)$ & $0.9617(1)$ & $0.9598(1)$ & $0.9598(1)$ & $0.9759(1)$ & $0.9809(1)$ \\
120 & $1/30$ & $0.9817(1)$ & $0.9738(1)$ & $0.9738(1)$ & $0.9738(1)$ & $0.9796(1)$ & $0.9877(1)$ \\
\bottomrule
\end{tabular}
\caption{Coverage of the $95 \%$ lower bound from the synthetic experiments. The associated uncertainties are binomial, $\sqrt{p(1-p)/N}$, with $p$ denoting the coverage and $N = 2 \times 10^6$. Entries with $p = 1.0000$ just mean all experiments produced a lower bound covering $t_{\rm true}$, and it is not a confidence bound; the binomial uncertainty is zero for these entries, which underestimates the real uncertainty. The lowest coverage is $p = 0.9536$ with uncertainty $0.0001$. The errors are rounded to one significant figure; if kept to two, $0.95$ would be $24$ standard deviations below the lowest coverage. The column labels are the same as in Table~\ref{tab:cov-cv}.}
\label{tab:cov-lcov}
\end{table}

\begin{table}[htbp]
\centering
\small
\begin{tabular}{rr|rrrrrr}
\toprule
\multicolumn{2}{c|}{} & \multicolumn{6}{c}{calibration factor of the $95\%$ bound} \\
\multicolumn{1}{c}{$n_{\rm IC}$} & \multicolumn{1}{c|}{$t_1/t_{\rm true}$} & \multicolumn{1}{c}{fixed} & \multicolumn{1}{c}{$0.1$} & \multicolumn{1}{c}{$0.5$} & \multicolumn{1}{c}{$0.8$} & \multicolumn{1}{c}{$n<3$} & \multicolumn{1}{c}{adv.} \\
\hline
32 & $10$ & $1.027(1)$ & $1.027(1)$ & $1.027(1)$ & $1.027(1)$ & $1.027(1)$ & $1.027(1)$ \\
32 & $10/3$ & $1.026(1)$ & $1.026(1)$ & $1.026(1)$ & $1.026(1)$ & $1.026(1)$ & $1.030(1)$ \\
32 & $1$ & $1.034(1)$ & $1.034(1)$ & $1.066(1)$ & $1.026(1)$ & $1.034(1)$ & $1.064(1)$ \\
32 & $1/3$ & $1.071(1)$ & $1.076(1)$ & $1.032(1)$ & $1.032(1)$ & $1.072(1)$ & $1.132(1)$ \\
32 & $1/10$ & $1.486(1)$ & $1.071(1)$ & $1.071(1)$ & $1.071(1)$ & $1.075(1)$ & $1.489(1)$ \\
32 & $1/30$ & $2.808$ & $1.343(1)$ & $1.343(1)$ & $1.343(1)$ & $1.343(1)$ & $2.808$ \\
\cline{1-8}
65 & $10$ & $1.014(1)$ & $1.014(1)$ & $1.014(1)$ & $1.014(1)$ & $1.014(1)$ & $1.014(1)$ \\
65 & $10/3$ & $1.013(1)$ & $1.013(1)$ & $1.013(1)$ & $1.013(1)$ & $1.013(1)$ & $1.015(1)$ \\
65 & $1$ & $1.017(1)$ & $1.017(1)$ & $1.023(1)$ & $1.014(1)$ & $1.017(1)$ & $1.035(1)$ \\
65 & $1/3$ & $1.040(1)$ & $1.040(1)$ & $1.016(1)$ & $1.016(1)$ & $1.040(1)$ & $1.075(1)$ \\
65 & $1/10$ & $1.205(1)$ & $1.046(1)$ & $1.039(1)$ & $1.039(1)$ & $1.216(1)$ & $1.171(1)$ \\
65 & $1/30$ & $1.383$ & $1.098(1)$ & $1.098(1)$ & $1.098(1)$ & $1.098(1)$ & $1.306(1)$ \\
\cline{1-8}
120 & $10$ & $1.008(1)$ & $1.008(1)$ & $1.008(1)$ & $1.008(1)$ & $1.008(1)$ & $1.008(1)$ \\
120 & $10/3$ & $1.007(1)$ & $1.007(1)$ & $1.007(1)$ & $1.007(1)$ & $1.007(1)$ & $1.008(1)$ \\
120 & $1$ & $1.009(1)$ & $1.009(1)$ & $1.010(1)$ & $1.008(1)$ & $1.009(1)$ & $1.021(1)$ \\
120 & $1/3$ & $1.018(1)$ & $1.018(1)$ & $1.009(1)$ & $1.009(1)$ & $1.018(1)$ & $1.049(1)$ \\
120 & $1/10$ & $1.019(1)$ & $1.022(1)$ & $1.018(1)$ & $1.018(1)$ & $1.020(1)$ & $1.126(1)$ \\
120 & $1/30$ & $1.190(1)$ & $1.022(1)$ & $1.022(1)$ & $1.022(1)$ & $1.115(1)$ & $1.312(1)$ \\
\bottomrule
\end{tabular}
\caption{Calibration factor measured from the synthetic experiments. Labelling the $95\%$ confidence lower bound by $L$, it is defined as $t_{\rm true}$ divided by the $95$th percentile in the distribution of $L$. The uncertainty interval comes from the one standard deviation interval of binomial statistics: the $100p$-th percentile has rank $p N \pm \sqrt{N p (1-p)}$; with $N = 2 \times 10^6$ and $p = 0.95$, the rank interval is about $\pm 308$; reading $L$ at the boundary of the $\pm 308$ range of ranks gives the uncertainty on the $95$th percentile; finally we propagate percentile uncertainty to the calibration factor. Entries without a quoted uncertainty have the same $L$ over this $\pm 308$ interval. The column labels are the same as in Table~\ref{tab:cov-cv}.}
\label{tab:cov-cal}
\end{table}

\FloatBarrier
\section{Reproducibility information}
Table~\ref{tab:hmc-metadata} shows the HMC meta-data. Every row consists of one HMC chain with a pre-fixed length, except for the three rows with the highest $\sigma$ for each $N \geq 15$. Among those, the ones with more than $10^4$ steps include a long chain with pre-fixed steps ($10^4$ for $N = 15,17,20,30$, $2 \times 10^4$ for $N = 40$, and $6 \times 10^4$ for $N=50$.) plus an independent extension that is cut off after the chain starts sampling emitted configurations; the others have only the shorter extension. Only chains with a pre-fixed length are used in calculating HMC averages in the main text (see Section~\ref{sec:matrixresults}); single-clump samples from the extension chains are included in the real time lifetime measurements. As described in the main text, we fix $\hat{J}$ and vary $\hat{T}$ for the $\sigma$-scan. For $N = 10$, we fix $\hat{J} = 30$, and vary $\hat{T}$ between $[0.26, 0.40]$; for $N = 15$, $\hat{J} = 61$, $\hat{T} \in [0.26, 0.40]$; for $N = 17$, $\hat{J} = 76$, $\hat{T} \in [0.26, 0.40]$; for $N = 20$, $\hat{J} = 100$, $\hat{T} \in [0.26, 0.42]$; for $N = 30$, $\hat{J} = 166$, $\hat{T} \in [0.20, 0.30]$; for $N = 40$, $\hat{J} = 337$, $\hat{T} \in [0.26, 0.40]$; for $N = 50$, $\hat{J} = 498$, $\hat{T} \in [0.26, 0.32]$. We also report the integrated correlation length associated with $|\mathcal{O}_2|$, $\lambda_{|\mathcal{O}_2|}$, estimated by the Madras-Sokal self-consistent windowing as described in Section~\ref{sec:hmcmethod}. For entries marked with an asterisk, the self-consistent window $W \geq 5 \lambda(W)$ is not reached within the lag range the single-blob record supports, so the sum is truncated at the longest measurable lag. The quoted uncertainty is the statistical error \eqref{eq:lambdaerr} of the truncated sum, and does not account for the missing tail. For rows with more than one chain, the listed $\lambda_{|\mathcal{O}_2|}$ are averages of $\lambda$ from each chain with errors added in quadrature; they are marked with an asterisk if any one chain is short. The uncertainties for the decay lifetime estimates quoted in the main text are computed using \eqref{eq:deff} and treating samples from different chains as uncorrelated, not directly using this average $\lambda_{|\mathcal{O}_2|}$.

\begin{table}[htbp]
\centering
\footnotesize
\begin{tabular}{rrrrrr|rrrrrr}
\toprule
$N$ & $\sigma$ & $\mathcal{N}$ & $\mathcal{N}_{\rm blob}$ & $P_{\rm acc}$ & $\lambda_{|\mathcal{O}_2|}$ & $N$ & $\sigma$ & $\mathcal{N}$ & $\mathcal{N}_{\rm blob}$ & $P_{\rm acc}$ & $\lambda_{|\mathcal{O}_2|}$ \\
 & & (steps) & (steps) & & (steps) &  & & (steps) & (steps) & & (steps) \\
\hline
10 & 1.061 & 10000 & 9998 & 0.9990 & $12.2 \pm 1.9$ & 20 & 1.138 & 10000 & 9952 & 0.9983 & $13.0 \pm 2.1$ \\
10 & 1.102 & 10000 & 9998 & 0.9994 & $15.3 \pm 2.7$ & 20 & 1.162 & 10000 & 9954 & 0.9983 & $16.0 \pm 2.9$ \\
10 & 1.148 & 10000 & 9998 & 0.9993 & $12.6 \pm 2.0$ & 20 & 1.187 & 10000 & 9953 & 0.9978 & $12.6 \pm 2.0$ \\
10 & 1.198 & 10000 & 9997 & 0.9986 & $19.2 \pm 3.8$ & 20 & 1.201 & 12000 & 11952 & 0.9978 & $17.5 \pm 3.0$ \\
10 & 1.225 & 12000 & 11996 & 0.9993 & $22.3 \pm 4.3$ & 20 & 1.214 & 10000 & 9955 & 0.9981 & $18.2 \pm 3.5$ \\
10 & 1.254 & 10000 & 9996 & 0.9996 & $16.2 \pm 2.9$ & 20 & 1.228 & 12000 & 11955 & 0.9985 & $18.6 \pm 3.3$ \\
10 & 1.284 & 12000 & 11979 & 0.9992 & $21.6 \pm 4.1$ & 20 & 1.243 & 10000 & 2444 & 0.9970 & $21.0 \pm 8.7$ \\
10 & 1.316 & 10000 & 9988 & 0.9995 & $19.8 \pm 3.9$ & 20 & 1.257 & 12000 & 2117 & 0.9971 & $17.5 \pm 7.1$ \\
10 & 1.350 & 12000 & 11982 & 0.9989 & $26.7 \pm 5.6$ & 20 & 1.273 & 10000 & 6053 & 0.9973 & $34 \pm 12$ \\
10 & 1.386 & 10000 & 9957 & 0.9997 & $19.8 \pm 3.9$ & 20 & 1.288 & 12000 & 4072 & 0.9961 & $19.6 \pm 6.1$ \\
10 & 1.424 & 12000 & 11883 & 0.9995 & $37.3 \pm 9.3$ & 20 & 1.304 & 10000 & 1057 & 0.9962 & $39 \pm 33$ \\
10 & 1.465 & 10000 & 9180 & 0.9996 & $26.4 \pm 6.4$ & 20 & 1.321 & 12000 & 104 & 0.9962 & $13 \pm 17^{*}$ \\\cline{1-6}
15 & 1.061 & 10000 & 9968 & 0.9985 & $10.1 \pm 1.4$ & 20 & 1.338 & 10000 & 4830 & 0.9978 & $20.2 \pm 5.8$ \\
15 & 1.102 & 10000 & 9968 & 0.9988 & $12.2 \pm 1.9$ & 20 & 1.374 & 11650 & 1978 & 0.9962 & $23 \pm 14$ \\
15 & 1.148 & 10000 & 9963 & 0.9992 & $15.2 \pm 2.6$ & 20 & 1.411 & 11620 & 2387 & 0.9969 & $25 \pm 12$ \\
15 & 1.198 & 10000 & 9967 & 0.9984 & $16.7 \pm 3.1$ & 20 & 1.452 & 12160 & 2947 & 0.9963 & $115 \pm 95^{*}$ \\\cline{7-12}
15 & 1.225 & 12000 & 11961 & 0.9986 & $17.0 \pm 2.9$ & 30 & 1.065 & 10000 & 9996 & 0.9964 & $10.0 \pm 1.4$ \\
15 & 1.254 & 10000 & 9958 & 0.9987 & $28.3 \pm 6.7$ & 30 & 1.121 & 10000 & 9995 & 0.9973 & $11.8 \pm 1.8$ \\
15 & 1.284 & 12000 & 11963 & 0.9984 & $23.9 \pm 4.8$ & 30 & 1.186 & 10000 & 9996 & 0.9963 & $14.5 \pm 2.5$ \\
15 & 1.316 & 10000 & 9173 & 0.9989 & $27.8 \pm 6.8$ & 30 & 1.259 & 10000 & 9996 & 0.9964 & $41 \pm 12$ \\
15 & 1.350 & 12000 & 7089 & 0.9991 & $34 \pm 11$ & 30 & 1.344 & 11650 & 4657 & 0.9963 & $59 \pm 49$ \\
15 & 1.386 & 11680 & 1969 & 0.9982 & $37 \pm 26$ & 30 & 1.391 & 1770 & 1718 & 0.9949 & $82 \pm 80$ \\
15 & 1.424 & 1650 & 1629 & 0.9994 & $52 \pm 42$ & 30 & 1.443 & 11740 & 1801 & 0.9961 & $42 \pm 33^{*}$ \\\cline{7-12}
15 & 1.465 & 11560 & 1978 & 0.9980 & $39 \pm 34$ & 40 & 1.053 & 20000 & 19969 & 0.9944 & $8.44 \pm 0.78$ \\\cline{1-6}
17 & 1.060 & 10000 & 9962 & 0.9981 & $10.3 \pm 1.5$ & 40 & 1.140 & 20000 & 19965 & 0.9931 & $11.8 \pm 1.3$ \\
17 & 1.102 & 10000 & 9962 & 0.9986 & $12.5 \pm 2.0$ & 40 & 1.190 & 20000 & 19988 & 0.9942 & $13.4 \pm 1.6$ \\
17 & 1.147 & 10000 & 9965 & 0.9988 & $14.0 \pm 2.3$ & 40 & 1.217 & 20000 & 19970 & 0.9946 & $18.1 \pm 2.4$ \\
17 & 1.198 & 10000 & 9962 & 0.9987 & $15.1 \pm 2.6$ & 40 & 1.245 & 20000 & 18301 & 0.9930 & $24.8 \pm 4.1$ \\
17 & 1.225 & 12000 & 11960 & 0.9986 & $22.5 \pm 4.4$ & 40 & 1.275 & 20000 & 8170 & 0.9802 & $24.3 \pm 5.9$ \\
17 & 1.253 & 10000 & 9958 & 0.9981 & $25.6 \pm 5.8$ & 40 & 1.307 & 20000 & 4935 & 0.9744 & $24.2 \pm 7.6$ \\
17 & 1.284 & 12000 & 11957 & 0.9982 & $18.8 \pm 3.3$ & 40 & 1.340 & 20000 & 2528 & 0.9686 & $29 \pm 14$ \\
17 & 1.316 & 10000 & 8281 & 0.9988 & $22.4 \pm 5.2$ & 40 & 1.376 & 21710 & 2347 & 0.9694 & $35 \pm 22$ \\
17 & 1.349 & 12000 & 486 & 0.9967 & $11.3 \pm 7.7$ & 40 & 1.414 & 1830 & 1790 & 0.9923 & $120 \pm 120^{*}$ \\
17 & 1.385 & 11950 & 1949 & 0.9972 & $14.9 \pm 8.7^{*}$ & 40 & 1.455 & 22250 & 2310 & 0.9673 & $26 \pm 15$ \\\cline{7-12}
17 & 1.424 & 1950 & 1858 & 0.9969 & $21 \pm 10$ & 50 & 1.245 & 60000 & 59984 & 0.9920 & $20.1 \pm 1.6$ \\
17 & 1.465 & 11980 & 1896 & 0.9978 & $30 \pm 17$ & 50 & 1.275 & 60000 & 14536 & 0.9816 & $24.0 \pm 4.4$ \\\cline{1-6}
20 & 1.013 & 5000 & 4954 & 0.9980 & $9.2 \pm 1.8$ & 50 & 1.307 & 60000 & 6001 & 0.9822 & $22.2 \pm 6.0$ \\
20 & 1.032 & 5000 & 4954 & 0.9976 & $11.4 \pm 2.5$ & 50 & 1.340 & 60000 & 3522 & 0.9829 & $48 \pm 29$ \\
20 & 1.071 & 10000 & 9954 & 0.9976 & $10.4 \pm 1.5$ & 50 & 1.376 & 61860 & 4360 & 0.9812 & $43 \pm 20$ \\
20 & 1.092 & 10000 & 9956 & 0.9973 & $9.5 \pm 1.3$ & 50 & 1.414 & 2310 & 2249 & 0.9913 & $57 \pm 40$ \\
20 & 1.114 & 10000 & 9956 & 0.9985 & $10.9 \pm 1.6$ & 50 & 1.455 & 3630 & 3392 & 0.9945 & $44 \pm 22$ \\
\bottomrule
\end{tabular}
\caption{HMC sampling meta-data. Every chain uses the same leap-frog integrator with $\delta \hat{t}=0.005$ and $n=40$ per HMC step. $\mathcal{N}$ is the total number of HMC steps, and $\mathcal{N}_{\rm blob}$ is the number of single-blob configurations. $P_{\rm acc}$ is the mean Metropolis acceptance rate after burn-in. $\lambda_{|\mathcal{O}_2|}$ is the integrated autocorrelation length of $|O_2|$ restricted to single-clump samples. Entries marked with an asterisk are truncated estimates with an unresolved tail.} \label{tab:hmc-metadata}
\end{table}

\begin{table}[htbp]
\centering
\small
\begin{tabular}{rrrrrr|rrrrrr}
\toprule
$N$ & $\sigma$ & $n_{\rm IC}$ & $n_{\rm em}$ & $\hat{t}_{\rm exp}$ & $\hat{t}_{\rm decay}$ & $N$ & $\sigma$ & $n_{\rm IC}$ & $n_{\rm em}$ & $\hat{t}_{\rm exp}$ & $\hat{t}_{\rm decay}$ \\
 & & & & ($\times 10^{3}$) & ($\times 10^{3}$) &  & & & & ($\times 10^{3}$) & ($\times 10^{3}$) \\
\hline
10 & 1.061 & 32 & 0 & 96.0 & $> 32$ & 20 & 1.214 & 117 & 5 & 1210 & $240^{+280}_{-120}$ \\
10 & 1.102 & 32 & 0 & 96.0 & $> 26$ & 20 & 1.243 & 70 & 10 & 1230 & $123^{+176}_{-65}$ \\
10 & 1.148 & 32 & 0 & 288 & $> 76$ & 20 & 1.273 & 109 & 24 & 1290 & $54^{+29}_{-18}$ \\
10 & 1.198 & 115 & 4 & 2680 & $670^{+1040}_{-360}$ & 20 & 1.304 & 26 & 19 & 316 & $16.6^{+11.0}_{-6.2}$ \\
10 & 1.254 & 116 & 21 & 2720 & $130^{+55}_{-37}$ & 20 & 1.338 & 31 & 19 & 161 & $8.5^{+3.0}_{-2.2}$ \\
10 & 1.316 & 110 & 56 & 1960 & $35.0^{+6.7}_{-5.6}$ & 20 & 1.374 & 53 & 50 & 118 & $2.36^{+0.75}_{-0.55}$ \\
10 & 1.386 & 32 & 15 & 229 & $15.3^{+6.1}_{-4.2}$ & 20 & 1.411 & 65 & 65 & 55.7 & $0.86^{+0.23}_{-0.18}$ \\
10 & 1.465 & 26 & 21 & 97.7 & $4.65^{+1.29}_{-0.99}$ & 20 & 1.452 & 64 & 64 & 27.9 & $0.44^{+0.16}_{-0.11}$ \\\cline{1-6}\cline{7-12}
15 & 1.061 & 32 & 0 & 96.0 & $> 30$ & 30 & 1.065 & 32 & 0 & 96.0 & $> 28$ \\
15 & 1.102 & 32 & 0 & 96.0 & $> 27$ & 30 & 1.121 & 32 & 0 & 96.0 & $> 26$ \\
15 & 1.148 & 32 & 0 & 288 & $> 71$ & 30 & 1.186 & 117 & 0 & 628 & $> 140$ \\
15 & 1.198 & 116 & 3 & 1610 & $540^{+1020}_{-310}$ & 30 & 1.259 & 117 & 12 & 596 & $50^{+41}_{-21}$ \\
15 & 1.254 & 116 & 11 & 1330 & $121^{+86}_{-47}$ & 30 & 1.344 & 81 & 79 & 193 & $2.45^{+0.69}_{-0.53}$ \\
15 & 1.316 & 114 & 54 & 1070 & $19.8^{+6.5}_{-4.8}$ & 30 & 1.391 & 48 & 48 & 39.4 & $0.82^{+0.35}_{-0.24}$ \\
15 & 1.386 & 48 & 43 & 151 & $3.50^{+1.44}_{-0.99}$ & 30 & 1.443 & 47 & 47 & 15.8 & $0.34^{+0.16}_{-0.10}$ \\\cline{7-12}
15 & 1.424 & 50 & 50 & 101 & $2.02^{+0.90}_{-0.60}$ & 40 & 1.053 & 32 & 0 & 96.0 & $> 30$ \\
15 & 1.465 & 53 & 53 & 63.7 & $1.20^{+0.60}_{-0.38}$ & 40 & 1.140 & 32 & 0 & 96.0 & $> 30$ \\\cline{1-6}
17 & 1.060 & 32 & 0 & 96.0 & $> 30$ & 40 & 1.190 & 32 & 0 & 288 & $> 89$ \\
17 & 1.102 & 32 & 0 & 96.0 & $> 27$ & 40 & 1.217 & 32 & 1 & 284 & $> 58$ \\
17 & 1.147 & 32 & 0 & 288 & $> 85$ & 40 & 1.245 & 32 & 1 & 285 & $> 60$ \\
17 & 1.198 & 117 & 1 & 1910 & $> 280$ & 40 & 1.275 & 31 & 7 & 247 & $35^{+30}_{-15}$ \\
17 & 1.253 & 117 & 22 & 1720 & $78^{+28}_{-20}$ & 40 & 1.307 & 31 & 18 & 181 & $10.1^{+3.8}_{-2.7}$ \\
17 & 1.316 & 111 & 78 & 1290 & $16.6^{+3.1}_{-2.6}$ & 40 & 1.340 & 31 & 29 & 105 & $3.63^{+1.40}_{-0.98}$ \\
17 & 1.385 & 48 & 46 & 136 & $2.95^{+0.69}_{-0.55}$ & 40 & 1.376 & 62 & 62 & 57.8 & $0.93^{+0.32}_{-0.23}$ \\
17 & 1.424 & 44 & 44 & 66.7 & $1.52^{+0.40}_{-0.31}$ & 40 & 1.414 & 47 & 47 & 19.2 & $0.41^{+0.18}_{-0.12}$ \\
17 & 1.465 & 48 & 48 & 29.6 & $0.62^{+0.17}_{-0.13}$ & 40 & 1.455 & 46 & 46 & 8.58 & $0.187^{+0.059}_{-0.044}$ \\\cline{1-6}\cline{7-12}
20 & 1.013 & 32 & 0 & 96.0 & $> 21$ & 50 & 1.245 & 128 & 1 & 894 & $> 170$ \\
20 & 1.032 & 32 & 0 & 96.0 & $> 21$ & 50 & 1.275 & 127 & 8 & 1020 & $128^{+94}_{-51}$ \\
20 & 1.071 & 32 & 0 & 96.0 & $> 30$ & 50 & 1.307 & 128 & 42 & 872 & $20.8^{+5.9}_{-4.5}$ \\
20 & 1.092 & 32 & 0 & 96.0 & $> 30$ & 50 & 1.340 & 105 & 105 & 447 & $4.26^{+1.26}_{-0.95}$ \\
20 & 1.114 & 32 & 0 & 96.0 & $> 27$ & 50 & 1.376 & 144 & 144 & 172 & $1.19^{+0.36}_{-0.27}$ \\
20 & 1.138 & 32 & 0 & 96.0 & $> 31$ & 50 & 1.414 & 60 & 60 & 20.1 & $0.335^{+0.112}_{-0.082}$ \\
20 & 1.162 & 32 & 0 & 288 & $> 91$ & 50 & 1.455 & 57 & 57 & 7.67 & $0.135^{+0.037}_{-0.028}$ \\
20 & 1.187 & 117 & 3 & 1710 & $570^{+1090}_{-330}$ &  &  &  &  &  &  \\
\bottomrule
\end{tabular}
\caption{Real-time emission rate results. All trajectories are evolved with time step $\delta \hat{t}=0.005$.  $n_{\rm IC}$ is the total number of trajectories out of which $n_{\rm{em}}$ emitted within the observation time. The observation cutoffs $\hat{t}^I_{\rm obs}$ differ between trajectories; the column $\hat{t}_{\rm exp}$ reports the total exposure time as defined in \eqref{eq:texp}. For rows with more than two emissions, the reported uncertainty range is the Garwood interval with $68.3\%$ nominal confidence level. For rows with less than or equal to two emissions, the reported lower bound is at $95 \%$ nominal confidence level. }
\label{tab:decay-metadata}
\end{table}

\begin{table}[htbp]
\centering
\small
\begin{tabular}{rrrlr|rrrlr}
\toprule
$N$ & $\sigma$ & $\hat{t}_{\rm obs}^{\max}$ & censored at $\hat{t}_{\rm obs}$ & $f_{\rm cens}$ & $N$ & $\sigma$ & $\hat{t}_{\rm obs}^{\max}$ & censored at $\hat{t}_{\rm obs}$ & $f_{\rm cens}$ \\
 & & ($\times 10^{3}$) & ($\times 10^{3}$) &  &  & & ($\times 10^{3}$) & ($\times 10^{3}$) &  \\
\hline
10 & 1.061 & 3 & 3\,(32) & 1.00 & 20 & 1.214 & 11 & 9\,(19), 11\,(93) & 0.99 \\
10 & 1.102 & 3 & 3\,(32) & 1.00 & 20 & 1.243 & 19 & 19\,(60) & 0.92 \\
10 & 1.148 & 9 & 9\,(32) & 1.00 & 20 & 1.273 & 14 & 9\,(11), 14\,(74) & 0.88 \\
10 & 1.198 & 27 & 9\,(21), 27\,(90) & 0.98 & 20 & 1.304 & 30 & 30\,(7) & 0.66 \\
10 & 1.254 & 30 & 9\,(20), 30\,(75) & 0.89 & 20 & 1.338 & 9 & 9\,(12) & 0.67 \\
10 & 1.316 & 30 & 9\,(13), 30\,(41) & 0.69 & 20 & 1.374 & 9 & 3\,(1), 9\,(2) & 0.18 \\
10 & 1.386 & 9 & 9\,(17) & 0.67 & 20 & 1.411 & 6.5 & -- & 0.00 \\
10 & 1.465 & 9 & 9\,(5) & 0.46 & 20 & 1.452 & 2.3 & -- & 0.00 \\\cline{1-5}\cline{6-10}
15 & 1.061 & 3 & 3\,(32) & 1.00 & 30 & 1.065 & 3 & 3\,(32) & 1.00 \\
15 & 1.102 & 3 & 3\,(32) & 1.00 & 30 & 1.121 & 3 & 3\,(32) & 1.00 \\
15 & 1.148 & 9 & 9\,(32) & 1.00 & 30 & 1.186 & 9 & 4\,(85), 9\,(32) & 1.00 \\
15 & 1.198 & 15 & 9\,(19), 15\,(94) & 0.98 & 30 & 1.259 & 9 & 4\,(78), 9\,(27) & 0.93 \\
15 & 1.254 & 13 & 9\,(20), 13\,(85) & 0.96 & 30 & 1.344 & 9 & 9\,(2) & 0.09 \\
15 & 1.316 & 14 & 9\,(16), 14\,(44) & 0.71 & 30 & 1.391 & 3.1 & -- & 0.00 \\
15 & 1.386 & 9 & 9\,(5) & 0.30 & 30 & 1.443 & 1.3 & -- & 0.00 \\\cline{6-10}
15 & 1.424 & 8.1 & -- & 0.00 & 40 & 1.053 & 3 & 3\,(32) & 1.00 \\
15 & 1.465 & 7.7 & -- & 0.00 & 40 & 1.140 & 3 & 3\,(32) & 1.00 \\\cline{1-5}
17 & 1.060 & 3 & 3\,(32) & 1.00 & 40 & 1.190 & 9 & 9\,(32) & 1.00 \\
17 & 1.102 & 3 & 3\,(32) & 1.00 & 40 & 1.217 & 9 & 9\,(31) & 0.98 \\
17 & 1.147 & 9 & 9\,(32) & 1.00 & 40 & 1.245 & 9 & 9\,(31) & 0.98 \\
17 & 1.198 & 18 & 9\,(21), 18\,(95) & 0.99 & 40 & 1.275 & 9 & 9\,(24) & 0.87 \\
17 & 1.253 & 18 & 9\,(20), 18\,(75) & 0.89 & 40 & 1.307 & 9 & 9\,(13) & 0.65 \\
17 & 1.316 & 24 & 9\,(11), 24\,(22) & 0.49 & 40 & 1.340 & 9 & 9\,(2) & 0.17 \\
17 & 1.385 & 9 & 9\,(2) & 0.13 & 40 & 1.376 & 3.6 & -- & 0.00 \\
17 & 1.424 & 5.2 & -- & 0.00 & 40 & 1.414 & 1.2 & -- & 0.00 \\
17 & 1.465 & 2.9 & -- & 0.00 & 40 & 1.455 & 0.7 & -- & 0.00 \\\cline{1-5}\cline{6-10}
20 & 1.013 & 3 & 3\,(32) & 1.00 & 50 & 1.245 & 19 & 3\,(96), 19\,(31) & 0.98 \\
20 & 1.032 & 3 & 3\,(32) & 1.00 & 50 & 1.275 & 22 & 4\,(93), 22\,(26) & 0.92 \\
20 & 1.071 & 3 & 3\,(32) & 1.00 & 50 & 1.307 & 23 & 4\,(72), 23\,(14) & 0.70 \\
20 & 1.092 & 3 & 3\,(32) & 1.00 & 50 & 1.340 & 22 & -- & 0.00 \\
20 & 1.114 & 3 & 3\,(32) & 1.00 & 50 & 1.376 & 6 & -- & 0.00 \\
20 & 1.138 & 3 & 3\,(32) & 1.00 & 50 & 1.414 & 2 & -- & 0.00 \\
20 & 1.162 & 9 & 9\,(32) & 1.00 & 50 & 1.455 & 1 & -- & 0.00 \\
20 & 1.187 & 16 & 9\,(21), 16\,(93) & 0.98 &  &  &  &  &  \\
\bottomrule
\end{tabular}
\caption{Observation windows for the real time decay rate estimate. $\hat{t}^{\rm max}_{\rm obs}$ labels the maximum observation time. The column "censored at $\hat{t}_{\rm obs}$" reports the times at which the observation stopped and in parentheses the number of censored trajectories; entries with a dash have no censored trajectories. The column $f_{\rm cens}$ lists the fraction of exposure contributed by censored trajectories.}
\label{tab:obs-windows}
\end{table}

\FloatBarrier
\section{Sensitivity checks on the emission criteria}\label{app:sens}

\begin{table}[htbp]
\centering
\small
\begin{tabular}{rr|rr|rr|rr|rr}
\toprule
\multicolumn{1}{c}{$N$} & \multicolumn{1}{c|}{$\sigma$} & \multicolumn{2}{c|}{$0.5$} & \multicolumn{2}{c|}{$2$} & \multicolumn{2}{c|}{$5$} & \multicolumn{2}{c}{$10$} \\[2pt]
 &  & \multicolumn{1}{c}{$B$} & \multicolumn{1}{c|}{$\frac{\delta n_{\rm em}}{n_{\rm em}}$} & \multicolumn{1}{c}{$B$} & \multicolumn{1}{c|}{$\frac{\delta n_{\rm em}}{n_{\rm em}}$} & \multicolumn{1}{c}{$B$} & \multicolumn{1}{c|}{$\frac{\delta n_{\rm em}}{n_{\rm em}}$} & \multicolumn{1}{c}{$B$} & \multicolumn{1}{c}{$\frac{\delta n_{\rm em}}{n_{\rm em}}$} \\
\midrule
20 & 1.187 & $1.000$ & $0$ & $-$ & $-0.67$ & $-$ & $-1.00$ & $-$ & $-1.00$ \\[2pt]
20 & 1.214 & $0.83^{+0.17}_{-0.23}$ & $+0.20$ & $1.25^{+0.51}_{-0.25}$ & $-0.20$ & $1.25^{+0.51}_{-0.25}$ & $-0.20$ & $1.25^{+0.51}_{-0.25}$ & $-0.20$ \\[2pt]
20 & 1.243 & $0.69^{+0.14}_{-0.16}$ & $+0.40$ & $1.26^{+0.26}_{-0.26}$ & $-0.20$ & $1.45^{+0.45}_{-0.31}$ & $-0.30$ & $1.46^{+0.45}_{-0.31}$ & $-0.30$ \\[2pt]
20 & 1.273 & $0.659^{+0.132}_{-0.092}$ & $+0.42$ & $1.045^{+0.051}_{-0.044}$ & $-0.04$ & $1.29^{+0.17}_{-0.13}$ & $-0.21$ & $1.45^{+0.24}_{-0.18}$ & $-0.29$ \\[2pt]
20 & 1.304 & $0.893^{+0.105}_{-0.095}$ & $+0.05$ & $1.18^{+0.13}_{-0.17}$ & $-0.11$ & $1.26^{+0.19}_{-0.19}$ & $-0.16$ & $1.26^{+0.19}_{-0.19}$ & $-0.16$ \\[2pt]
20 & 1.338 & $0.875^{+0.122}_{-0.089}$ & $+0.05$ & $1.070^{+0.098}_{-0.066}$ & $-0.05$ & $1.077^{+0.097}_{-0.067}$ & $-0.05$ & $1.084^{+0.099}_{-0.068}$ & $-0.05$ \\[2pt]
20 & 1.374 & $0.869^{+0.079}_{-0.070}$ & $+0.02$ & $1.035^{+0.023}_{-0.021}$ & $-0.02$ & $1.155^{+0.097}_{-0.061}$ & $-0.08$ & $1.253^{+0.137}_{-0.077}$ & $-0.14$ \\[2pt]
\midrule
30 & 1.259 & $1.000$ & $0$ & $1.001$ & $0$ & $1.001$ & $0$ & $1.002^{+0.001}_{-0.001}$ & $0$ \\[2pt]
30 & 1.344 & $0.994^{+0.001}_{-0.001}$ & $0$ & $1.019^{+0.014}_{-0.013}$ & $-0.01$ & $1.052^{+0.029}_{-0.025}$ & $-0.04$ & $1.133^{+0.065}_{-0.044}$ & $-0.10$ \\[2pt]
30 & 1.391 & $0.985^{+0.003}_{-0.005}$ & $0$ & $1.022^{+0.005}_{-0.006}$ & $0$ & $1.081^{+0.036}_{-0.035}$ & $-0.04$ & $1.228^{+0.089}_{-0.069}$ & $-0.15$ \\[2pt]
30 & 1.443 & $0.973^{+0.008}_{-0.009}$ & $0$ & $1.052^{+0.025}_{-0.024}$ & $-0.02$ & $1.188^{+0.087}_{-0.055}$ & $-0.11$ & $1.307^{+0.099}_{-0.078}$ & $-0.17$ \\[2pt]
\midrule
40 & 1.275 & $1.000$ & $0$ & $1.000$ & $0$ & $1.001^{+0.001}_{-0.001}$ & $0$ & $1.001^{+0.001}_{-0.001}$ & $0$ \\[2pt]
40 & 1.307 & $0.999$ & $0$ & $1.001$ & $0$ & $1.002^{+0.001}_{-0.001}$ & $0$ & $1.002^{+0.001}_{-0.001}$ & $0$ \\[2pt]
40 & 1.414 & $0.984^{+0.003}_{-0.007}$ & $0$ & $1.066^{+0.047}_{-0.028}$ & $-0.04$ & $1.131^{+0.062}_{-0.051}$ & $-0.09$ & $1.200^{+0.097}_{-0.064}$ & $-0.13$ \\[2pt]
\midrule
50 & 1.275 & $1.000$ & $0$ & $1.000$ & $0$ & $1.000$ & $0$ & $1.000$ & $0$ \\[2pt]
50 & 1.307 & $1.000$ & $0$ & $1.000$ & $0$ & $1.000$ & $0$ & $1.001$ & $0$ \\[2pt]
50 & 1.340 & $0.999$ & $0$ & $1.001$ & $0$ & $1.003$ & $0$ & $1.003$ & $0$ \\[2pt]
50 & 1.414 & $0.984^{+0.004}_{-0.003}$ & $0$ & $1.009^{+0.003}_{-0.002}$ & $0$ & $1.024^{+0.008}_{-0.004}$ & $0$ & $1.033^{+0.007}_{-0.005}$ & $0$ \\[2pt]
50 & 1.455 & $0.977^{+0.004}_{-0.007}$ & $0$ & $1.033^{+0.010}_{-0.007}$ & $0$ & $1.067^{+0.019}_{-0.011}$ & $0$ & $1.090^{+0.023}_{-0.013}$ & $0$ \\[2pt]
\bottomrule
\end{tabular}
\caption{Sensitivity check on the MLE lifetime with confirmation interval length. We show the change in measured decay time if the interval is changed to $0.5$, $2$, $5$, and $10$ time units. The fractional change in the number of emitted trajectories is $\delta n_{\rm em}/n_{\rm em}$. The column $B$ is $\hat{t}_{\rm decay}$ under the different confirmation interval divided by that with $1$ time unit of confirmation. Entries with a dash have $n_{\rm em}$ fallen below three. Entries quoted without an error bar have one smaller than the last decimal place shown. }
\label{tab:hold-ratio}
\end{table}

\begin{table}[htbp]
\centering
\small
\begin{tabular}{rr|rrl|rrl|rrl}
\toprule
\multicolumn{1}{c}{$N$} & \multicolumn{1}{c|}{$\sigma$} & \multicolumn{3}{c|}{$\mathfrak{r}=5$} & \multicolumn{3}{c|}{$\mathfrak{r}=7$} & \multicolumn{3}{c}{$\mathfrak{r}=10$} \\[2pt]
 &  & \multicolumn{1}{c}{$B$} & \multicolumn{1}{c}{$q$} & \multicolumn{1}{c|}{$1/A$} & \multicolumn{1}{c}{$B$} & \multicolumn{1}{c}{$q$} & \multicolumn{1}{c|}{$1/A$} & \multicolumn{1}{c}{$B$} & \multicolumn{1}{c}{$q$} & \multicolumn{1}{c}{$1/A$} \\
\midrule
20 & 1.187 & $-$ & $0$ & $1.71$--$\infty$ & $-$ & $0$ & $1.71$--$\infty$ & $-$ & $0$ & $1.71$--$\infty$ \\[2pt]
20 & 1.214 & $1.25^{+0.42}_{-0.25}$ & $0$ & $1.26$--$1.29$ & $1.26^{+0.42}_{-0.25}$ & $0$ & $1.42$--$1.45$ & $1.69^{+1.36}_{-0.47}$ & $0$ & $2.41$--$3.23$ \\[2pt]
20 & 1.243 & $1.45^{+0.44}_{-0.31}$ & $0$ & $1.15$--$1.47$ & $1.71^{+0.69}_{-0.40}$ & $0$ & $1.37$--$2.21$ & $1.72^{+0.69}_{-0.40}$ & $0$ & $1.59$--$2.85$ \\[2pt]
20 & 1.273 & $1.29^{+0.18}_{-0.13}$ & $0$ & $1.12$--$1.29$ & $1.46^{+0.24}_{-0.18}$ & $0$ & $1.23$--$1.60$ & $1.68^{+0.36}_{-0.24}$ & $0$ & $1.45$--$2.27$ \\[2pt]
20 & 1.304 & $1.19^{+0.13}_{-0.17}$ & $11$ & $1.17$--$1.19$ & $1.29^{+0.19}_{-0.19}$ & $16$ & $1.34$--$1.46$ & $1.34^{+0.20}_{-0.19}$ & $16$ & $1.55$--$1.71$ \\[2pt]
20 & 1.338 & $1.079^{+0.105}_{-0.068}$ & $0$ & $1.14$--$1.17$ & $1.27^{+0.24}_{-0.14}$ & $0$ & $1.42$--$1.65$ & $1.71^{+0.48}_{-0.23}$ & $0$ & $1.82$--$2.92$ \\[2pt]
20 & 1.374 & $1.189^{+0.100}_{-0.066}$ & $10$ & $1.15$--$1.27$ & $1.72^{+0.19}_{-0.15}$ & $30$ & $1.40$--$2.22$ & $3.93^{+1.31}_{-0.62}$ & $64$ & $1.66$--$5.41$ \\[2pt]
20 & 1.411 & $1.331^{+0.087}_{-0.095}$ & $18$ & $1.15$--$1.42$ & $2.78^{+0.35}_{-0.52}$ & $54$ & $1.43$--$3.58$ & $5.53^{+0.54}_{-0.69}$ & $72$ & $1.61$--$9.97$ \\[2pt]
20 & 1.452 & $1.372^{+0.089}_{-0.072}$ & $16$ & $1.27$--$1.52$ & $3.47^{+0.38}_{-0.30}$ & $58$ & $1.65$--$4.34$ & $5.86^{+0.58}_{-0.35}$ & $69$ & $1.85$--$9.26$ \\[2pt]
\midrule
30 & 1.259 & $1.002^{+0.001}_{-0.001}$ & $0$ & $1.04$ & $1.006^{+0.002}_{-0.002}$ & $0$ & $1.12$ & $1.52^{+0.71}_{-0.26}$ & $0$ & $1.25$--$1.94$ \\[2pt]
30 & 1.344 & $1.107^{+0.045}_{-0.032}$ & $8$ & $1.07$--$1.16$ & $1.304^{+0.079}_{-0.072}$ & $18$ & $1.19$--$1.50$ & $1.88^{+0.33}_{-0.12}$ & $34$ & $1.37$--$2.64$ \\[2pt]
30 & 1.391 & $1.241^{+0.093}_{-0.073}$ & $15$ & $1.13$--$1.32$ & $2.17^{+0.46}_{-0.23}$ & $46$ & $1.35$--$2.75$ & $11.9^{+8.6}_{-3.4}$ & $90$ & $1.44$--$14$ \\[2pt]
30 & 1.443 & $1.412^{+0.123}_{-0.094}$ & $21$ & $1.18$--$1.53$ & $4.68^{+1.64}_{-0.96}$ & $72$ & $1.44$--$6.84$ & $-$ & $96$ & $1.52$--$37$ \\[2pt]
\midrule
40 & 1.275 & $1.002^{+0.001}_{-0.001}$ & $0$ & $1.02$ & $1.005^{+0.002}_{-0.002}$ & $0$ & $1.04$ & $1.18^{+0.50}_{-0.17}$ & $0$ & $1.08$--$1.26$ \\[2pt]
40 & 1.307 & $1.003^{+0.001}_{-0.001}$ & $0$ & $1.02$ & $1.013^{+0.003}_{-0.004}$ & $0$ & $1.08$ & $1.16^{+0.11}_{-0.13}$ & $0$ & $1.16$--$1.31$ \\[2pt]
40 & 1.340 & $1.125^{+0.092}_{-0.078}$ & $7$ & $1.02$--$1.14$ & $1.192^{+0.102}_{-0.084}$ & $7$ & $1.08$--$1.26$ & $1.289^{+0.129}_{-0.094}$ & $10$ & $1.17$--$1.46$ \\[2pt]
40 & 1.376 & $1.145^{+0.047}_{-0.055}$ & $10$ & $1.07$--$1.19$ & $1.467^{+0.201}_{-0.086}$ & $24$ & $1.19$--$1.64$ & $2.60^{+0.70}_{-0.25}$ & $53$ & $1.35$--$3.44$ \\[2pt]
40 & 1.414 & $1.204^{+0.085}_{-0.057}$ & $11$ & $1.13$--$1.27$ & $1.97^{+0.23}_{-0.17}$ & $38$ & $1.34$--$2.42$ & $6.3^{+2.5}_{-1.3}$ & $79$ & $1.49$--$8.89$ \\[2pt]
40 & 1.455 & $1.396^{+0.138}_{-0.084}$ & $17$ & $1.24$--$1.55$ & $2.48^{+0.45}_{-0.26}$ & $43$ & $1.59$--$3.47$ & $-$ & $100$ & $1.74$--$\infty$ \\[2pt]
\midrule
50 & 1.275 & $1.000$ & $0$ & $1.00$ & $1.14^{+0.26}_{-0.14}$ & $0$ & $1.02$--$1.16$ & $1.34^{+0.42}_{-0.34}$ & $12$ & $1.03$--$1.38$ \\[2pt]
50 & 1.307 & $1.001$ & $0$ & $1.02$ & $1.030^{+0.038}_{-0.025}$ & $0$ & $1.06$--$1.09$ & $1.088^{+0.062}_{-0.051}$ & $0$ & $1.12$--$1.21$ \\[2pt]
50 & 1.340 & $1.007^{+0.001}_{-0.001}$ & $0$ & $1.02$ & $1.096^{+0.033}_{-0.034}$ & $7$ & $1.05$--$1.13$ & $1.136^{+0.034}_{-0.036}$ & $8$ & $1.10$--$1.20$ \\[2pt]
50 & 1.376 & $1.046^{+0.020}_{-0.013}$ & $2$ & $1.06$--$1.08$ & $1.148^{+0.031}_{-0.023}$ & $6$ & $1.18$--$1.27$ & $1.286^{+0.037}_{-0.033}$ & $10$ & $1.34$--$1.52$ \\[2pt]
50 & 1.414 & $1.074^{+0.016}_{-0.012}$ & $0$ & $1.15$ & $1.281^{+0.071}_{-0.052}$ & $3$ & $1.43$--$1.50$ & $1.593^{+0.109}_{-0.064}$ & $5$ & $1.82$--$1.98$ \\[2pt]
50 & 1.455 & $1.181^{+0.047}_{-0.023}$ & $0$ & $1.29$ & $1.552^{+0.136}_{-0.066}$ & $0$ & $1.83$ & $2.21^{+0.28}_{-0.14}$ & $0$ & $2.68$ \\[2pt]
\bottomrule
\end{tabular}
\caption{Sensitivity check on the lifetime with threshold $\lambda_1/\lambda_2$. We vary the threshold from $4$ to $5,7,10$. The column $B$ is the ratio of the MLE lifetime $\hat{t}_{\rm decay}(\mathfrak{r})$ to those with threshold $\mathfrak{r} = 4$. Entries with a dash have fewer than three trajectories emitting under the higher threshold. The error bars come from paired bootstrap, and do not include systematic bias due to high $q$. Values of $B$ quoted without an uncertainty interval have one narrower than the last digit shown. The column $q$ records the percentage of trajectories that crossed the $\mathfrak{r} = 4$ threshold but not the higher threshold before their evolution is cut off early; this ratio does not include trajectories that were stopped by the pre-set observation time ceiling. The column $1/A$ reports only one number if all trajectories that crossed the $\mathfrak{r}=4$ threshold also reached the higher threshold; otherwise, the lower and upper bounds are listed as an interval.}
\label{tab:threshold-ratio}
\end{table}

\FloatBarrier
\section{Kaluza--Klein emission and the grey-body barrier}\label{app:greybody}
The collective quadrupolar mode studied in the main text and the emission of a D0-brane are distinct channels. A probe massless scalar in the eleven-dimensional geometry provides a tractable transmission problem for the latter: non-zero momentum on the M-theory circle becomes D0 charge upon reduction. The probe does not replace the gravitational perturbation calculation in Appendix~\ref{app:11d}. For the non-rotating background, an analysis of D0-brane emission is given in~\cite[Appendix~A]{Lin:2013jra}.

For the axisymmetric solution, introduce the angle $\varphi=\phi+\Omega_Ht/\mathcal A$, which is non-rotating in the local Myers--Perry coordinates, and write the probe as
\begin{equation}
 \Psi_{\rm probe}=e^{-{\rm i}\omega t+{\rm i}m_\varphi\varphi+{\rm i}nz/L_{11}}
 R(r)X(u)Y_\ell(S^6),\qquad u=\sin\theta.
 \label{eq:probeansatz}
\end{equation}
Here $\ell(\ell+5)$ is the $S^6$ eigenvalue and $n$ is an integer for a circle of period $2\pi L_{11}$. Under Eq.~\eqref{eq:Carroll}, the frequency and longitudinal momentum in the local product geometry are
\begin{equation}
 \varpi=\mathcal A\omega,\qquad
 k_Z=\frac{n}{\mathcal A L_{11}}-\mathcal A\omega.
 \label{eq:probefrequencies}
\end{equation}
The asymptotic radial momentum is consequently
\begin{equation}
 k_\infty^2=\varpi^2-k_Z^2
 =\frac{2n}{L_{11}}(\omega-\omega_{\rm th}),\qquad
 \omega_{\rm th}=\frac{nM_0}{L_{11}\mathcal R^7}.
 \label{eq:thresholdenergy}
\end{equation}
For $n>0$, propagation to infinity requires $\omega>\omega_{\rm th}$. This threshold and the frequency convention must be specified together.

We focus on $\ell=0$ to avoid the additional $S^6$ centrifugal barrier. The wave equation separates as
\begin{align}
 \frac1{r^6}\frac{{\rm d}}{{\rm d}r}
 \left(r^6\Delta_rR'\right)+U_rR&=\Lambda R,\nonumber\\
 (1-u^2)X''+\left(\frac1u-8u\right)X'
 -\left(\frac{m_\varphi^2}{u^2}+a^2k_\infty^2u^2\right)X&=-\Lambda X,
 \label{eq:probeequations}
\end{align}
where
\begin{equation}
 U_r=k_\infty^2(r^2+a^2)
 +\frac{(am_\varphi-2M_0\varpi/r^5)^2}{\Delta_r}
 +\frac{2M_0\varpi^2}{r^5}.
 \label{eq:probeUr}
\end{equation}
Regularity on the angular interval quantises $\Lambda$. At $a^2k_\infty^2=0$,
\begin{equation}
 \Lambda=(|m_\varphi|+2q)(|m_\varphi|+2q+7),
 \qquad q=0,1,\ldots.
 \label{eq:probeLambda}
\end{equation}
For $k_\infty^2>0$, the positive spheroidal potential raises the angular eigenvalue, so these are lower bounds on the corresponding branches.

With $R=r^{-3}\Delta_r^{-1/2}F$, the radial equation becomes
\begin{equation}
 -F''+V_rF=0,\qquad
 V_r=\frac{\Lambda-U_r}{\Delta_r}+\frac6{r^2}
 +\frac{3\Delta_r'}{r\Delta_r}
 +\frac{\Delta_r''}{2\Delta_r}
 -\frac{(\Delta_r')^2}{4\Delta_r^2}.
 \label{eq:schro}
\end{equation}
When there are two turning points enclosing a barrier, the leading WKB estimate is
\begin{equation}
 \mathcal P_{\rm trans}\sim
 \exp\left[-2\int_{r_1}^{r_2}\sqrt{V_r}\,{\rm d}r\right].
 \label{eq:WKB}
\end{equation}
This estimates the barrier exponent; it does not determine the prefactor or the full emission rate. In a superradiant window it should be interpreted through the magnitude of the tunnelling amplitude, with the horizon flux sign treated separately.

The potentially enhanced modes have negative co-rotating horizon frequency,
\begin{equation}
 \varpi-m_\varphi\Omega_H<0,
 \qquad\text{or}\qquad
 \omega<m_\varphi\Omega,\quad \Omega=\Omega_H/\mathcal A.
\end{equation}
This condition must overlap the propagation condition in Eq.~\eqref{eq:thresholdenergy}. In particular, a finite azimuthal number does not suffice at large $N$ when the Kaluza--Klein momentum grows with $N$.

To see the scaling, hold $\rho_h=r_+/\mathcal R$, $\kappa=a/r_+$, and $l_{11}=NL_{11}/\mathcal R$ fixed, with $n>0$ fixed. Define
\begin{equation}
 y=\frac r{r_+},\qquad
 q_N=\frac{nr_+}{\mathcal A L_{11}}
 =\frac{n\rho_h^{9/2}\sqrt{1+\kappa^2}}{l_{11}}N,
 \quad \widehat\omega=r_+\varpi,
 \quad \omega_0=r_+\Omega_H=\frac\kappa{1+\kappa^2}.
\end{equation}
The overlap window is $q_N/2<\widehat\omega<m_\varphi\omega_0$. Thus $m_\varphi$ must be of order $N$ or larger. Write
\begin{equation}
 m_\varphi=\left\lceil\frac{q_N}{2\omega_0}\right\rceil+\delta m,
 \qquad
 \widehat\omega=\frac{q_N}{2}
 +\left(m_\varphi\omega_0-\frac{q_N}{2}\right)\xi,
 \quad 0<\xi<1.
\end{equation}
For the following estimates we keep $\xi$ away from the endpoints of the window and consider a barrier with separated turning points.

When $1\lesssim\delta m\ll N$, the inner turning point approaches the horizon while the outer one scales as $y_2\sim\sqrt{N/\delta m}$. At fixed $y>1$, the leading potential is positive and of order $N^2$. Its long-distance tail is
\begin{equation}
 r_+^2 V_r\sim
 \frac{q_N^2(1+\kappa^2)^2}{4\kappa^2 y^2}
\end{equation}
before the outer turning point is reached. For example, using the leading angular lower bound, its coefficient at finite $y$ is positive because the numerator is proportional to
\begin{equation}
 (1+\kappa^2)\left[y^2-y^{-5}
 +\kappa^2(y^2-y^{-3})\right],
\end{equation}
which is positive for $y>1$. The logarithmic tail gives
\begin{equation}
 2\int_{r_1}^{r_2}\sqrt{V_r}\,{\rm d}r
 =\frac{q_N(1+\kappa^2)}{2\kappa}
 \log\frac{N}{\delta m}+O(N).
 \label{eq:nearminimumWKB}
\end{equation}
Replacing the angular eigenvalue by its lower bound gives a conservative leading barrier estimate in this regime. For $\delta m=O(N^\alpha)$ with $0\leq\alpha<1$, it implies an exponent of order $N\log N$.

For $\delta m=O(N)$, the frequency, longitudinal momentum, and angular momentum are all of order $N$. The leading potential is of order $N^2$ and, for fixed mode ratios with a non-degenerate barrier, the turning points remain at finite $y$. The WKB exponent is then of order $N$. Its coefficient depends on the angular eigenvalue and the mode ratios and must be determined from the full potential.

For $N\ll\delta m\ll N^2$, the leading angular barrier is instead of order $(\delta m)^2$. Its outer turning point scales as $\sqrt{\delta m/N}$, and the angular eigenvalue approaches $m_\varphi^2$ at leading order. The tail contribution therefore scales as
\begin{equation}
 2\int_{r_1}^{r_2}\sqrt{V_r}\,{\rm d}r
 =\delta m\log\frac{\delta m}{N}+O(\delta m).
 \label{eq:largemWKB}
\end{equation}
The exponent in this regime is controlled by $\delta m$, rather than by $N$ alone. These estimates show why thermodynamic enhancement does not by itself remove the large-$N$ transmission barrier for the probe channel. They do not establish stability of the rotating solution: the collective $m_\varphi=2$ gravitational mode is a separate instability, and neither its nonlinear evolution nor a complete D0 emission rate follows from this probe calculation.

\bibliographystyle{ourbst}
\bibliography{./references}
\end{document}